\documentclass[10pt,letterpaper]{report}
\usepackage[utf8]{inputenc}
\usepackage{amsmath}
\usepackage{amsfonts}
\usepackage{mathrsfs}
\usepackage{amssymb}
\usepackage{float}
\restylefloat{table}
\usepackage{multicol}
\usepackage{epsfig}
\usepackage{graphicx}
\usepackage{epstopdf}
\usepackage{xcolor}
\usepackage{atbegshi}% http://ctan.org/pkg/atbegshi
\AtBeginDocument{\AtBeginShipoutNext{\AtBeginShipoutDiscard}}
\usepackage{hyperref}
\hypersetup{%
  colorlinks=false,
}
\usepackage{apacite}
\usepackage{textcomp}
\usepackage[margin=1.0in]{geometry}
\usepackage{booktabs}
\usepackage{pdflscape}
\usepackage[tableposition=top]{caption}

\usepackage[hang,flushmargin]{footmisc}
\usepackage{epigraph}

\usepackage{setspace}
\usepackage{calligra}
\DeclareMathAlphabet{\mathcalligra}{T1}{calligra}{m}{n}
\DeclareFontShape{T1}{calligra}{m}{n}{<->s*[2.2]callig15}{}

\begin{document}

\begin{titlepage}

\begin{figure*}
\vspace*{-3cm}

\end{figure*}
   
\author{
\\{\it Written by} 
\\
\\Patrik Pirkola 
\\
\\
\\{\it Advised by}
\\
\\Dr. Horbatsch
\\
\\Thesis submitted in partial 
\\fulfilment of the requirements for the Degree of
\\
\\MASTER OF SCIENCE
\\
\\GRADUATE PROGRAM IN PHYSICS AND ASTRONOMY
\\YORK UNIVERSITY
\\TORONTO
\\
\\
\\
\\
\\
\\
\\
\\
\\
\\
\\
\\
\\
\\
\\
\\
\\
\\
\\
\\\copyright ~Patrik Pirkola, 2021}
\date{}
   
\title{Exterior Complex Scaling Approach for Atoms and Molecules in Strong DC Fields}

\end{titlepage}

\makeatletter
\newenvironment{tablehere}
  {\def\@captype{table}}
  {}
  \newenvironment{figurehere}
  {\def\@captype{figure}}
  {}

\maketitle

\clearpage
\noindent I'd like to thank my mother for instilling in me the drive to understand the universe, my uncle for helping channel that drive into physics, my sister for helping me with my mathematics courses when I needed it and with some of the proofs in this thesis, and my friends for being good sounding boards for various technical problems I had in writing this thesis. I'd also like to thank my current advisor for putting up with me over the years and my previous advisors, Dr. Hall for offering me my first research position, Dr. Haslam for helping me with the spectral integrator I used in this thesis and getting me an account on one of the multi-core computers in the math department on which I did most of my computations, and Dr. Stone for helping me figure out how to use his and Dr. Wood's code for calculating Wigner coefficients. 
\\~\\I am gracious for the ability to pursue physics, because above all the sciences it relies most heavily on the rule of nature and what it says. Thus the physicist is measured by how well they, by theoretical or experimental exercises, can commune with nature. Because of this physics is the last sane place on earth. 
\newpage
\noindent 
{\it If you wish to make an apple pie from scratch, you must first invent the universe.}
\\Carl Sagan

%\\~\\ {\it Physics, it's well important, 'innit?
%\\Ali G
%\\~\\ {\it But I've got a job to do, too. Where I'm going, you can't follow. What I've got to do, you can't be any part of. Ilsa, I'm no good at being noble, but it doesn't take much to see that the problems of three little people don't amount to a hill of beans in this crazy world. Someday you'll understand that.}
%\\Rick Blaine, {\it Casablanca}
\begin{abstract}
We perform a short review of the history of quantum mechanics, with a focus on the historical problems with describing ionization theoretically in the context of quantum mechanics. The essentials of the theory of resonances are presented. The exterior complex scaling method for finding resonance parameters within the context of the Schr\"{o}dinger equation is detailed. We explain how this is implemented for a numerical solution using a finite element method for the scaled variable. Results for the resonance parameters of a one-dimensional hydrogen model in an external direct current (DC) electric field are presented as proof of the independence of the theory from the scaling angle. We apply the theory to the real hydrogen atom in a DC field and present results which agree with literature values. The resonance parameters for singly ionized helium are also presented. Using a model potential energy for the water molecule, we solve for the energy eigenvalues. We then solve for the resonance parameters of the water molecule in a DC field and compare to literature results. Our widths for the valence orbital are shown to agree well with the so-called "coupled-cluster singles and doubles with perturbative triples excitations" method.

\end{abstract}
\tableofcontents

\chapter{Motivations}
\epigraph{If I have seen further it is by standing on ye shoulders of Giants.}{Sir Isaac Newton}
\section*{Overview}
In $\S \ref{section:1point1}$ we introduce a history of the ideas which were required to introduce the possibility of ionization of atoms and molecules within a theoretical framework based on quantum mechanics.
We also discuss some elements of the Stark shift which is the foundation for ionization in constant or direct current (DC) electric fields. We introduce the idea of perturbation theory, which can be used to solve for the Stark shift, and explain why it fails in describing tunnelling.
In $\S \ref{section:1point2}$ we motivate the idea of resonances which are required to formally describe the tunnelling of an electron from an atom/molecule. 
In $\S \ref{section:1point3}$ we discuss the exterior complex scaling (ECS) method for obtaining resonances by scaling the radial variable into the complex plane.
In $\S \ref{section:1point4}$ we discuss boundary conditions, definitions of integrals with scaling, and our integration methods.
In $\S \ref{section:1point5}$ we discuss the implementation of a finite element method which allows the explicit admission of the wave function discontinuity which arises at the point of scaling. In $\S \ref{section:1point6}$ we solve the resonance problem of 1D hydrogen. In $\S \ref{section:1point7}$ we discuss our conclusions.
%\section*{Citations}
%Books:
%\\~\\Sitenko - Scattering Theory
%\\~\\Taylor - Scattering Theory
%\\~\\Press et al - Numerical Recipes 
%\section*{Notes}
%\section*{Done}
%- Chebyshev integrator
%\\~\\ - Parallel integrator
%\\~\\ - Create $f_{im}$ from $h_{m}$ 
%\\~\\ - Generalize to global H matrix
%\\~\\ - Solved eigenvalue problem
%\section*{To Do}
%- Add symmetry to program
%\subsection{Major goals}
%** Write Introduction
%\\~\\*** Write ECS section
%\\~\\*** Work out finite element method
%\subsection{Minor goals}
%- Finish plot with widths
\section{Introduction}\label{section:1point1}
After Newton, {\it non-relativistic classical} notions dominated our understanding of physics. By non-relativistic, we mean particles moving at speeds much less than the speed of light. By classical, we mean that given a particle $A$'s position and momentum at any given moment, that same information for any particle interacting with it, along with the forces between them, we can establish all future evolution of such a particle {\it A} \cite{arminhenning2006}. Such a universe could be called a {\it clockwork universe} as all the parts in the universe would move automatically in a predictable order, like the mechanism of a clock. It was only later, at the beginning of the twentieth century, thanks to the heroic efforts of Einstein, Schr{\"o}dinger, Bohr, Born, Heisenberg, Dirac, and others that we began to understand how this theory failed. It was replaced by what became known as {\it quantum theory}, from the Latin {\it quantus} ("how great"), referring to the smallness of the smallest fixed energy that very light particles (usually fundamental to the universe) were found to have in certain circumstances that the theory described \cite{milonni94, bohm89}. 
\\~\\We could no longer expect the universe to be so well behaved. It turned out that things like a particle's position and momentum were somewhat naive concepts, and were only fixed before measurement at what was later known at the {\it classical scale}, involving heavy objects. Underneath this universe of expected perfect precision lay a swamp of fogginess. That swamp was where the fundamental particles  like the electron, really lived: where their position and momentum were not known before measurement, are up to chance, and are never seen simultaneously with perfect precision. Instead, if one knew a particle's position perfectly, its momentum would become totally uncertain, and vice versa. This was so at odds with thinking at the beginning of the twentieth century, that even Einstein, one of the earlier pioneers for quantum theory, rejected this inevitable conclusion of the theory. In a letter to Max Born in December of 1926, Einstein wrote, {\it The theory says a lot, but does not really bring us any closer to the secret of the "Old One." I, at any rate, am convinced that He is not playing at dice} \cite{A05}. He and others also tried to debunk quantum theory by showing it had certain paradoxes when looked at from a classical point of view. This became 
the famous "EPR" paradox. The paradox was named after a paper by Einstein, Podolsky, and Rosen. They showed that the states of two particles could be correlated, in the sense that the measurement of one could determine the other non-locally \cite{brunner2014, E35}. Today the phenomenon they discovered is more generally called quantum entanglement.
\\~\\Before this radical change in thinking, Dalton's theory of atoms established the beginning of the modern understanding of chemical elements in the nineteenth century. Dalton proposed that each chemical element was a unique, tiny hard ball, or atom, that could interact classically with other elements through an atmosphere of caloric, but could not be split \cite{gros17, Ro78}. Since these elements could not be split, it was impossible to imagine a process such as ionization based on this model. It turned out to be true that an atom could be imagined as a tiny hard ball in certain circumstances, but it actually consists of a core of positive charge, surrounded by negatively charged electrons some distance away. During ionization an electron can be ejected from a chemical element. This requires some external energy to be provided for the electron to escape the attractive force of the center of the element, or the nucleus. The simplest nucleus is just a single proton, the basic unit of positive charge inside an atom.
Although the proper mechanism for ionization was qualitatively described by Thomson in 1899 \cite{ach91}, the earliest mathematical theory that even incorporated this possibility was known as the Bohr model, in which electrons have circular orbits around the nucleus. It received its namesake from Niels Bohr who developed the model and published a three-part series on it in 1913 \cite{Bohr19131, Bohr19132, Bohr19133}. %His model took the classical framework in which electrons and a nucleus could interact, and modified it to form a stable atom, by applying a {\it quantum} restriction on the angular momentum of the electron, meaning, that the electron had to zip around the nucleus at fixed velocities, which translated into fixed energies and distances from the nucleus \cite{shankar95} \cite{bohr1913}. In other words, the electrons orbited the nucleus steadily in circles (or more specifically ellipses in the Bohr-Sommerfeld model). Without this restriction, the electron was classically doomed to spiral into the nucleus and admit infinite radiation while doing so! Despite its success in explaining early spectroscopic observations (measurements of the energy differences between the fixed levels by means of radiation from the atom) this theory, which became known as a part of the {\it old quantum theory}, was usurped.
\\~\\The electron and nucleus act as tiny hard balls part of the time. However, we now know that, at least as far as an atom is concerned, the electron doesn't behave this way at all in a stable configuration. Rather, in a stable configuration, the electron has some probability to be found somewhere around the atom. The electron is usually quite close to it, considering the electron is really supposed to be bound to the nucleus. These stable configurations are often called {\it clouds of probability} and replaced the circular orbits. One can quantify exactly how probable it is to find the electron in any small cubic (or otherwise shaped) region around the nucleus. 
\\~\\Today we understand that electrons "orbit" around the nucleus in clouds of probability at fixed energy levels. The specific geometry of the cloud at a given energy level is called an {\it eigenstate}, and the energy level is called an {\it eigenenergy}. These words are derived from the type of mathematical object that defines the basic theory of non-relativistic quantum mechanics, the {\it eigenvalue} problem known as the time-independent Schr{\"o}dinger equation, given in compact form as $\hat{H} \Psi = E \Psi$, where $\hat{H}$ is an {\it operator} also known as the {\it Hamiltonian operator}. The Hamiltonian acts on an eigenstate $\Psi$ to return the eigenenergy $E$ multiplied by the eigenstate. For a one-electron system, the quantity 
\begin{equation*}
P(x_o,y_o,z_o) = \int_{x_o-\varepsilon}^{x_o+\varepsilon}\int_{y_o-\varepsilon}^{y_o+\varepsilon}\int_{z_o-\varepsilon}^{z_o+\varepsilon}|\Psi(x,y,z)|^2dV,
\end{equation*}
 which is basically the square of the eigenstate (called a {\it probability density}) multiplied by a small volume, quantifies the the probability of finding a particle in that same region surrounding the point $(x_o,y_o,z_o)$ in space \cite{scherrer}. In this case, the number $\varepsilon$ controls the size of a box containing the region.
\\~\\ In the same century as Dalton, Mendeleev developed the periodic table of elements, which organized the elements in similarity groups. Eventually the modern periodic table of elements developed, in which the columns of the table are groups of atoms which share the same number of electrons at the highest energy level. This level is called the {\it valence level}. When one ionizes an atom, the first electron to exit is from this energy \cite{pet17}. This can be done with lasers or a strong enough DC electric field \cite{laso16}. In the case of the DC electric field at low field strengths this effect is important in the history of quantum theory, and is known as the {\it Stark effect} \cite{scherrer}.
\\~\\The Stark effect is a classic problem in quantum mechanics that involves perturbation calculations in the realm of Schr{\"o}dinger theory. Perturbation theory involves the expansion of a solution in terms of a known quantity plus higher-order corrections. %That is to say, suppose one knows $E_o$ and $\Psi_o$ for the Hamiltonian $\hat{H}_o$ and wants to solve for $E$ and $\Psi$ for $\hat{H} = \hat{H}_o + \epsilon\hat{H}_p$, where it is understood that $\epsilon$ is small. Then one assumes $E = E_o + \epsilon E_1 + \epsilon^2 E_2 + \epsilon^3 E_3 +h.o.c.$ and $\Psi = \Psi_o + \epsilon\Psi_1 + \epsilon^2 \Psi_2 + \epsilon^3 \Psi_3 +h.o.c.$, where $h.o.c.$ are all the corrections higher than order $\epsilon^3$ \cite{scherrer}. The term $\hat{H}_o$ is understood to be the first approximation to the Hamiltonian, usually just the potential energy of the nucleus/nuclei (the energy that supplies an attractive force on an electron) plus the kinetic energy of the electron, which is the energy that the electron has intrinsically because of its momentum \cite{scherrer}.
\\~\\The first approximation one usually deals with for the case of hydrogen is the simple problem with only the Coulomb potential (the potential energy of a single proton). This potential leads to energy levels that only depend on one quantum number, that is, the quantum number $n$, because of the symmetries of the Hamiltonian of hydrogen. One finds, from early experiments such as the detection of the anomalous Zeeman shift, that this cannot be the only potential involved. In the anomalous Zeeman effect, the electron energy levels begin to split based on the quantum number $l$. This is because in the electron rest frame there is a magnetic field generated by the orbit of the nucleus perpendicular to the plane of orbit. The electron has an intrinsic magnetic moment due to spin which couples to the magnetic field and produces an energy shift based on the electron's quantum number $l$, which is the electron's angular momentum from the proton's rest frame. This effect is called spin-orbit coupling \cite{sakurai17}.
\\~\\This is generally dealt with by considering that the potential energy for spin-orbit coupling has an effect of generating an energy shift which is significantly smaller than the unperturbed energy (the $n$ dependent energy).  The energy of the spin-orbit coupling sits under the umbrella of what are called perturbing energies $\epsilon\hat{H}_p$ where $\epsilon$ is a small real number and $\hat{H}_p$ is  generally a {\it Hermitian} operator, meaning its eigenvalue problem only produces real eigenvalues, or more specifically eigenenergies \cite{sakurai17}. Thus the Zeeman effect became another classic problem in perturbation theory.
\\~\\One will additionally note that if $\hat{H}_p$ is not dependent on time, like in the Stark shift, one, under the suitable conditions, uses what is called time-independent perturbation theory. However, one can find a realistic problem with a strong enough external electric field (such that ionization is allowed) that prevents us from using time-independent perturbation theory to calculate the Stark shift. The first issue is obviously that the external electric potential energy might not be so small in comparison to the unperturbed Hamiltonian. The second is that using standard perturbation theory with a perturbation of the form $\epsilon \hat{H}_p = -eF_oz$ as in the Stark shift \cite{scherrer}, one expects it produces real shifts in the eigenvalues if $F_o$ is real (we will soon show that complex eigenvalues are required if one is to describe ionization in a time-independent formalism). 
\\~\\More generally and concretely, as long as $\epsilon \hat{H}_p$ is Hermitian, then the first ($E_1$) and second order ($E_2$) energy corrections in the non-degenerate case (all eigenvalues of the unperturbed operator $\hat{H}_o$ are unique) can be shown to always be real \cite{mitnotes}. This is because $E_1$ is the expectation value of a Hermitian operator, and $E_2$ is the sum of square-norms (which are always real) of matrix elements divided by energy differences between eigenvalues of $\hat{H}_o$ \cite{scherrer}. %Reality of energy corrections can be shown for all orders for the degenerate case \cite{byron71}. 
\\~\\This makes sense since if $\epsilon \hat{H}_p$ is Hermitian, $\hat{H}_o + \epsilon \hat{H}_p$ must be as well. Therefore the exact solution has real eigenvalues \cite{byron71}. Obviously this fact combined with the reality of the first two energy corrections provided by non-degenerate perturbation theory does not prove all orders of correction are real and totally exclude the use of time-dependent perturbation theory. Neither have we discussed degenerate perturbation theory.
\\~\\However, we know that if the external electric field is strong enough, the magnitude of the imaginary part of the eigenvalue is comparable to the real part (see $\S \ref{section:2point5}$). For this case this excludes the use of time-independent perturbation theory since the corrections must be small in comparison to the unperturbed energy. Additionally, the restriction that the exact result for the perturbed energy be real excludes any possible imaginary corrections from having any likely physical meaning in regards to ionization (since the exact problem does not describe ionization anyway). Something else must take the place of time-independent perturbation theory.
\\~\\The states that are produced for real eigenvalues are fixed to have a time-evolution of the form $\exp[-iEt/\hbar]$, since the basic assumption that transforms the time-dependent Schr{\"o}dinger equation to a time-independent eigenvalue problem is that $\Psi(\vec{r},t) = \Psi(\vec{r})\exp[-iEt/\hbar]$. The square-norm of the time-depenence is then $\exp[-iEt/\hbar]^*\exp[-iEt/\hbar]=\exp[iEt/\hbar]\exp[-iEt/\hbar]=1$, so the probability density and subsequently probability of the state have no time-evolution and can be called static. Therefore, one cannot describe the ionizing effect that a strong electric field would have on the atom since the time-evolution of such a state is approximately, $\exp[-i(E-i\frac{\Gamma}{2})t/\hbar]$ \cite{sit91}.
\\~\\Why is ionization even possible? The interesting effect of a strong DC field is that it deforms the effective potential in such a way as to transform, though only in the direction of the field one of the walls of the Coulomb potential into a hill which has an opening on one side. Classically, the electron can overcome the potential energy hill given sufficient energy. This is not the only way.
%\begin{figure*} [!h]
%\centering
%\includegraphics[width=250pt]{fig2.eps}
%\vspace{0.2cm}
%\caption{\label{fig:fig2}
%Diagram of the potential energy of 1D Hydrogen with a DC electric field. The dotted lines represent %places where one can introduce a CAP or complex scaling (discussed more below).}
%\end{figure*}
\\~\\Quantum mechanics tells us that deformation of the wall allows the wave function of the electron ($e^{-}$) to stretch out meaningfully into the new open region, in the sense that the $e^{-}$ is allowed to tunnel through the barrier. The wave function that appears on the other side of the barrier oscillates somewhat like a free (quantum) particle. As it goes farther down into negative region of the electric potential the oscillations become faster. This can be understood from the classical total energy: given energy conservation, as the potential energy becomes more negative the positive energy lost becomes kinetic energy. 
\\~\\One finds the so-called resonance parameters by transforming the time-independent Schr{\"o}dinger equation for the problem involving the external field using a complex absorbing potential (CAP) or exterior complex scaling (ECS) to make the problem non-Hermitian. If $H\Psi = E\Psi$ produces complex eigenvalues, then time-dependence of the form $\exp[-i(E-i\frac{\Gamma}{2})t/\hbar]$ is enforced without solving the time-dependent form of Schr{\"o}dinger's equation or using time-dependent perturbation theory.\footnote{This produces an assumption that the time-dependence of a resonance is exponential. However, it can be shown to be approximately true under a certain definition of a resonance wave packet \cite{sit91}. Additionally, in Chapter 3 we show that the solution for a resonance using ECS under the time-dependent form of the Schr{\"o}dinger equation can be well-fit to an exponential outside of a turn-over region where ionization begins, confirming the eigenvalue model's usefulness.} This is the striking feature of non-Hermitian quantum mechanics: a state that has meaningful time evolution, in the sense that it actually decays, can be solved for using a formalism that was created to produce states in equilibrium. One may note that both methods (CAP and ECS) actually frame the problem as a normalizable state at each instant of time, since the nature of the methodologies effectively damps the oscillations on the right side in the ionization region. Thus in time, one does not solve for outgoing states, as much as one describes the decaying square-norm of the intermediate bound state, which is described by $\sim\exp[-\Gamma t/\hbar]$. An important real world application of this is the interest of the final chapter of this thesis: the ionization of water.
\section{Resonances}\label{section:1point2}
%** some background on scattering
%\\~\\** perhaps mention s-i functions

%** general stuff about resonance
Resonances are of interest in any situation that involves an intermediate quantum mechanical state that eventually decays. In particular, a cancer therapy known as radiation therapy involves the ionization of water surrounding cancer cells in the human body. One way of doing this is to position kilovoltage x-ray radiation sources at different angles in such a way that they all coalesce at the site of the cancer cells. Water in the region is ionized, and hydroxyl radicals are formed which damage the cancer DNA, causing cell death. In standard radiation therapy, about 50 to 70 percent of DNA damage is mediated by hydroxyl radicals upon x-ray irradiation \cite{hill14,roselli14,baskar2012}.
\\~\\The ionization process in a laser field can be understood to involve three steps: first the electron is excited, that excited state decays, and then the free electron either stays free or returns to the original atom/molecule \cite{feynman1}. In static fields the intermediate state also decays. The complex scaling method gives complex eigenvalues which give a time-evolution that allows for decay \cite{razavy2003,r82}. Therefore it only describes the decay of the intermediate quasi-stationary state and does not involve or allow for a description of anything else. In this sense it is somewhat phenomenological, for two reasons: (1) the complex scaled Hamiltonian is not derived from a fundamental axiom of quantum theory but simply imposed by the co-ordinate transformation, and (2) the eigenstate describing the intermediate state decays exponentially but does not transform into some outgoing state. Reason (2) is fundamentally at odds with the axiom of norm conservation in quantum mechanics. Norm conservation demands that a particle like an electron should have a probability of 100 percent to be found somewhere. 
\\~\\Furthermore, since the eigenstate is statistical in nature its decay doesn't indicate how long a single electron takes to ionize but indicates how an ensemble, meaning a very large amount of atomic/molecular systems each with its own ionizing electron, would take to completely ionize. However, since one deals with a large number of water molecules in the human body in the case of radiation therapy, this is a good model after all.
\\~\\In our work we will be using a method proposed by Scrinzi \cite{s10,se93} to solve for the resonant parameters of hydrogen in a DC electric field in both a static (Chapter 2) and time-dependent formalism (Chapter 3). We will also briefly look at how to carry our methodology with the analogue problem for water. We will expand upon previous work in the field \cite{laso16} that solved such a problem using a model potential energy that they derived from a self-consistent field calculation. We will be using a specific model potential energy for the water molecule which was derived by another group \cite{ice11}. In this work, we solve the non-Hermitian eigenvalue problem for water in a DC electric field using this model potential (Chapter 4). In our appendix, we discuss the details of self-consistent field calculations. Since we use a DC field, these results are not directly applicable to x-ray radiation.
%\\~\\In regards to how realistic this formulation is for a real-world problem like radiation therapy, one first notes that the DC electric field is a good approximation for sufficiently low frequency radiation. This is because locally in time, a sinusoidal wave, which is how the potential energy of a laser behaves, has the same potential energy as a DC electric field. This would prove to be a good approximation for x-ray radiation, which is used in radiation therapy \cite{hill14}, if the ionization time scale of water is much shorter than the inverse frequency of x-ray radiation. In Chapter 4 we find decay widths of order $10^{-3}~au$ to $10^{-2}~au$ for the valence orbital of water up to a field strength of $0.14~au$, which can be related to a time-scale of order (at the lower range) $10^{-15}$ seconds. The inverse frequency of x-ray radiation is from $10^{-16}$ to $10^{-20}$ seconds. Therefore the DC field approximation is not consistent with x-ray radiation with the field strengths we computed. The approximation may be consistent with higher DC field strengths as ionization would occur faster in those situations. Otherwise the use of an AC field in modelling is required.
\\~\\The mathematical formulation of the general form of a resonance is not overly difficult. Suppose that an intermediate state of a scattering problem that we are interested in has a low decay probability, meaning it is generally long-lived. The state can be called {\it quasi-stationary}. Generally for a specific scenario (say, an atom or molecule with one or more electrons) there is a set of {\it quasi-discrete} states that have broadened energy levels $E_r$ with widths $\Gamma_r$. These make up the spectrum of the complex-scaled Hamiltonian. The widths are related to characteristic decay time by the usual relationship $\Gamma_r = \hbar/\tau_r$ \cite{sit91}. In general we can approximate the decaying state by using the following basis states which are valid wave functions outside the interaction region \cite{sit91},
\begin{equation}
\psi_E(r,t)\sim \frac{1}{r}\bigg(e^{-ikr}-S(k)e^{ikr}\bigg)\exp\left[-i\frac{Et}{\hbar}\right],
\end{equation}
where $S(k)$ is called the {\it scattering matrix}.
We note that an actual resonant state associated with a pole in the fourth quadrant has the following momentum and energy \cite{sit91}:
\begin{equation}
k_r = \sigma_r + i\lambda_r,
\end{equation}
and,
\begin{equation}
E_r = \frac{\hbar^2}{2\mu}(\sigma^2_r -\lambda^2_r),
\end{equation}
where $\mu$ is the mass (or reduced mass) of the particle that occupies the state.
The resonance width is given by
\begin{equation}
\Gamma_r = -\frac{2\hbar^2}{\mu}\sigma_r\lambda_r > 0.
\end{equation}
In summary the complex eigenvalue of such a state is 
\begin{equation}
E = E_r - i\frac{\Gamma_r}{2}.
\end{equation}
The above assumes that $\sigma_r$ is positive and $\lambda_r$ is negative such that $\Gamma_r$ is postive \cite{sit91}.
\\~\\The game is then to project our trial wave packet describing the resonant state onto these states, but only in the domain in energy space with width $\Delta$ surrounding the resonant energy $E_r$. The width satisfies the condition that $\Gamma_r << \Delta << D$ where $D$ is the average distance between associated quasi-discrete levels of the problem.
\begin{figure*}
\centering
\includegraphics[width=300pt]{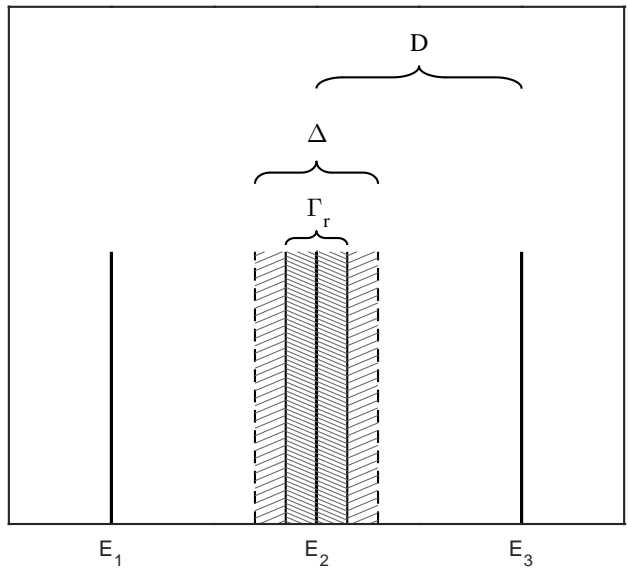}
\vspace{0.2cm}
\caption{\label{fig:fig1}
Diagram of the various widths involved in the construction of a wave packet describing a resonant state. The dark lines represent the sequence of various resonance positions in the quasi-discrete spectrum.}
\end{figure*}
\\~\\We find a wave packet localized inside the interaction region given by \cite{sit91}
\begin{equation}
\psi(r,t) = \int_{E_r-\Delta/2}^{E_r+\Delta/2}dE~a(E)\psi_E(r,t).
\end{equation}
After some work one finds the result that 
\begin{equation}
\psi(r,t)\sim \exp\left[i\sigma r-\frac{i}{\hbar}E_rt-\frac{\Gamma_r}{2\hbar}\left(t-\frac{r}{v_r}\right)\right]
\end{equation}
where $v_r = \hbar \sigma/\mu$ \cite{sit91}.
Simply looking at the time-dependent portion, we get,
\begin{equation}
\psi(t)\sim \exp\left[-\frac{i}{\hbar}E_rt-\frac{\Gamma_r}{2\hbar}t\right].
\end{equation}
Thus the wave function has exponential decay. One might imagine that the eigenvalue problem for such a state would not be Hermitian, as the eigenvalue $E_r-i\frac{\Gamma_r}{2}$ is clearly complex. In fact to solve for the resonance parameters $E_r$ and $\Gamma_r$ by an eigenvalue problem, one would have to somehow make the problem manifestly non-Hermitian.
\\~\\Two such methods have been developed in somewhat recent times. The first is the method of complex absorbing potentials (CAPs), in which a somewhat arbitrary imaginary valued potential is introduced into the Hamiltonian \cite{ja85}. This produces a complex eigenvalue as expected. After using the Riss-Meyer method \cite{rm93} for removing the artefact of the CAP using perturbation theory, one finds a good approximation to the resonant parameters. However, the eigenvalues this method produces are dependent on the strength of the CAP, which is why the corrective method of Riss and Meyer is required. This in turn requires the eigenvalue problem to be computed many times, which can be computationally expensive. To solve this problem, exterior complex scaling (ECS) was introduced. In addition to providing complex eigenvalues, both methods also provide the ability to model the wave function without too much difficulty. This is because the strong oscillations that appear outside the potential well due to the external field become damped (see Fig. $\ref{fig:ECSind}$).
\\~\\ECS involves scaling the radial variable of the Hamiltonian. Consider for example a 1D problem in $x$. We can continue a variable $x$ into the complex plane outside of the so-called "scaling radius" $x_o$ in the following way \cite{s10}:
\begin{equation}\label{eq:scale}
x \rightarrow \tilde{x} = \begin{cases}
    x, ~|x|<x_o,\\
    e^{i\xi}(x+ x_o)- x_o, ~ x<-x_o,\\
     e^{i\xi}(x- x_o)+ x_o,~ x>x_o.\\
  \end{cases}
\end{equation}
\begin{figure*}
\centering
\includegraphics[width=300pt]{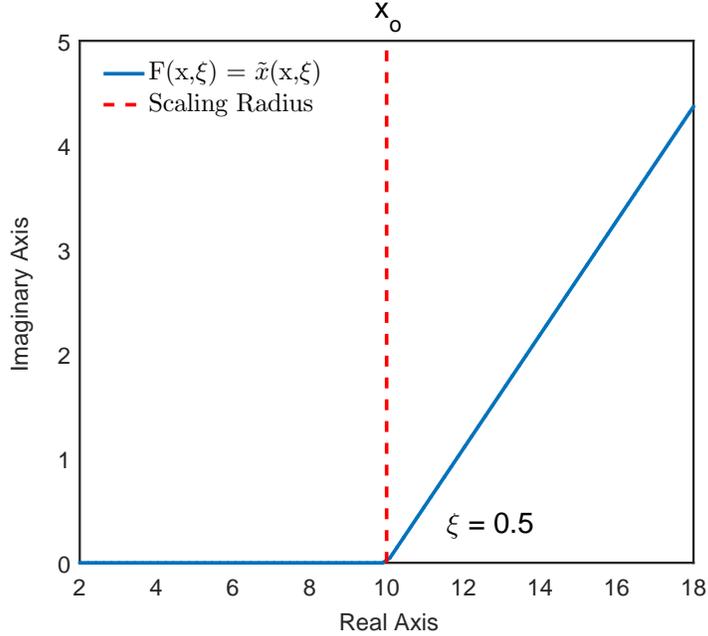}
\vspace{0.2cm}
\caption{\label{fig:zvsx}
Plot of $\tilde{x}$ in the first sector. Here the scaling radius is positioned at $x = 10~au$. For a 1D hydrogen model in an external field $F = -eF_o x$, with $F_o=0.11$ this sits somewhat past the peak of the potential barrier, meaning that the potential well region where the bound part of the wave function sits is unaffected. The scaling angle is 0.5 $rad$, so the path is neither too real nor too imaginary. For the most part, the results are not changed by a change in the scaling radius.}
\end{figure*}
One can see an example of what this looks like in Fig. $\ref{fig:zvsx}$ below.
\\~\\For a 3D problem, the radial axis is scaled, this time outside for $r>r_o$,
\begin{equation}\label{eq:scale3}
r \rightarrow \tilde{r} = \begin{cases}
    r, ~r<r_o,\\
     e^{i\xi}(r-r_o)+ r_o,~ r>r_o.\\
  \end{cases}
\end{equation}
This procedure also produces complex eigenvalues, and if one implements the method correctly the results are not very dependent on the strength of the scaling, the scaling angle $\xi$ (see $\S \ref{section:1point6point1}$ and $\S \ref{section:2point5}$). Therefore the problem can be solved more easily: for only one scaling angle.
\\~\\In this thesis we end up mixing multiple conventions for indices, following both the conventions of linear algebra, where $i$ and $j$ represent the rows and columns of a matrix respectively, and the convention where $i$ represents an subinterval/interval (we will use these words interchageably) in $x$ or $r$ for the finite element method \cite{se93}. Additionally, $m$ represents the order of basis function in the finite element approach. One will note wherever $i$ and $j$ appear together (often representing $bra$ and $ket$ in Hamiltonian matrix elements), we are following linear algebra convention, and wherever $i$ and $m$ appear together, representing the subinterval and order of basis function, we are following the convention of our reference \cite{se93}. Additionally, $i$ represents the unit imaginary number. This should be clear in context. Solutions are given in combined gaussian and atomic units. In particular $e^2/4\pi\epsilon_o \rightarrow e^2$, and further, $\hbar = m_e = e = 1$. For simplicity, we will just refer to this as atomic units.
\section{Transformations Due to Exterior Complex Scaling}\label{section:1point3}
\subsection{Scaling Path and Measure}
Exterior complex scaling can be understood to be a limit of a specific family of paths (which define the movement of the real co-ordinate variable in the problem into the complex plane) in the smooth exterior scaling method. One parametrization, which leads to the ECS path in a specific limit is,
\begin{equation}
F(x,\xi,\lambda) = x + \bigg(e^{i\xi}-1\bigg)\bigg[x+\frac{1}{2\lambda}\ln\bigg(\frac{\cosh[\lambda (x-x_o)]}{\cosh[\lambda(x+x_o)]}\bigg)\bigg].
\end{equation}
Taking $\lambda \rightarrow \infty$ one finds the ECS path \cite{Moi98}, and re-writing $F$ as $\tilde{x}$, we have,
\begin{equation*}
x \rightarrow \tilde{x} = \begin{cases}
    x, ~|x|<x_o,\\
    e^{i\xi}(x+ x_o)- x_o, ~ x<-x_o,\\
     e^{i\xi}(x- x_o)+ x_o, ~ x> x_o.\\
  \end{cases}
\end{equation*}
If we define
\begin{equation}
f(x,\xi,\lambda) = \frac{\partial F(x,\xi,\lambda)}{\partial x},
\end{equation}
then the integration measure along the ECS path is \cite{Moi98}
\begin{equation}
d\tilde{x} = f(x,\xi,\lambda)dx.
\end{equation}
\subsection{Operators}
We want to calculate the transformation of first and second derivative operators under ECS. 
\\~\\Formally, in $N$ dimensions, if we let
\begin{equation}
\tilde{x}(\vec{x}) = \tilde{x}(x_1....x_N),
\end{equation}
then the standard transformation applies for derivatives, namely the chain rule \cite{stewart5e},
\begin{equation}
\frac{\partial u}{\partial \tilde{x}_i} = \sum_{k=1}^{N}\frac{\partial u}{\partial x_k}\frac{\partial x_k}{\partial \tilde{x}_i}.
\end{equation}
We note that for our problems only one variable will be scaled. Still, the problem of explicitly calculating $\partial x/\partial \tilde{x}$ is not all that straight forward.
\subsubsection{Finding $\partial x/\partial \tilde{x}$}
Although it is harder to explicitly calculate $\partial x / \partial \tilde{x}$ one can find $\partial \tilde{x} / \partial x$ easily as it is defined in the previous subsection. For the chain rule this produces,
\begin{equation} 
\frac{\partial u}{\partial x} = \frac{\partial u}{\partial \tilde{x}}\frac{\partial \tilde{x}}{\partial x} =\frac{\partial u }{\partial \tilde{x}}f(x,\xi,\lambda).
\end{equation}
We can use this result to define $\partial x / \partial \tilde{x}$ properly. Multiplying through by $f^{-1}(x,\xi,\lambda)$ we get,\footnote{Here $f^{-1}$ denotes the reciprocal of the function and not the inverse function.}
\begin{equation}
\frac{\partial u}{\partial x}f^{-1}(x,\xi,\lambda) = \frac{\partial u}{\partial \tilde{x}}\frac{\partial \tilde{x}}{\partial x}f^{-1}(x,\xi,\lambda) =\frac{\partial u }{\partial \tilde{x}}.
\end{equation}
Since
\begin{equation}
\frac{\partial u}{\partial \tilde{x}}\frac{\partial \tilde{x}}{\partial x}f^{-1}(x,\xi,\lambda) =\frac{\partial u }{\partial \tilde{x}},
\end{equation} 
we can do a re-write under limits. Suppose $\lim\limits_{h \to 0}g(x,\xi,\lambda,h)$ = $f^{-1}(x,\xi,\lambda)$, which is a true assumption,\footnote{If this is true this demands $f^{-1}(x,\xi,\lambda)$ be under the class of functions which can be expressed as limits of other functions. However, this is not necessarily true for $f^{-1}(x,\xi,\lambda)$ {\it a priori}. However, this assumption produces an answer for $f^{-1}(x,\xi,\lambda)$. Since $f^{-1}(x,\xi,\lambda)$ must be the inverse of $\partial \tilde{x}/\partial x$ and inverses of functions are unique, our solution for $f^{-1}(x,\xi,\lambda)$ will be the only solution, meaning it definitely can be expressed as a limit of another function.} then,
\begin{equation}
\frac{\partial u}{\partial \tilde{x}}\lim\limits_{h \to 0}\frac{F(x+h,\xi,\lambda)-F(x,\xi,\lambda)}{h}\lim\limits_{h \to 0}g(x,\xi,\lambda,h) =\frac{\partial u }{\partial \tilde{x}}.
\end{equation} 
The multiplication of the limits must be one, to satisfy the previous equation, so,
\begin{equation}
\left[\lim\limits_{h \to 0}\frac{F(x+h,\xi,\lambda)-F(x,\xi,\lambda)}{h}\right]^{-1}=\lim\limits_{h \to 0}g(x,\xi,\lambda,h).
\end{equation} 
By the limit quotient rule \cite{stewart5e} we can move the inverse to the argument of the limit,\footnote{\begin{equation*}\lim\limits_{h \to 0}[k(x,h)]^{-1}=\lim\limits_{h \to 0}\frac{1}{k(x,h)}=\frac{\lim\limits_{h \to 0}1}{\lim\limits_{h \to 0}k(x,h)}=\frac{1}{\lim\limits_{h \to 0}k(x,h)}=\left[\lim\limits_{h \to 0}k(x,h)\right]^{-1}
\end{equation*}}
\begin{equation}
\lim\limits_{h \to 0}\left[\frac{F(x+h,\xi,\lambda)-F(x,\xi,\lambda)}{h}\right]^{-1}=\lim\limits_{h \to 0}g(x,\xi,\lambda,h),
\end{equation} 
and so,
\begin{equation}
\lim\limits_{h \to 0}g(x,\xi,\lambda,h)= \lim\limits_{h \to 0}\frac{h}{F(x+h,\xi,\lambda)-F(x,\xi,\lambda)} = \frac{\partial x}{\partial \tilde{x}}.
\end{equation}
So it must be that
\begin{equation}
f^{-1}(x,\xi,\lambda)=\frac{\partial x}{\partial \tilde{x}}.
\end{equation} 
\subsubsection{The First Derivative}
The chain rule therefore produces
\begin{equation}
\frac{\partial u}{\partial \tilde{x}} = \frac{\partial u}{\partial x}\frac{\partial x}{\partial \tilde{x}} =f^{-1}(x,\xi,\lambda)\frac{\partial u }{\partial x}.
\end{equation}
Or for the ECS case,
\begin{equation}
\frac{\partial u}{\partial \tilde{x}} = \frac{\partial u}{\partial x}\frac{\partial x}{\partial \tilde{x}} =e^{-i\xi}\frac{\partial u }{\partial x}, ~x > x_o.
\end{equation}
\subsubsection{The Second Derivative}
At the second derivative level the transformation is derived from applications of the chain rule and the product rule:
\begin{equation}
\frac{\partial }{\partial \tilde{x}} \frac{\partial u}{\partial \tilde{x}} = \frac{\partial }{\partial \tilde{x}}\bigg[\frac{\partial u}{\partial x}\frac{\partial x}{\partial \tilde{x}}\bigg],
\end{equation}
or,
\begin{equation}
\frac{\partial^2 u}{\partial \tilde{x}^2} = \bigg(\frac{\partial^2 u}{\partial x\partial x}\frac{\partial x}{\partial \tilde{x}}\bigg)\frac{\partial x}{\partial \tilde{x}}+\frac{\partial u}{\partial x}\frac{\partial ^2 x}{\partial \tilde{x}^2}.
\end{equation}
For the scaling this becomes,
\begin{equation}
\frac{\partial^2 u}{\partial \tilde{x}^2} = \bigg[f^{-2}(x,\xi,\lambda)\frac{\partial^2 u}{\partial x^2}-f^{-3}(x,\xi,\lambda)f_x(x,\xi,\lambda)\frac{\partial u}{\partial x}\bigg].
\end{equation}
In the ECS case, $f_x=0$ so
\begin{equation}
\frac{\partial^2 u}{\partial \tilde{x}^2} = f^{-2}(x,\xi,\lambda)\frac{\partial^2 u}{\partial x^2},
\end{equation}
and that is,
\begin{equation}
\frac{\partial^2 u}{\partial \tilde{x}^2} = e^{-2i\xi}\frac{\partial^2 u}{\partial x^2}.
\end{equation}
\section{Further Considerations}\label{section:1point4}
Some of the things we discuss here will be revisited, but for now we present a treatment of some basic technical concerns. This section can be considered as a reference for some of the finer points of our method.
\subsection{Boundary Conditions}
First, there is the concern of boundary conditions for wave functions. Considering a basic expansion of radial functions and spherical harmonics, we can define the wave function multiplied by $r$ under ECS as,
\begin{equation}
\Psi (\tilde{r},\theta,\phi)\tilde{r} = \psi(\tilde{r},\theta,\phi) = \sum_{i,l,m}c_{ilm} g_i(r)Y_{lm}(\theta,\phi),
\end{equation}
where it must be that the coefficients $c_{ilm}$ make sure the right side of the equation is really scaled.
One therefore expects that $\psi(\tilde{r},\theta,\phi)$ goes to zero at $r=0$ (since $\tilde{r} = r$ at the origin). Note that generally one expects hydrogen-like wave functions to go like \cite{scherrer}
\begin{equation}
\psi(r,\theta,\phi) \sim r^{l+1} \dots
\end{equation}
so the boundary conditions at $r=0$ should respect this for hydrogen-like systems using the ECS path as the scaling applies outside a certain radius. We find that the finite element method, which we will discuss shortly, only requires $\psi(\tilde{r}=0,\theta,\phi) =0$ as the method is "smart" enough to find the right dependence and not produce spurious eigenvalues for hydrogen for example. One therefore expects that the eigenstates of a hydrogen-like system naturally pick a single $l$ value as the angular dependence of these eigenstates is a single spherical harmonic (ideally, although not necessarily in a computational setting, but true to a very close approximation). That is to say the solutions should be of the form,
\begin{equation}
\psi_{lm}(r,\theta,\phi) = Y_{lm}(\theta,\phi)\sum_{i}c_{ilm} g_i(r).
\end{equation}
\subsection{Definitions of Integrals}
For the radial integrals, one might expect
\begin{equation}\label{eq:scaleint}
\int_{0}^{\infty}(~)r^2dr \rightarrow \int_{0}^{\infty}(~)\tilde{r}^2d\tilde{r}.
\end{equation}
The discontinuity at the scaling radius is in the form of a factor $e^{i\xi}$ applied to calculations of the matrix elements outside of the scaling radius \cite{s10}. This shows up as $d\tilde{r} = e^{i\xi}dr$, in which case it can be interpreted as the Jacobian for the transformation of the $r$ variable.
\\~\\One still must consider the $\tilde{r}^2$ part. Additionally, the calculation of expectation values introduces additional concerns. 
\\~\\If one assumes that we are solving for $\psi(\tilde{r},\theta,\phi,t)=\tilde{r}\Psi(\tilde{r},\theta,\phi,t)=\sum c_{ilm}(t)g_{i}(r)Y_{lm}(\theta,\phi)$, consider an expectation value defined the following way,
\begin{equation}
\langle \hat{\mathcal{O}} \rangle = \int_{0}^{2\pi}\int_{0}^{\pi}\int_{0}^{\mathcal{R}} \psi^*(\tilde{r},\theta,\phi,t) \hat{\mathcal{O}}\psi(\tilde{r},\theta,\phi,t)dr \sin \theta d\theta d\phi,
\end{equation}
that is,
\begin{equation}\label{eq:expectationvalue}
\langle \hat{\mathcal{O}} \rangle = \int_{0}^{2\pi}\int_{0}^{\pi}\int_{0}^{\mathcal{R}} \bigg(\tilde{r} \Psi(\tilde{r},\theta,\phi,t)\bigg)^* \hat{\mathcal{O}}~\bigg(\tilde{r}\Psi(\tilde{r},\theta,\phi,t)\bigg)dr \sin \theta d\theta d\phi.
\end{equation}
This is how we will define expectation values in this thesis. Here $\mathcal{R}$ (in atomic units) is the outer limit of the radial box. It is real valued since the radial integration measure is $dr$. %For $\mathcal{R}\rightarrow \infty$ the scaled variable $\tilde{r}$ is inevitably evaluated at complex infinity. 
\\~\\We clarify here that expectation values defined this way do not include $e^{i\xi}$ tacked onto $dr$, or otherwise they would not be real-valued for Hermitian operators. For example, a square normalizable wave function should have a norm of 1 (expectation value of the identity operator) in the case of a true bound state. The norm of a complex scaled resonant state should always be real and less than or equal to 1.\footnote{Since they are square integrable they can be normalized to unity at any time step. This feature is related to the fact that they are basically eigenstates of a complex scaled Hamiltonian with complex energies. The only mathematical difference between a complex scaled state and an eigenstate of the same non-scaled Hamiltonian is the factor of $\exp[-\Gamma t /2\hbar]$.}
\\~\\To show what would happen if the factor $e^{i\xi}$ was included, which we note by the subscript, consider the expectation value of the identity operator given as,
\begin{align}
\begin{split}
\langle \hat{\mathcal{I}}\rangle_{e^{i\xi}} = \int_{0}^{2\pi}\int_{0}^{\pi}\int_{0}^{\infty } \bigg(\tilde{r}\Psi(\tilde{r},\theta,\phi,t)\bigg)^*\bigg(\tilde{r}\Psi(\tilde{r},\theta,\phi,t)\bigg)e^{i\xi}dr \sin \theta d\theta d\phi\\= \int_{0}^{2\pi}\int_{0}^{\pi}\int_{0}^{\infty } \bigg(\Psi(\tilde{r},\theta,\phi,t)\bigg)^*\bigg(\Psi(\tilde{r},\theta,\phi,t)\bigg)e^{i\xi}|r|^2dr \sin \theta d\theta d\phi.
\end{split}
\end{align}
The integrand, excluding the new complex factor $e^{i\xi}$ should be real valued. However, the complex factor gives us a non-real norm. Thus, we have shown the complex factor in the integral breaks the Hermiticity of an Hermitian operator.
\\~\\Additionally, if $\tilde{r}^2$ was in the Jacobian it would also generally turn the expectation values of Hermitian operators complex. The combination $\tilde{r}^2e^{i\xi}$ is also complex and must be avoided in expectation values. These considerations justify the inclusion of $\tilde{r}$ in the wave function we ultimately solve for rather than in the definition of the expectation value. We also must exclude $e^{i\xi}$ from expectation values.
\\~\\It follows the useful definition for the radial integral denoted by a modified symbol for Dirac brackets, $\{~ \} $, is $\int_{0}^{\infty}dr$. We denote the full 3D integral as $\{~ \} \left[~~\right] =\int_{0}^{2\pi}\int_{0}^{\pi}\int_{0}^{\infty }dr \sin \theta d\theta d\phi$ where the square brackets contain the standard spherical part. 
\\~\\The definition for expectation values developed here is found to work.\footnote{This gives us,
\begin{equation}
\int_{0}^{2\pi}\int_{0}^{\pi}\int_{0}^{\mathcal{R}} \psi(\tilde{r},\theta,\phi,t)^* \psi(\tilde{r},\theta,\phi,t)dr \sin \theta d\theta d\phi \propto exp [-\Gamma t/\hbar],
\end{equation}
which, when later we do time-dependent ECS in $\S \ref{section:3point4}$ agrees with our previous numerical results in $\S \ref{section:2point5}$.}
\\~\\Looking at the definition of the matrix elements instead of the expectation values, since the adjusted wave function (scaled by the radial co-ordinate) is
\begin{equation}
\psi(\tilde{r},\theta,\phi,t)=\tilde{r}\Psi(\tilde{r},\theta,\phi,t)=\sum_{i,l,m} c_{ilm}g_{i}(r)Y_{lm}(\theta,\phi),
\end{equation}
then, matrix elements are everywhere defined as
\begin{equation}
\langle g_{i}Y_{lm}| \hat{H}| g_{i'}Y_{l'm'}\rangle = \int_{0}^{2\pi}\int_{0}^{\pi}\int_{0}^{\mathcal{\infty}} \bigg( g_{i}(r)Y_{lm}(\theta,\phi) \bigg) \hat{H} \bigg( g_{i'}(r)Y_{l'm'}(\theta,\phi)\bigg) f(r,\xi,\lambda)dr \sin\theta d\theta d\phi,
\end{equation}
where $f(r,\xi,\lambda) = 1$ inside the scaling radius and $f(r,\xi,\lambda)=e^{i\xi}$ outside.
\\~\\So we recognize that $g_{i}(r)$ inside the scaling radius contains the multiplicative factor $r$ and the whole function $\psi(\tilde{r},\theta,\phi,t)$ contains $\tilde{r}$. When the expectation value of the identity operator is taken, this makes up the factor $|\tilde{r}|^2$ in the integral. It cannot be in the Jacobian of the matrix elements, since it is part of the entire wave function. A second explanation for why the factor $e^{i\xi}$ doesn't show up in the definition of the integral in expectation values is that it comes from the factor $e^{i\xi/2}$ that ultimately should be part of the solved wave function (see Eq. \ref {eq:psixi}).
\subsection{Chebyshev Integrator and Evaluation of Coulomb Singularities}
An integrator that in principle finds the solutions to matrix elements by representing the integrands by Chebyshev series can be devised as long as some conditions are met. Chebyshev polynomials have a close relationship with cosines through a simple change of variables, so a Chebyshev series can be recast as a cosine series. Very smooth, even-symmetry periodic functions are best described with a cosine series since the coefficients of the series fall off very quickly at the higher orders for such functions \cite{byron71}. The cosine series can be called an {\it optimal representation} for such functions. Such functions meet and go beyond the sufficient conditions for a cosine series representation. For functions that are non-periodic on an interval $[a,b]$, a cosine transform in the function variable can allow for a representation with a cosine series. If one maps the variable $x = \cos(\rho)$ then $f(x) \rightarrow f(\rho)$. One can see this work for the example $f(x)=x$ since $f(\rho) =\cos\rho$. In this case the resulting function is smooth and periodic.
\subsubsection{Smoothness and Periodicity After a Change of Variables}\label{sec:smoothper}
More to the point, consider a function of the type (which we will use)
\begin{equation}
f_M(x) = \sum_{m=1}^{M}c_mh_m(x)
\end{equation}
over $x \in [a,b]$, where $h_m$ are monomials and $c_m$ are constants. The final functions we integrate are more complicated than $f_M(x)$, but we must prove smoothness of $f_M(x)$ under a change of variables first. Inevitably, since we are using monomials as a basis, $f_M(x)$ is smooth over this interval before a change of variables but not necessarily periodic.
\\~\\Before we do a transformation of variables to $\cos\rho$ we need to cast $x$ as a function of a new variable $z$ with the domain $[-1,1]$ so that it maps to the range of $\cos\rho$ with the domain $\rho \in [\pi, 0]$.\footnote{This domain allows one to use the symmetry properties of integrals. See Eq. \ref{eq:symmetry} and Eq. \ref{eq:bigthing}.}
\\~\\This can be done using the transformation
\begin{equation}
x=\frac{b-a}{2}z + \frac{b+a}{2}.
\end{equation}
Then we can write,
\begin{equation}
z = \cos \rho.
\end{equation} 
The expansion above can then be written as
\begin{equation}
f_M\bigg(\frac{b-a}{2}\cos \rho + \frac{b+a}{2}\bigg) = f_M(\rho) = \sum_{m=1}^{M}c_mh_m\bigg(\frac{b-a}{2}\cos \rho + \frac{b+a}{2}\bigg).
\end{equation}
This must now be periodic in the $\rho$ variable since 
\begin{equation}
x(2\pi) = \bigg(\frac{b-a}{2}\cos (2\pi) + \frac{b+a}{2}\bigg) = x(0) = \bigg(\frac{b-a}{2}\cos (0) + \frac{b+a}{2}\bigg) = b.
\end{equation}
So $f_M(\rho)$ (in fact, this is true for any function under this change of variables) is evaluated at the same $x$ value at the two sides of a period. 
\\~\\It is also even, since,
\begin{equation}
x(-\pi) = \bigg(\frac{b-a}{2}\cos (-\pi) + \frac{b+a}{2}\bigg) = x(\pi) = \bigg(\frac{b-a}{2}\cos (\pi) + \frac{b+a}{2}\bigg) = a.
\end{equation}
Again, this applies to any function under this change of variables.
\\~\\The function $f_M(\rho)$ is also smooth so a Fourier series is an optimal representation (see $\S \ref{sec:smooth}$).
\subsubsection{Fourier Basics}
Consider the Fourier series given by
\begin{equation}
f(x) = \frac{a_0}{2} + \sum_{n=1}^{\infty}\bigg(a_n\cos(nx) + b_n \sin(nx)\bigg),
\end{equation}
with coefficients given by \cite{byron71}
\begin{equation}
a_n = \frac{1}{\pi}\int_{-\pi}^{\pi}f(x)\cos(nx)dx,~n \geq 0, ~n \in \mathbb{Z},
\end{equation}
\begin{equation}
b_n = \frac{1}{\pi}\int_{-\pi}^{\pi}f(x)\sin(nx)dx,~n \geq 1,~n \in \mathbb{Z}.
\end{equation}
Or, as long as $f(x)$ is even-symmetric\footnote{The product of two even functions is even, since
\begin{equation}
f_e(-x)g_e(-x) = f_e(x)g_e(x).
\end{equation}
Then one can use standard rule for integrals, which states that if the integrand of an integral is even, the symmetric limits $[-\pi,\pi]$ can be reduced to the half-domain $[0,\pi]$, as long as the resulting integral is multiplied by two
\cite{stewart5e}.}
\begin{equation}\label{eq:symmetry}
a_n = \frac{2}{\pi}\int_{0}^{\pi}f(x)\cos(nx)dx,~n \geq 0,~n \in \mathbb{Z}.
\end{equation}
One finds that an even-symmetry periodic function described using a Fourier representation immediately reduces to a cosine series because of (a) a basic property of integrals and (b) the properties of even and odd functions. In other words, $b_n = 0$.
\\~\\To show this, consider the integral between an even and odd function, given by $f_e(x)$ for even and $f_o(x)$ for odd (standing in for sine) respectively:
\begin{equation}
\int_{-\pi}^{\pi}f_e(x)f_o(x)dx.
\end{equation}
However, $f_e(-x)f_o(-x)= -f_e(x)f_o(x)$ so the product is odd. The integral of an odd function over $[-\pi,\pi]$ must be 0.
%By splitting the integral up into two segments, $[-\pi, 0]$ and $[0, \pi]$, this can be written as
%\begin{equation}\label{eq:oddeven}
%\int_{-\pi}^{0}f_e(x)f_o(x)dx+\int_{0}^{\pi}f_e(x)f_o(x)dx.
%\end{equation}
%We can re-write the first integral in such a way that it matches the integration domain of the second integral. We do this by using the feature that $\int_{-\pi}^{0}\rightarrow -\int_{0}^{-\pi}$ \cite{stewart5e}. This gives,
%\begin{equation}
%-\int_{0}^{-\pi}f_e(x)f_o(x)dx+\int_{0}^{\pi}f_e(x)f_o(x)dx.
%\end{equation}
%We can change the form of the first integral using the substitution $x = -\tilde{x}$, 
%\begin{equation}
%(-)(-)\int_{0}^{\pi}f_e(-\tilde{x})f_o(-\tilde{x})d\tilde{x}=\int_{0}^{\pi}f_e(\tilde{x})(-f_o(\tilde{x}))d\tilde{x}.
%\end{equation}
%Where $\tilde{x}$ can be now recognized as a dummy variable. We can re-write it as just $x$,
%\begin{equation}
%-\int_{0}^{\pi}f_e(\tilde{x})f_o(\tilde{x})d\tilde{x}=-\int_{0}^{\pi}f_e(x)f_o(x)dx.
%\end{equation}
%Thus the first integral in Eq. \ref{eq:oddeven} is just the negative of the second. 
So we see that 
\begin{equation}
b_n = \frac{1}{\pi}\int_{-\pi}^{\pi}f(x)\sin(nx)dx = 0
\end{equation}
if $f(x)$ is an even function since $\sin(nx)$ is odd.
\subsubsection{Chebyshev Polynomials}
The Chebyshev polynomials have the unique property that (and therefore can be defined strictly by),
\begin{equation}
T_n(\cos(\rho)) = \cos(n\rho).
\end{equation}
So if we take a Chebyshev series for a function defined on $z \in [-1,1]$,
\begin{equation}
f(z) = \frac{C_0}{2} + \sum_{n=1}^{\infty}C_nT_n(z),
\end{equation}
and do a $z = \cos(\rho)$ substitution,
\begin{equation}
f(\cos \rho) = f(\rho) = \frac{C_0}{2} + \sum_{n=1}^{\infty} C_n
\cos(n\rho).
\end{equation}
More to the point, the function $f_M(x)$ defined on $x \in [a,b]$ can be transformed by a change of variables 
\begin{equation}
x = \bigg(\frac{b-a}{2}\cos \rho + \frac{b+a}{2}\bigg),
\end{equation}
taking $f_M(x)$ to $f_M(\rho)$. The cosine series both exists and is optimal for $f_M(\rho)$ since it is even-symmetric, periodic and smooth.
The Chebyshev coefficients are given by the equations
\begin{equation}
C_n = \frac{2}{\pi}\int_{0}^{\pi}f_M(\rho)\cos(n\rho)d\rho,~n \geq 0,~n \in \mathbb{Z}.
\end{equation}
Since this is just the solution to a cosine Fourier series, the solutions to the Chebyshev series of $f_M(x)$ are readily available.
To calculate a matrix element by integration we transform the co-ordinates of the integrand this way and use an FFT (Fast Fourier Transform) to evaluate the Chebyshev coefficients. All that is left is to evaluate integrals which are well known. To describe the process in all its mathematical glory, consider the integral of $f_M(x)$, first in the $x$ variable, then transformed to the $z$ variable which maps the domain to $[-1,1]$,\footnote{Where $x$ is related to $z$ by \begin{equation*}
x=\frac{b-a}{2}z + \frac{b+a}{2}
\end{equation*} 
and
\begin{equation*}
dx = \frac{b-a}{2}dz.
\end{equation*}} then $z=\cos\rho$ for the final integral\footnote{Where $z$ becomes \begin{equation*}
z=\cos\rho
\end{equation*}
and
\begin{equation*}
dz=-\sin\rho d\rho.
\end{equation*}} before using the Chebyshev representation. 
\newpage
\noindent 
The integration bounds are $\rho \in [\pi,0]$ for the $d\rho$ integral since $\cos \pi = -1$ and $\cos 0 = 1$. However, these can be flipped. The process of changing variables proceeds as follows:
\begin{equation}\label{eq:bigthing}
\int_{a}^{b}f_M(x)dx=\int_{-1}^{1}f_M(z)\left[\frac{b-a}{2}\right]dz=\int_{\pi}^{0}f_M(\rho)\left[\frac{b-a}{2}\right](-\sin\rho) d\rho=\int_{0}^{\pi}f_M(\rho)\left[\frac{b-a}{2}\right](\sin\rho) d\rho.
\end{equation}
Then under the Chebyshev representation (the basis functions $h_m$ don't play a role here, since we use a cosine basis for $f_M$ with the knowledge that it is optimal),
\begin{equation}
\int_{0}^{\pi}f_M(\rho)\left[\frac{b-a}{2}\right](\sin\rho) d\rho \approx \int_{0}^{\pi}\left(\frac{C_0}{2}+\sum_{n=1}^{N}C_n\cos (n\rho) \right)\left[\frac{b-a}{2}\right](\sin\rho) d\rho,
\end{equation}
or,
\begin{align}\label{eq:intcheb}
\begin{split}
\int_{0}^{\pi}\left(\frac{C_0}{2}+\sum_{n=1}^{N}C_n\cos (n\rho) \right)\left[\frac{b-a}{2}\right](\sin\rho) d\rho \\ = \left[\frac{b-a}{4}\right]C_0\int_{0}^{\pi} (\sin\rho) d\rho+\left[\frac{b-a}{2}\right]\sum_{n=1}^{N}C_n\int_{0}^{\pi}\cos (n\rho) (\sin\rho) d\rho \\ = \left[\frac{b-a}{2}\right]C_0 +\left[\frac{b-a}{2}\right]\sum_{n=1}^{N}C_n\int_{0}^{\pi}\cos (n\rho) (\sin\rho) d\rho.
\end{split}
\end{align}
This is,
\begin{equation}\label{eq:intcheb2}
\int_{a}^{b}f_M(x)dx \approx \left[\frac{b-a}{2}\right]C_0 +\left[\frac{b-a}{2}\right]\sum_{n=1}^{N}C_n\left[\frac{\cos(n\pi) + 1}{1 - n^2}\right].
\end{equation}
For odd $n$, $\cos (n\pi) + 1=0$, so we only need even $n$, given by,
\begin{equation}\label{eq:intcheb3}
\left[\frac{\cos(n\pi) + 1}{1 - n^2}\right]= \left[\frac{2}{1 - n^2}\right].
\end{equation}
So we can re-write the integral of $f_M(x)$ as,
\begin{equation}\label{eq:intcheb5}
\int_{a}^{b}f_M(x)dx \approx \left[\frac{b-a}{2}\right]C_0 +\left[\frac{b-a}{2}\right]\sum_{n=2}^{N}C_n\left[\frac{2}{1 - n^2}\right],~n~{\rm even}.
\end{equation}
This can be written as,
\begin{equation}\label{eq:intcheb6}
\int_{a}^{b}f_M(x)dx \approx \left[\frac{b-a}{2}\right]C_0-\bigg[b-a\bigg]\sum_{k=1}^{(N-1)/2}C_k\bigg[(2k+1)(2k-1)\bigg]^{-1},
\end{equation}
where $2k = n ~{\rm even}$ and so $(2k+1)(2k-1) = (n+1)(n-1) = - (1-n^2)$.
$C_0$ and $C_k$ can be solved by a Fast Fourier Transform (FFT) since they are essentially Fourier coefficients under the change of variables.
\subsubsection{Mathematical Singularities and Other Loose Ends}\label{sec:loose}
We note that the Coulomb singularity at $r=0$ for a $1/r$ type function is missed by a choice of grid that starts sampling the integrand function a small distance $\eta$ away from the left boundary of any interval.
\\~\\We must admit that $f_M(x)$ is not the full integrand which we use our integration method for. 
\\~\\The integrands we consider contain functions of the same type as $f_M(x)$, which we call $f_{im}(r)$. These will be discussed in detail in $\S \ref{sec:fim}$. However, in regard to the optimality of the representation, the integrands $f_{im}'(r)f_{i'm'}'(r)$, $f_{im}(r)f_{i'm'}(r)/\tilde{r}$, and $f_{im}(r)f_{i'm'}(r)/\tilde{r}^{2}$ are all integrated using the method we have detailed, and are smooth. Therefore a Fourier series representation is optimal (see $\S \ref{sec:smooth}$). 
\\~\\We only use this specific integration method for the radial part of the integral. 
\section{Numerical Implementation}\label{section:1point5}
\subsection{Finite Element Method}\label{section:1point5point1}
Given some eigenvalue problem
\begin{equation}
\hat{H}\psi = E\psi,
\end{equation}
we seek to find an approximate solution $\tilde{\psi}$ for $\psi$ by a finite element method. From now on we will drop the hat on the Hamiltonian operator. 
\\~\\We consider a 1D problem. The domain of the $x$ co-ordinate $[x_{min}, x_{max}]$ is broken into $N$ equally spaced subintervals with label $i$ each with domain $[x_{i-1},x_i]$ \cite{se93}. Each subinterval $i$ has associated with it a set of $M$ basis vectors with labels $m$ that are zero outside the subinterval. The basis elements are named $f_{im}$. 
\\~\\The game is then to project the approximate solution $\tilde{\psi}$ onto the entire set of basis functions,
\begin{equation}
\tilde{\psi}(x) = \sum_{i,m}c_{im}f_{im}(x).
\end{equation}
The method then rests in finding the coefficients $c_{im}$. Before we do this, we must determine the nature of the $f_{im}(x)$ we will use. In general the above process will not give a continuous function at the boundaries. That is to say $\tilde{\psi}(x_i-\delta)= \tilde{\psi}(x_i+\delta)$ will not generally be enforced in the limit that $\delta \rightarrow 0$. To avoid this problem, consider a construction of $f_{im}(x)$ where the first basis function for any subinterval is equal to 1 at the left boundary, the second basis function (labelled by $m=2$) for any subinterval is equal to 1 at the right boundary, and all other basis functions are zero at the boundaries \cite{se93}. No explicit care is taken to ensure that the first derivatives of $\tilde{\psi}$ are continuous at the boundaries. The boundary conditions can be expressed as,
\begin{equation}
f_{i,1}(x_{i-1}) = 1,
\end{equation}
\begin{equation}
f_{i,2}(x_{i}) = 1,
\end{equation}
\begin{equation}
f_{i,m}(x_{i-1}) = f_{i,m}(x_{i})=0,~ m \neq 1,2.
\end{equation}
These three conditions can be summarized as a matrix of boundary values
\begin{equation}
B_f= \left(\begin{tabular}{c c c c}
$1 $ &0&0&\dots\\
0 & 1 &0&  \dots\\

\end{tabular}\right),
\end{equation}
where the first row represents values of the basis functions at $x_{i-1}$ and the second row represents values of the basis functions at $x_{i}$. The order of the basis function $m$ is represented by the column index. With these conditions, we additionally demand  \cite{se93}
\begin{equation}
c_{i+1,1} = c_{i,2}.
\end{equation}
In matrix formalism we make the last element of the $ith$ sub-matrix $c_{i,2}$. With this condition, one allows the last element of the $ith$ sub-matrix and first element of the $(i+1st)$ sub-matrix to overlap such that they add together (see Eq. \ref{referencemat} in the appendix for example). 
In the case of ECS, there is a discontinuity that is implemented by requiring that \cite{s10}
\begin{equation} \label{eq:psixi}
\psi(x_o-0) = e^{i\xi/2}\psi(x_o+0),
\end{equation}
where $x_o$ is the scaling radius. There is also a first derivative discontinuity which we omit since it need not be enforced along with the first derivative continuities at boundaries \cite{se93}.
\\~\\We note that in application this is equivalent to the Jacobian produced by the co-ordinate scaling ($e^{i\xi}$), as the two factors of $e^{i\xi/2}$ in Scrinzi's work are produced when forming the inner product of two states outside the scaling radius. In that approach these factors are not conjugated along with the basis functions when inner products are taken \cite{s10}. 
\\~\\Note that the Jacobian itself has a discontinuity, as it quickly jumps from $1$ to $e^{i\xi}$. Strictly speaking the basis functions (since they are only the chosen representation) need not be multiplied by any complex factor to enforce the discontinuity, as it is only necessary that it should show up in the final wave function (see Eq. \ref {eq:psixi}). So only the Jacobian for the scaling should be introduced. This interpretation is more satisfying (as far as matrix formalism is concerned) than Scrinzi's explanation, where complex factors are introduced through the basis functions. 
%We can now approach from the left or right of the right boundary of an interval $i$. 
%\begin{figure*} 
%\centering
%\includegraphics[width=300pt]{fig2_boundary.eps}
%\vspace{0.2cm}
%\caption{\label{fig:fig2} ** replace this plot with Eq. 8,9,10 plot}
%\end{figure*}
\subsection{Defining the $f_{im}$ from $h_{m}$}\label{sec:fim}
The exact form of $f_{im}$ will be constructed from the fundamental functions $h_m$ that will be defined later. These fundamental functions are defined on the domain $[0,1]$ so they must be transformed to satisfy the conditions of the problem \cite{se93}. First we construct a matrix of the boundary values of $h_m$,
\begin{equation}
B_h= \left(\begin{tabular}{c c c c}
$h_1(0)$ &$h_2(0)$&$h_3(0)$&\dots\\
$h_1(1)$ &$h_2(1)$&$h_3(1)$&\dots\\

\end{tabular}\right).
\end{equation}
We now want to find a matrix $W$ that will map $B_h$ to $B_f$, defined by $B_h W = B_f$, or
\begin{equation}
\left(\begin{tabular}{c c c c}
$h_1(0)$ &$h_2(0)$&$h_3(0)$&\dots\\
$h_1(1)$ &$h_2(1)$&$h_3(1)$&\dots\\

\end{tabular}\right)\cdot W= \left(\begin{tabular}{c c c c}
$1 $ &0&0&\dots\\
0 & 1 &0&  \dots\\

\end{tabular}\right).
\end{equation}
If we define $B_f = (\bold{1},0)$ where $\bold{1}$ is the $2\times2$ identity matrix and $B_h = (P,Q)$ where $P$ is a matrix containing the first two columns of $B_h$, one solution is
\begin{equation}
W = \left(\begin{tabular}{c c }
$P^{-1}$ &$-P^{-1}\cdot Q$\\
$0$ &$\bold{1}$\\
\end{tabular}\right),
\end{equation}
where $\bold{1}$ now denotes the $(\rm dim(W)-2)\times(\rm dim(W)-2)$ identity matrix. 
Using this result it can be shown that the $f_{im}$ that satisfy boundary conditions given by $B_f$ can be constructed from the fundamental functions $h_m$ in the following way:
\begin{equation}
f_{im}(x) = \sum_{m'}W_{m'm}h_{m'}\left(\frac{x-x_{i-1}}{x_{i}-x_{i-1}}\right).
\end{equation}
The only difference for a 3D problem is that the analogous variable, $r$, has a domain of $[0,r_{max}]$. One can then treat the angular parts with global basis functions since the scaling radius has no effect on the angular variables (in particular no discontinuities need to be enforced).
%\subsection{The Specifics of $h_m$ and Computing $f_{im}$}
%Using the

%\subsection{Chebyshev Integration}
%To calculate the matrix elements, we employ an integration technique using the Chebyshev coefficients of the desired integrand. Let us consider the Chebyshev polynomials of the first kind,
\section{Test Case (1D Hydrogen Model)}\label{section:1point6}
We use a 1D hydrogen model given as,
\begin{equation}
\left[\frac{-\hbar^2}{2m_e}\frac{d^2}{dx^2} - \frac{e^2}{4\pi \epsilon_o}\frac{1}{ \sqrt{1+x^2}}\right]\psi(x) = E\psi(x).
\end{equation}
Using the scaling defined in Eq. \ref{eq:scale} it can be shown that Schr{\"o}dinger's equation for the 1D hydrogen model outside the scaling radius becomes \cite{s10}
\begin{equation}
\left[\frac{-\hbar^2}{2m_e}e^{-2i\xi}\frac{d^2}{dx^2} - \frac{e^2}{4\pi \epsilon_o}\frac{1}{ e^{i\xi}(x- x_o)+ x_o}\right]\psi(x)= E\psi(x).
\end{equation}
Here we have assumed that outside the scaling radius $-\frac{1}{ \sqrt{1+x^2}}$ can be approximated by $-1/\tilde{x}$. Inside the scaling radius Schr{\"o}dinger's equation remains as
\begin{equation*}
\left[\frac{-\hbar^2}{2m_e}\frac{d^2}{dx^2} - \frac{e^2}{4\pi \epsilon_o}\frac{1}{ \sqrt{1+x^2}}\right]\psi(x)= E\psi(x).
\end{equation*}
 We have purposefully omitted an external electric field, which will be considered later.
\\~\\To calculate the matrix elements we will need the Jacobian for the co-ordinate scaling, given by $d\tilde{x}/dx = e^{i\xi}$.
This makes the Hamiltonian matrix elements outside the scaling radius look like,
\begin{equation}
H_{ij}=e^{i\xi}\frac{\hbar^2}{2m_e}e^{-2i\xi}\langle \frac{d}{dx}i|\frac{d}{dx}j\rangle - e^{i\xi}\langle i|\frac{e^2}{4\pi \epsilon_o}\frac{1}{ e^{i\xi}(x- x_o)+ x_o}|j\rangle,
\end{equation}
where $\langle~\rangle = \int_{-\infty}^{\infty}dx$. The symmetric form of the kinetic energy is derived in \S \ref{section:5point1}.
\\~\\The overlap matrix also has a change outside the scaling radius due to the Jacobian given by
\begin{equation}
S_{ij}=e^{i\xi}\langle i|j\rangle.
\end{equation}
Simplifying the Hamiltonian matrix elements,
\begin{equation}
H_{ij}=\frac{\hbar^2}{2m_e}e^{-i\xi}\langle \frac{d}{dx}i|\frac{d}{dx}j\rangle - \langle i|\frac{e^2}{4\pi \epsilon_o}\frac{1}{ (x- x_o)+ e^{-i\xi}x_o}|j\rangle,
\end{equation}
and then expanding the denominator of the potential energy,
\begin{equation}
H_{ij}=\frac{\hbar^2}{2m_e}e^{-i\xi}\langle \frac{d}{dx}i|\frac{d}{dx}j\rangle- \langle i|\frac{e^2}{4\pi \epsilon_o}\frac{1}{ (x- x_o)+  x_o\cos\xi-ix_o\sin\xi }|j\rangle.
\end{equation}
We rationalize the potential energy, giving us,
\begin{equation}
H_{ij}=\frac{\hbar^2}{2m_e}e^{-i\xi}\langle \frac{d}{dx}i|\frac{d}{dx}j\rangle - \frac{e^2}{4\pi \epsilon_o}\langle i|\frac{x-x_o+x_o\cos\xi +ix_o\sin\xi}{ (x-x_o+x_o\cos\xi )^2+x_o^2\sin^2\xi}|j\rangle,
\end{equation}
and split the real and imaginary parts,
\begin{equation}
H_{ij}=\frac{\hbar^2}{2m_e}e^{-i\xi}\langle \frac{d}{dx}i|\frac{d}{dx}j\rangle - \frac{e^2}{4\pi \epsilon_o}\langle i|\frac{x-x_o+x_o\cos\xi}{ (x-x_o+x_o\cos\xi )^2+x_o^2\sin^2\xi}|j\rangle-\frac{e^2}{4\pi \epsilon_o}i\langle i|\frac{x_o\sin\xi}{ (x-x_o+x_o\cos\xi )^2+x_o^2\sin^2\xi}|j\rangle.
\end{equation}
Here we have kept the real and imaginary parts together for the kinetic energy part to save space.
\\~\\We calculate the kinetic and potential energy matrices separately. In the case of the kinetic energy, we further separate the calculation into a matrix to the left of the scaling radius and a matrix to the right of the scaling radius which are then added. For the potential energy matrix, we also further separate the calculations the same way. For the matrices after the scaling radius, we separate the real and imaginary parts. Essentially there will be one matrix for KE and PE each before scaling, and two matrices for KE and PE each (separating real and imaginary parts) after scaling. All the matrices are then combined appropriately. In particular the sub-matrices are combined in the same way as they were for the Hermitian case.
\\~\\We must also include a DC electric field; it must be scaled outside $x_o$ as well, this is simply 
\begin{equation}
DC_{ij}=-e^{i\xi}\langle i|eF_o (e^{i\xi}(x-x_o)+x_o)| j\rangle,
\end{equation}
and inside it is,
\begin{equation}
DC_{ij}=-\langle i|eF_o x| j\rangle,
\end{equation}
where $eF_o$ is the field strength multiplied by charge.
We also separate the real and imaginary parts for the $DC_{ij}$ matrix. 
\\~\\First we multiply through by $e^{i\xi}$,
\begin{equation}
DC_{ij}=-\langle i|eF_o (e^{2i\xi}(x-x_o)+e^{i\xi}x_o)| j\rangle,
\end{equation}
and simplifying,
\begin{equation}
DC_{ij}=-\bigg\langle i~\bigg|eF_o \bigg(\bigg[x-x_o\bigg]\cos2\xi + x_o\cos\xi\bigg)\bigg|~j\bigg\rangle - i\bigg\langle i~\bigg|eF_o \bigg( \bigg[x-x_o\bigg]\sin2\xi+x_o\sin\xi \bigg)\bigg| ~j\bigg\rangle.
\end{equation}
We chose as an example $F_o = 0.11~au$ since this is still in the so-called "under-the-barrier" regime which allows for a tunnelling solution. At a higher field strength the energy of the potential barrier for the electron becomes so low that the energy of the ground state is above it, which is not a tunnelling problem for obvious reasons. This is the so-called "over-the-barrier" regime.\subsection{Results}\label{section:1point6point1}

 \setlength{\tabcolsep}{40pt}
\bgroup
\def\arraystretch{1.4}
\begin{table}[!h]
\begin{center}
 \begin{tabular}{|c c c|} 
\hline
\hline
  $x_o$ & $\xi$ & $E$  \\ [0.5ex] 
 \hline\hline

 9.8 & $0.5 ~rad$& -0.713019302601830 - 0.006368222805639i \\ 
 \hline
 9.8 & $1.0~rad$ &  -0.713019302601829 - 0.006368222805638i
 \\
 \hline
9.8 & $\pi/2~ rad$& -0.713019302601829 - 0.006368222805641i   \\
 \hline
 \hline

\end{tabular}
\caption{Energy eigenvalues for the first resonance with an external DC field ($F_o = 0.11~au$) as found by ECS for various $\xi$.  The resonant parameters are related to the energy by $E = E_r-i\frac{\Gamma_r}{2}$.}
\end{center}
\end{table}
 \setlength{\tabcolsep}{6pt}

\begin{landscape}
\begin{table}[!h]
 \begin{center}

\begingroup
 \fontsize{8pt}{10pt}\selectfont
\begin{tabular}{ccccc}
\toprule
 & (-10, 100)$\times100$, $x_o = 9.8~au$& (-30, 120)$\times100$, $x_o = 10.5~au$& (-30, 120)$\times136$, $x_o = 9.7059~au$ & (-10, 120)$\times118$, $x_o = 9.8305~au$ \\
 \midrule
6 & -0.713019296791160 - 0.006368223548749i & -0.713021990854758 - 0.006364794164144i &-0.713019343902138 - 0.006368982408592i&-0.713019310482031 - 0.006367983118111i\\
\midrule
7 & -0.713019302282175 - 0.006368222846273i &  -0.713022155411564 - 0.006364771319910i&&-0.713019316040830 - 0.006367982546801i\\
\midrule
8 & -0.713019302592545 - 0.006368222807059i & -0.713022156359239 - 0.006364771249782i &&-0.713019316322346 - 0.006367982511521i\\
\midrule
9 & -0.713019302601829 - 0.006368222805638i & -0.713022156713583 - 0.006364771200781i&&-0.713019316333220 - 0.006367982509889i \\
\midrule
10 & -0.713019302602111 - 0.006368222805606i & &&-0.713019316333470 - 0.006367982509856i  \\
\midrule
11 & -0.713019302602131 - 0.006368222805604i &   &&-0.713019316333484 - 0.006367982509854i\\
\midrule
12 & -0.713019302602131 - 0.006368222805605i &   \\
\midrule
13 & -0.713019302602131 - 0.006368222805602i &   \\
\midrule
14 & -0.713019302602129 - 0.006368222805602i &   \\
\midrule
15 & -0.713019302602631 - 0.006368222806240i &   \\
\midrule
16 & -0.713019302438137 - 0.006368222886472i &  \\
\midrule
17 & -0.713019302773472 - 0.00636822269606732i &   \\
\midrule
18 & -0.713019313437927 - 0.006368354547714i &   \\
\bottomrule

\end{tabular}
\caption{Intervals are shown on the top in the form $(start, end)\times subintervals$, and number of basis functions per subinterval are shown on the left. Values of the first resonant eigenvalue are shown inside in atomic units. Various settings are tried. Noteworthy: The first column indicates strong convergence towards stability when using about 14 basis functions per subinterval.}
\endgroup

\end{center}
\end{table}
\end{landscape}

\section{Conclusion}\label{section:1point7}
The resonant widths attained with ECS using three very different values of the scaling angle agree with each other up to 13 decimal places and show change in the resonant position for only the last two decimal places. One could say that the ECS method is very self-consistent in the sense that it does not depend on the scaling angle $\xi$ up to the thirteenth decimal place in the real and imaginary part of the eigenvalue. The method shows convergence towards stability at the machine precision (specifically double precision) level around 14 basis functions per subinterval.% The instability that arises beyond this number is likely due to the weakness of a finite-precision evaluation of the effective difference between higher order polynomials. In other words, the overlap matrix elements between higher-order polynomials are too close to unity (citation).
\newpage
\chapter{A 3D Time-independent Model for Resonances}
\epigraph{We struggle so hard to hold on to these things that we know are gonna disappear eventually. And that's really noble.}{Lily Adrin, {\it How I Met Your Mother}}
\section*{Overview}
In $\S \ref{section:2point1}$ we discuss our notation.
In $\S \ref{section:2point2}$ we discuss the conversion of our 1D ECS algorithm to fully 3D problems (specifically hydrogen, and later singly ionized helium as well). The major thing to note is that, when going from the 1D hydrogen model to 3D hydrogen, the adjusted wave function, which is the analogue for $\psi(x)$, must now go to zero at $r = 0$. In $\S \ref{section:2point3}$ we discuss the use of a spherical basis (Legendre polynomials) with the quantum number $m=0$ fixed to establish azimuthal symmetry. In $\S \ref{section:2point4}$ we discuss the generalization for $m \neq 0$. In $\S \ref{section:2point5}$ we list our results including a figure of our results for singly ionized helium helium (Fig. \ref{fig:decay}). In $\S \ref{section:2point6}$ we discuss our conclusions.
\section{Notation}\label{section:2point1}
In this chapter, when discussing our previous work in 1D, we use angular brackets to refer to the integral across $x$, that is, $\langle~\rangle = \int_{x_{min}}^{x_{max}}dx$. For 3D problems, $\langle~\rangle$ represents the volume integral with no radial Jacobian. We separate radial and angular parts as $\{ ~\}$ and $[~~]$ respectively,  such that $\{ ~\}[~~] =\int_0^{\infty}...dr \int_0^{2\pi} \int_0^{\pi}...~ \sin\theta d\theta d\phi$. When writing matrix elements with $\varphi_{iml}(r,\theta)$ or $\varphi_{iml}(r,\theta,\phi)$ (including $r\rightarrow \tilde{r}$), such as $\langle \varphi_{iml}|\mathcal{O}|\varphi_{i'm'l'}\rangle$, the brackets $\langle ~ \rangle = \int_0^{\infty}dr \int_0^{2\pi} \int_0^{\pi} \sin\theta d\theta d\phi$. The same integrals are later written with $f_{im}$ functions, and the angular parts are taken care of by Wigner $3j$ coefficients. These take the form $W_{3j}^{2}(...)\{ f_{im}|\mathcal{O}|f_{i'm'}\}=W_{3j}^{2}(...)\{ i|\mathcal{O}|j\}$. The definition $\{ ~ \} =\int_0^{\infty}dr$ is again used.
\\~\\Additionally, since we retain the notation from our previous chapter, where $m$ indicates the order of the basis function, when $f_{im}$ appears together with spherical harmonics $Y_{lm}$, the conflict shall be solved by writing $Y_{ln}$. Everywhere else spherical harmonics shall be written $Y_{lm}$.
\\~\\Solutions are given in combined gaussian and atomic units. In particular $e^2/4\pi\epsilon_o \rightarrow e^2$, and further, $\hbar = m_e = e = 1$. For simplicity, we will just refer to this as atomic units.
\section{Converting a 1D Algorithm for Radial Problems}\label{section:2point2}
We will employ finite elements as a basis for the radial part of a 3D problem in spherical-polar co-ordinates. The finite elements are called $f_{im}$ where $i$ indicates the subinterval on which they are non-zero and $m$ indicates the order of the basis function.
\\~\\The adjusted wave function $\psi$ has a boundary condition at $r=0$ which requires it to go to zero since $\psi =\Psi r$.
\\~\\This requires us to change the $B_f$ matrix from $\S \ref{section:1point5}$ to implement the new boundary condition on the  main axis (which now $r$ compared to $x$ from before). We will call the new matrix $B_f^*$, given as
\begin{equation}
B_f^*= \left(\begin{tabular}{c c c c}
$0 $ &0&0&\dots\\
0 & 1 &0&  \dots\\

\end{tabular}\right).
\end{equation} 
Referring to $\S \ref{section:1point5}$, one can note that only one element has changed in $B_f^*$ (the first element in the first row), and thus the only column of the $W$ matrix that needs to change is the first one. However, we have to change the first column in such a way as to not affect the first element of the second row of $B_f^*$. Assuming we are only using three fundamental functions (adding more will not affect the result),
\begin{equation}
W^* = \left(\begin{tabular}{c c c c}
$W^*_{11}$ & $P^{-1}_{12}$&$-P^{-1}_{11}Q_1$$-P^{-1}_{12}Q_2$\\
$W^*_{21}$ & $P^{-1}_{22}$&$-P^{-1}_{21}Q_1$$-P^{-1}_{22}Q_2$\\
$0$ & $0$&$1$\\
\end{tabular}\right).
\end{equation}
Adding more fundamental functions does not affect the result since this will only add more zeroes to the first (and second) column.
\\~\\Now we only have to solve (in general)
\begin{equation}
\sum_i W^*_{i,1}h_i(0)=0,
\end{equation}
and,
\begin{equation}
\sum_i W^*_{i,1}h_i(1)=0,
\end{equation}
implying,
\begin{equation}
\sum_i W^*_{i,1}h_i(0)=\sum_i W^*_{i,1}h_i(1).
\end{equation}
For the case of three fundamental functions,
\begin{equation}
 W^*_{11}h_1(0)+ W^*_{21}h_2(0)+ 0h_3(0)= W^*_{11}h_1(1)+ W^*_{21}h_2(1)+ 0h_3(1),
\end{equation}
or,
\begin{equation}
 W^*_{11}h_1(0)+ W^*_{21}h_2(0)= W^*_{11}h_1(1)+ W^*_{21}h_2(1),
\end{equation}
re-arranging,
\begin{equation}
 W^*_{11}\bigg(h_1(0)-h_1(1)\bigg)= W^*_{21}\bigg(h_2(1)-h_2(0)\bigg),
\end{equation}
and finally,
\begin{equation}
 W^*_{11}= W^*_{21}\left[\frac{h_2(1)-h_2(0)}{h_1(0)-h_1(1)}\right].
\end{equation}
So,
\begin{equation}
\sum_i W^*_{i,1}h_i(0)=W^*_{21}\left[\frac{h_2(1)-h_2(0)}{h_1(0)-h_1(1)}\right]h_1(0)+W^*_{21}h_2(0)=0,
\end{equation}
or,
\begin{equation}
W^*_{21}\bigg[\frac{h_2(1)-h_2(0)}{h_1(0)-h_1(1)}h_1(0)+h_2(0)\bigg]=0.
\end{equation}
One solution is $W^*_{11}= W^*_{21}=0$.
\\~\\Once again this is already generalized to $M$ fundamental functions since all the extra fundamental functions will be multiplied by 0.
\\~\\Using this result it can be shown that the $f_{1,m}$ that satisfy boundary conditions given by $B_f^*$ can be constructed from the fundamental functions $h_m$ in the following way 
\begin{equation}
f_{1,m}(r) = \sum_{m'}W^*_{m'm}h_{m'}\bigg(\frac{r-r_{0}}{r_{1}-r_{0}}\bigg)
\end{equation}
where $r_0$ is the $0^{th}$ boundary (the far left boundary) and is not to be confused with the scaling radius.
\\~\\The same effect can be achieved by simply deleting the first row and column since these are the only parts of the matrix that depend on $f_{1,1}$. In fact, in practice it is simplest to do this. %\\~\\One might believe that this procedure only works for potentials that are finite at $r=0$, like the (quantum) harmonic oscillator. One example is the 3D hydrogen problem which has a $1/r$ potential that blows up at $r=0$. Since inner products involving this potential must be calculated numerically, this could lead to problems. However as long as one does not evaluate the potential at its singularity (i.e., a Coulomb potential at $r=0$), there is no issue. For example, a spectral integrator which avoids evaluations at $r=0$ should not produce any difficulties. Since we use a Chebyshev integrator (mentioned in our previous chapter) we avoid this issue.
\\~\\The fundamental functions we will use are given by monomials defined over $r \in [0,1]:$
\begin{align}
h_m(r) = 1,~m=1,
\\
h_m(r) = \frac{r^{m-1}}{m-1},~m>1.
\end{align}
\section{Separation of Variables and Coupled Equations for an External Field ($m = 0$)}\label{section:2point3}
We start with Schr{\"o}dinger's equation for the 3D hydrogen atom in an external DC electric field in spherical-polar co-ordinates for only $m=0$ states (not to be confused with the mass $m$ in the equation, but the quantum number $m$). The mass of the proton is assumed to be infinite so that it doesn't add kinetic energy to the system. Schr{\"o}dinger's equation is
\begin{equation}
\left[\frac{-\hbar^2}{2m_e}\nabla^2 - \frac{e^2}{4\pi \epsilon_o r} - eF_or\cos\theta\right]\Psi(r,\theta) = E\Psi(r,\theta).
\end{equation}
Here $\Psi$ is not written as a function of $\phi$ due to the azimuthal symmetry of the problem. It is well known that the solutions to hydrogen only have $\phi$ dependence when $m\neq0$ since the form of the $\phi$ dependence (demanded by the spherical harmonics which are naturally part of the eigenfunctions of the hydrogen Hamiltonian) is $e^{im\phi}$ \cite{scherrer}.
The Laplacian in spherical coordinates is
\begin{equation}\label{eq:Lap}
\nabla^2 = \frac{1}{r^2}\frac{\partial}{\partial r}r^2\frac{\partial}{\partial r} + \frac{1}{r^2 \sin \theta}\frac{\partial}{\partial \theta}\sin \theta \frac{\partial}{\partial \theta} + \frac{1}{r^2\sin^2\theta}\frac{\partial ^2}{\partial \phi^2}.
\end{equation}
The second and third term are related to the $L^2$ operator by 
\begin{equation}
\frac{1}{r^2 \sin \theta}\frac{\partial}{\partial \theta}\sin \theta \frac{\partial}{\partial \theta} + \frac{1}{r^2\sin^2\theta}\frac{\partial ^2}{\partial \phi^2} = \frac{L^2}{2m_er^2}
 \rightarrow \frac{\hbar^2l(l+1)}{2m_er^2},
 \end{equation}
where the arrow indicates the eigenvalue of the operator which we will use in place of the operator since we are acting on spherical harmonics (i.e., for $m=0$, Legendre polynomials).
%\begin{equation}
% \frac{1}{r^2}\frac{\partial}{\partial r}%r^2\frac{\partial}{\partial r}R(r) = \frac{1}{r}\frac{\partial ^2}{\partial r^2}rR(r)
%\end{equation}
\\~\\We will use a partial wave expansion with basis functions of the form
\begin{equation}
\Psi_l = R_l(r)P_l(\cos\theta),
\end{equation}
where $P_l$ are Legendre polynomials related to the spherical harmonics by \cite{byron71}
\begin{equation}\label{eq:sphere}
P_l(\cos\theta) = \sqrt{\frac{4\pi}{2l +1}}Y_{l,0}(\theta,\phi),
\end{equation}
and are thus orthogonal on the sphere like spherical harmonics \cite{byron71}.
We then have for the approximate wave function $\tilde{\Psi}(r,\theta)$,
\begin{equation}\label{eq:psi1}
\tilde{\Psi}(r,\theta) = \sum_{l=0}^{L}a_lR_l(r)P_l(\cos\theta).
\end{equation}
Suppose for now $L\rightarrow \infty$ so we can write down the exact eigenvalue problem.
\\~\\Putting this all together the exact eigenvalue problem is,
\begin{align}
\begin{split}
\frac{-\hbar^2}{2m_e}\frac{1}{r^2}\frac{\partial}{\partial r}r^2\frac{\partial}{\partial r} \sum_{l=0}^{L}a_lR_l(r)P_l(\cos\theta) + \frac{-\hbar^2}{2m_e}\bigg(\frac{1}{r^2 \sin \theta}\frac{\partial}{\partial \theta}\sin \theta \frac{\partial}{\partial \theta}\bigg)\sum_{l=0}^{L}a_lR_l(r)P_l(\cos\theta)\\-\frac{e^2}{4\pi \epsilon_o r} \sum_{l=0}^{L}a_lR_l(r)P_l(\cos\theta) - eF_or\cos\theta  \sum_{l=0}^{L}a_lR_l(r)P_l(\cos\theta) = E \sum_{l=0}^{L}a_lR_l(r)P_l(\cos\theta).
\end{split}
\end{align}
Using the fact that $P_l$ are related to the spherical harmonics and are therefore eigenstates of the $L^2$ operator,
\begin{align}
\begin{split}
\frac{-\hbar^2}{2m_e}\frac{1}{r^2}\frac{\partial}{\partial r}r^2\frac{\partial}{\partial r} \sum_{l=0}^{L}a_lR_l(r)P_l(\cos\theta) + \sum_{l=0}^{L}\left[\frac{\hbar^2l(l+1)}{2m_er^2}\right]a_lR_l(r)P_l(\cos\theta)\\-\frac{e^2}{4\pi \epsilon_o r} \sum_{l=0}^{L}a_lR_l(r)P_l(\cos\theta) - eF_or\cos\theta  \sum_{l=0}^{L}a_lR_l(r)P_l(\cos\theta) = E \sum_{l=0}^{L}a_lR_l(r)P_l(\cos\theta).
\end{split}
\end{align}
%Expanding in low order,
%\begin{align}
%\begin{split}
%\frac{-\hbar^2}{2m}\frac{1}{r^2}\frac{\partial}{\partial r}r^2\frac{\partial}{\partial r} (R_0(r)+R_1(r)\cos\theta) + \frac{\hbar^2}{2mr^2}(R_0(r)+2R_1(r)\cos\theta)\\-\frac{e^2}{4\pi \epsilon_o r}(R_0(r)+R_1(r)\cos\theta) - F_or\cos\theta  (R_0(r)+R_1(r)\cos\theta)\\ - E (R_0(r)+R_1(r)\cos\theta)=0
%\end{split}
%\end{align}
If we define $u(r) = rR(r)$ and note that \cite{scherrer}
\begin{equation}
\frac{1}{r^2}\frac{\partial }{\partial r}r^2\frac{\partial R}{\partial r} = \frac{1}{r}\frac{\partial ^2}{\partial r^2}rR(r).
\end{equation}
%Then,
%\begin{align}
%\begin{split}
%\frac{-\hbar^2}{2mr}\frac{\partial^2}{\partial r^2}(rR_0(r)+rR_1(r)\cos\theta) + \frac{\hbar^2}{2mr^2}(R_0(r)+2R_1(r)\cos\theta)\\-\frac{e^2}{4\pi \epsilon_o r}(R_0(r)+R_1(r)\cos\theta) - F_or\cos\theta  (R_0(r)+R_1(r)\cos\theta)\\ - E (R_0(r)+R_1(r)\cos\theta)=0
%\end{split}
%\end{align}
%And,
%\begin{align}
%\begin{split}
%\frac{-\hbar^2}{2m}\frac{\partial^2}{\partial r^2}(u_0(r)+u_1(r)\cos\theta) + \frac{\hbar^2}{2mr^2}(u_0(r)+2u_1(r)\cos\theta)\\-\frac{e^2}{4\pi \epsilon_o r}(u_0(r)+u_1(r)\cos\theta) - F_or\cos\theta  (u_0(r)+u_1(r)\cos\theta)\\ - E (u_0(r)+u_1(r)\cos\theta)=0
%\end{split}
%\end{align}
Then,
\begin{align}
\begin{split}
\frac{-\hbar^2}{2m_e}\frac{\partial^2}{\partial r^2} \sum_{l=0}^{L}a_lu_l(r)P_l(\cos\theta) + \sum_{l=0}^{L}\left[\frac{\hbar^2l(l+1)}{2m_er^2}\right]a_lu_l(r)P_l(\cos\theta)\\-\frac{e^2}{4\pi \epsilon_o r} \sum_{l=0}^{L}a_lu_l(r)P_l(\cos\theta) - eF_or\cos\theta  \sum_{l=0}^{L}a_lu_l(r)P_l(\cos\theta) = E \sum_{l=0}^{L}a_lu_l(r)P_l(\cos\theta).
\end{split}
\end{align}
From this we can go ahead with the Ritz method \cite{mac33} to turn this intro a matrix problem with matrix elements of the form 
\begin{equation}
H_{ll'} = \langle u_lP_l|H|u_{l'}P_{l'}\rangle,
\end{equation}
which is a sparse matrix since the Legendre polynomials are orthogonal. 
The overlap matrix is,
\begin{equation}
S_{ll'} = \langle u_lP_l|u_{l'}P_{l'} \rangle.
\end{equation}
This is diagonal. 
Since the eigenvalue problem will be solved by an approximate wave function $\tilde{\Psi}(r,\theta)$ expressed as expansion with coefficients $\vec{c}$, and the $u(r)=rR(r)$ themselves expressed in a finite element basis, the working solution will be (here $L$ is now finite along with $I$ and $M$),
\begin{equation}
\tilde{\Psi}(r,\theta) = \sum_{l=0}^{L}a_l\frac{1}{r}(\sum_{i,m}^{I,M}c_{im}f_{im})_{l}P_l(\cos\theta).
\end{equation}
One can absorb the coefficients $a_l$ into $c_{im}$, and drop the subscript $l$ on the inner sum, giving,
\begin{equation}
\tilde{\Psi}(r,\theta) = \sum_{l=0}^{L}\frac{1}{r}(\sum_{i,m}^{I,M}c_{iml}f_{im})P_l(\cos\theta).
\end{equation}
This can be written as
\begin{equation}\label{eq:psi2}
\tilde{\Psi}(r,\theta) = \sum_{i,m,l}^{I,M,L}\frac{1}{r}c_{iml}f_{im}P_l(\cos\theta).
\end{equation}
Now, the matrix elements of $H$ can be expanded with the surface structure of Eq. \ref{eq:psi1}
\begin{align}
\begin{split}
H_{ll'} = \bigg\langle u_lP_l~\bigg|\frac{-\hbar^2}{2m_e}\frac{\partial^2}{\partial r^2}\bigg|~u_{l'}P_{l'}\bigg\rangle+ \bigg\langle u_lP_l~\bigg|\frac{\hbar^2l'(l'+1)}{2m_er^2}\bigg|~u_{l'}P_{l'} \bigg\rangle\\+~ \bigg\langle u_lP_l~\bigg|\frac{-e^2}{4\pi \epsilon_o r}\bigg|~u_{l'}P_{l'} \bigg\rangle+ \bigg\langle u_lP_l~\bigg|-eF_or\cos\theta ~\bigg|~u_{l'}P_{l'} \bigg\rangle.
\end{split}
\end{align}
While concentrating a bit on the surface structure we will change notation, and square brackets will denote the angular parts of the inner product, while angled brackets will denote the radial part of the inner product. Since we are using the substitution $u(r)=rR(r)$ the radial integral does not carry an $r^2$ term, it is contained in $u(r)$. 
\\~\\Additionally we take the DC potential energy, which we will just call $\Phi$, and express it as a product so we can move the dependency around to the appropriate integrals. So just the electric field part is,
\begin{align}
\begin{split}
DC_{ll'} =  \langle u_l P_l|\Phi|u_{l'}P_{l'} \rangle
 = \langle u_l P_l|\Phi_r\Phi_{\theta}|u_{l'}P_{l'} \rangle \\= \{ u_l |\Phi_r|u_{l'}\}[ P_{l} |\Phi_{\theta}|P_{l'}].
\end{split}
\end{align}
More explicitly this can be written as,
\begin{align}
\begin{split}
DC_{ll'} =   \{ u_l |-eF_or|u_{l'} \}[ P_{l} |P_1|P_{l'}].
\end{split}
\end{align}
Where $\{ ~ \}$ contains the entire radial part of the integral and $[~]$ contains the entire angular part.
The computation of the integral $[ P_{l} |P_1|P_{l'}]$ will essentially be handled by computing the Wigner $3j$ coefficients. For this we use a program originally developed by Stone and Wood \cite{stone80} for which the source code is provided on Wood's personal web page.\footnote{http://www-stone.ch.cam.ac.uk/wigner.shtml}
\\~\\The general relationship between the triple integral of spherical harmonics and the $3j$ coefficients (we will soon be more specific to Legendre polynomials) is given by
\begin{align}\label{eq:legpoly1}
\begin{split}
\int_{0}^{2\pi}\int_{0}^{\pi}Y_{l_1m_1}(\theta,\phi)Y_{l_2m_2}(\theta,\phi)Y_{l_3m_3}(\theta,\phi)\sin\theta d\theta d\phi\\=\sqrt{\frac{(2l_1+1)(2l_2+1)(2l_3+1)}{4\pi}}
\left(\begin{tabular}{c c c }
$l_1$ &$l_2$&$l_3$\\
$m_1$ &$m_2$&$m_3$\\
\end{tabular}\right)\left(\begin{tabular}{c c c }
$l_1$ &$l_2$&$l_3$\\
$0$ &$0$&$0$\\
\end{tabular}\right).
\end{split}
\end{align}
So if $m=0$ for all the spherical harmonics then by Eq. \ref{eq:sphere} we re-write the above is,
\begin{align}
\begin{split}
\int_{0}^{2\pi}\int_{0}^{\pi}\sqrt{\frac{2l_1 +1}{4\pi}}P_{l_1}(\cos\theta) \sqrt{\frac{2l_2 +1}{4\pi}}P_{l_2}(\cos\theta) \sqrt{\frac{2l_3 +1}{4\pi}}P_{l_3}(\cos\theta) \sin\theta d\theta d\phi\\=\sqrt{\frac{(2l_1+1)(2l_2+1)(2l_3+1)}{4\pi}}
\left(\begin{tabular}{c c c }
$l_1$ &$l_2$&$l_3$\\
$0$ &$0$&$0$\\
\end{tabular}\right)^2.
\end{split}
\end{align}
Written with Wigner coefficients this is,
\begin{align}
\begin{split}
\int_{0}^{2\pi}\int_{0}^{\pi}P_{l_1}(\cos\theta) P_{l_2}(\cos\theta) P_{l_3}(\cos\theta) \sin\theta d\theta d\phi\\=
4\pi\left(\begin{tabular}{c c c }
$l_1$ &$l_2$&$l_3$\\
$0$ &$0$&$0$\\
\end{tabular}\right)^2
=4\pi W_{3j}^{2}(l_1,l_2,l_3).
\end{split}
\end{align}
Now what we really want is the deep structure of Eq. \ref{eq:psi2} when creating the matrix $H$ so that we can use our FEM techniques. This just amounts to deciding how to order the blocks associated with various quantum numbers $l$ since we have already dealt with the structure associated with the subintervals $i$ and order $m$ in the 1D hydrogen model. Recall that the global structure of that model's Hamiltonian looks like
\begin{equation}
H_{1D} = \left(\begin{tabular}{c c c c c}
$\langle f_{11}|H|f_{11}\rangle$&$\langle f_{11}|H|f_{13}\rangle$&$\langle f_{11}|H|f_{12}\rangle$&$0$&$0$\\
$\langle f_{13}|H|f_{11}\rangle$&$\langle f_{13}|H|f_{13}\rangle$&$\langle f_{13}|H|f_{12}\rangle$&$0$&$0$\\
$\langle f_{12}|H|f_{11}\rangle$&$\langle f_{12}|H|f_{13}\rangle$&$\langle f_{12}|H|f_{12}\rangle+\langle f_{21}|H|f_{21}\rangle$&$\langle f_{21}|H|f_{23}\rangle$&$\langle f_{21}|H|f_{22}\rangle$\\
$0$&$0$&$\langle f_{23}|H|f_{21}\rangle$&$\langle f_{23}|H|f_{23}\rangle$&$\langle f_{23}|H|f_{22}\rangle$\\
$0$&$0$&$\langle f_{22}|H|f_{21}\rangle$&$\langle f_{22}|H|f_{23}\rangle$&$\langle f_{22}|H|f_{22}\rangle$\\
\end{tabular}\right)
\end{equation}
at low order. The structure of this Hamiltonian is such that one row represents dot products of the vector $|f_{im}\rangle$ with all basis vectors in the representation (sans the operator $H$). Since basis vectors are taken to be zero outside of their own subinterval $\langle f_{im}|H|f_{i'm'}\rangle$ is zero unless $i=i'$. If one looks at the subscripts of each ket vector in each column going towards the right, the rows correspond to equations that mix the basis vectors (look at the the operation $\underline{H}\vec{c}$ where $\underline{H}$ is the Hamiltonian matrix).
\\~\\Similarly, if we need an extra index the rows will be dot products of the vector $|f_{im}P_l\rangle=|\varphi_{iml}\rangle$ with all basis vectors. For clear organization we choose the dot products to cycle in blocks going towards the right, the first block is dot products with $|f_{i'm'}P_0\rangle$, the second block is dot products with $|f_{i'm'}P_1\rangle$, and etc.
\\~\\The full matrix then looks like
\begin{equation}
H= \left(\begin{tabular}{c c c c}
$H^{00}$ &$H^{01}$&$H^{02}$&\dots\\
$H^{10}$ &$H^{11}$&$H^{12}$&\dots\\
$H^{20}$ &$H^{21}$&$H^{22}$&\dots\\
\vdots & \vdots & \vdots &$\ddots$\\
\end{tabular}\right).
\end{equation}
Where $H^{ll'}$ refers to the quantum numbers $l$ of the two states involved in the matrix element $\langle \varphi_{iml}| H |\varphi_{i'm'l'}\rangle$.
\\~\\In a low-order expansion, the sub-matrices look like,
\begin{equation}
H^{00} = \left(\begin{tabular}{c c c c c}
$\langle \varphi_{110}|H|\varphi_{110}\rangle$&$\langle \varphi_{110}|H|\varphi_{130}\rangle$&$\langle \varphi_{110}|H|\varphi_{120}\rangle$&$0$&$0$\\
$\langle \varphi_{130}|H|\varphi_{110}\rangle$&$\langle \varphi_{130}|H|\varphi_{130}\rangle$&$\langle \varphi_{130}|H|\varphi_{120}\rangle$&$0$&$0$\\
$\langle \varphi_{120}|H|\varphi_{110}\rangle$&$\langle \varphi_{120}|H|\varphi_{130}\rangle$&$\langle \varphi_{120}|H|\varphi_{120}\rangle+\langle \varphi_{210}|H|\varphi_{210}\rangle$&$\langle \varphi_{210}|H|\varphi_{230}\rangle$&$\langle \varphi_{210}|H|\varphi_{220}\rangle$\\
$0$&$0$&$\langle \varphi_{230}|H|\varphi_{210}\rangle$&$\langle \varphi_{230}|H|\varphi_{230}\rangle$&$\langle \varphi_{230}|H|\varphi_{220}\rangle$\\
$0$&$0$&$\langle \varphi_{220}|H|\varphi_{210}\rangle$&$\langle \varphi_{220}|H|\varphi_{230}\rangle$&$\langle \varphi_{220}|H|\varphi_{220}\rangle$\\
\end{tabular}\right),
\end{equation}
for example.
\\~\\We should make sure that the continuity condition is still satisfied. In the 1D problem, we had $c_{i,2} = c_{i+1,1}$ to enforce continuity. Now we have multiple equalities. Just in the above we see $c_{120}=c_{210}$. It is not hard to see the continuity condition would still hold. Consider the effective basis functions we must look at. In only 1D, we wanted $c_{i,2}f_{i,2} = c_{i+1,1}f_{i+1,1}$ at the boundary of two intervals. Now we want $\psi_{i,2} = \psi_{i+1,1}$ at the boundary, defined by
\begin{equation}
\psi_{i,2}(r,\theta)=\bigg(\sum_{l}c_{i,2,l}P_l(\cos\theta)\bigg)f_{i,2}(r),
\end{equation}
and,
\begin{equation}
\psi_{i+1,1}(r,\theta)=\bigg(\sum_{l}c_{i+1,1,l}P_l(\cos\theta)\bigg)f_{i+1,1}(r).
\end{equation}
But as long as $c_{i,2,l}=c_{i+1,1,l}$ (which it must because of the structure of the matrix) then 
\begin{equation}
\sum_{l}c_{i,2,l}P_l(\cos\theta) = \sum_{l}c_{i+1,1,l}P_l(\cos\theta).
\end{equation}
So since $f_{i,2} = f_{i+1,1}$ at the boundary by construction, so does $\psi_{i,2} = \psi_{i+1,1}$.
\\~\\Now we focus on setting up for the computation of the Hamiltonian matrix elements. The matrix element $H^{ll'}_{ij}$ can be written as,
\begin{equation}\label{eq:mH}
H^{ll'}_{im,i'm'} =4\pi W_{3j}^{2}(0,l,l')\{ f_{im}|\tilde{H}_{l'}|f_{i'm'}\}-4\pi eF_o W_{3j}^{2}(1,l,l')\{ f_{im}|r|f_{i'm'}\},
\end{equation}
where $\tilde{H}_{l'}$ is the Hamiltonian without the external electric field. It retains the subscript ${l'}$ from acting on $|\varphi_{i'm'l'}\rangle$.
\\~\\Again the integral represented by $\{ ~\} $ here has a similar definition as in the 1D case, it is the integral $\int_0^{\infty} dr$.
\\~\\The solution will be of the form (here we show a low order expansion for local radial basis functions)
\begin{equation}\label{eq:longc}
\vec{c} = \left(\begin{tabular}{c}
$c_{110}$\\
$c_{130}$\\
$c_{120}=c_{210}$\\
$c_{230}$\\
$c_{220}$\\
$c_{111}$\\
$c_{131}$\\
$c_{121}=c_{211}$\\
$c_{231}$\\
$c_{221}$\\
\vdots \\
\end{tabular}\right).
\end{equation}
To compute the matrix elements we treat term 1 and term 2 from Eq. \ref{eq:mH} separately. We compute the matrix $\{ f_{im}|\tilde{H}_{l'}|f_{i'm'}\}$ once for every $l'$ then copy it into the appropriate blocks $H^{ll'}$ to get the form of the global matrix associated with term 1. We then multiply the blocks by $4\pi W_{3j}^{2}(0,l,l')$ according to the block (labeled by $H^{ll'}$). We then go ahead and do a similar procedure for the second term starting with computing the matrix $\{ f_{im}|r|f_{i'm'}\}$. The two global matrices are added.
%\\~\\One may note that in the case that there is no external electric field, 
%\begin{equation}\label{eq:mH2}
%H^{ll'}_{im,i'm'} =4\pi W_{3j}^{2}(0,l,l')\langle f_{im}|\tilde{H}_{l'}|f_{i'm'}\rangle
%\end{equation}
%The term $W_{3j}^{2}(0,l,l')$ implies a $\delta_{ll'}$ since it is always zero for $l\neq l'$.
%So we whave
%\begin{equation}
%H= \left(\begin{tabular}{c c c c}
%$H^{00}$ &$0$&$0$&\dots\\
%$0$ &$H^{11}$&$0$&\dots\\
%$0$ &$0$&$H^{22}$&\dots\\
%\vdots & \vdots & \vdots &$\ddots$\\
%\end{tabular}\right)
%\end{equation}
\\~\\In the above discussion we have neglected to explicitly discuss the scaling of the radial variable. Of course, we will not be able to predict the resonant parameters until we do this. Since we have separated the integrals, only the $\{ f_{im}|\tilde{H}_{l'}|f_{i'm'}\}$ and $\{ f_{im}|r|f_{i'm'}\}$ parts are affected by the scaling. Thus we can treat these two matrices in the same way as we treated the problem in the 1D case. Once they have been scaled appropriately, all the other operations we discussed above are applied. The scaling is
\begin{equation}\label{eq:scale4}
r \rightarrow \tilde{r} = \begin{cases}
    r, ~r<r_o,\\
     e^{i\xi}(r- r_o)+ r_o, ~ r>r_o.\\
  \end{cases}
\end{equation}
We will use real-valued basis functions, so the scaling is only applied to the Jacobian and Hamiltonian. 
%Is sufficient since the wave function is expected to quickly goes to zero outside of the scaling radius, and $\tilde{r}^2 \approx r^2$ if one is not too far to the right of the scaling radius. One can see if this assumption is consistent by observing the behaviour of the wave function once it has been solved for.
\\~\\The matrix elements in the Hamiltonian matrix must, with the transformation applied, be of the form
\begin{equation}
\{ ~|H(r)|~\} \rightarrow e^{i\xi}\{~|H(\tilde{r})|~\},
\end{equation}
so the integral represented by $\{ ~\} $ remains as $\int_{0}^{\infty}dr$.
\\~\\Since there is quite a lot of similarity between the 1D problem and the 3D problem when done in the way we have described above, we can carry over much of the same transformations we got out of the scaling. The total Hamiltonian outside the scaling radius acting on $\varphi_{i'm'l'}$ will be,
\small
\begin{equation}
H\varphi_{i'm'l'} = \frac{-\hbar^2}{2m_e}e^{-2i\xi}\frac{\partial^2}{\partial r^2}\varphi_{i'm'l'} - \frac{e^2}{4\pi \epsilon_o }\frac{1}{ e^{i\xi}(r- r_o)+ r_o}\varphi_{i'm'l'}+\frac{\hbar^2l'(l'+1)}{2m_e(e^{i\xi}(r- r_o)+ r_o)^2}\varphi_{i'm'l'} - eF_o\cos\theta(e^{i\xi}(r- r_o)+ r_o)\varphi_{i'm'l'}.
\end{equation}
\normalsize
Inside it is,
\begin{equation}
H\varphi_{i'm'l'} = \frac{-\hbar^2}{2m_e}\frac{\partial^2}{\partial r^2}\varphi_{i'm'l'} - \frac{e^2}{4\pi \epsilon_or}\varphi_{i'm'l'}+\frac{\hbar^2l'(l'+1)}{2m_er^2}\varphi_{i'm'l'} -eF_or\cos\theta\varphi_{i'm'l'}.
\end{equation}
If we split $\varphi_{i'm'l'}(r,\theta)$ into $f_{im}(r)P_l(\cos\theta)$ (the indices $i,m$ and $i',m'$ will be suppressed under $i$ and $j$ respectively), we have the following definitions for the matrix elements outside the scaling radius. 
\subsection{Calculating Matrix Elements Outside The Scaling Radius}
Here $\{ i |$ and $| j \}$ are $\{ f_{im} |$ and $| f_{i'm'}\}$. Here the large square brackets denote collection of terms and do not represent any integrals. 
For the kinetic energy the matrix elements are,
\begin{equation}
KE_{ij}^{ll'}=4\pi W_{3j}^{2}(0,l,l')\cdot\frac{\hbar^2}{2m_e}e^{-i\xi}\{ \frac{d}{dr}i|\frac{d}{dr}j\}.
\end{equation}
One can see the appendix for the derivation of the symmetric form of the kinetic energy matrix element.
\\~\\For the potential energy the matrix elements are,
\begin{equation}
PE_{ij}^{ll'}=4\pi W_{3j}^{2}(0,l,l')\cdot \bigg[- \frac{e^2}{4\pi \epsilon_o}\{ i|\frac{r-r_o+r_o\cos\xi}{ (r-r_o+r_o\cos\xi )^2+r_o^2\sin^2\xi}|j\}-i\frac{e^2}{4\pi \epsilon_o}\{ i|\frac{r_o\sin\xi}{ (r-r_o+r_o\cos\xi )^2+r_o^2\sin^2\xi}|j\}\bigg].
\end{equation}
There is now additionally an angular momentum piece which is given by,
\begin{equation}
L_{ij}^{ll'}=4\pi W_{3j}^{2}(0,l,l')\cdot e^{i\xi}\{ i|\frac{\hbar^2l'(l'+1)}{2m(e^{i\xi}(r- r_o)+ r_o)^2}|j\}.
\end{equation}
That will have to be split into real and imaginary parts. First we expand,
\begin{equation}
L_{ij}^{ll'}=4\pi W_{3j}^{2}(0,l,l')\cdot e^{i\xi}\{ i|\frac{\hbar^2l'(l'+1)}{2m(e^{i\xi}(r- r_o)+ r_o)(e^{i\xi}(r- r_o)+ r_o)}|j\}.
\end{equation}
This can be written as,
\begin{equation}
L_{ij}^{ll'}=4\pi W_{3j}^{2}(0,l,l')\cdot e^{i\xi} \{ i|\frac{\hbar^2l'(l'+1)}{2m(e^{i2\xi}r^2+ (-2e^{i2\xi}+2e^{i\xi})rr_o + (e^{i2\xi}-2e^{i\xi}+1)r_o^2)}|j\}.
\end{equation}
So,
\begin{equation}
L_{ij}^{ll'}=4\pi W_{3j}^{2}(0,l,l')\cdot  \{ i|\frac{\hbar^2l'(l'+1)}{2m(e^{i\xi}r^2+ (-2e^{i\xi}+2)rr_o + (e^{i\xi}-2+e^{-i\xi})r_o^2)}|j\}.
\end{equation}
Then the denominator is,
\begin{equation}
2m(\cos(\xi)r^2 + i\sin(\xi)r^2  -2\cos(\xi)rr_o  -i2\sin(\xi)rr_o  + 2rr_o + \cos(\xi)r_o^2 + i\sin(\xi)r_o^2-2r_o^2 +\cos(\xi) r_o^2-i\sin(\xi) r_o^2).
\end{equation}
Grouping real and imaginary terms of the denominator,
\begin{align}
\begin{split}
2m\bigg[\bigg(\cos(\xi)r^2-2\cos(\xi)rr_o+ 2rr_o+2\cos(\xi)r_o^2-2r_o^2\bigg)  \\+i\bigg(\sin(\xi)r^2-2\sin(\xi)rr_o\bigg)\bigg]=2m\bigg(\mathfrak{R}+i\mathfrak{I}\bigg).
\end{split}
\end{align}
Then,
\begin{equation}
L_{ij}^{ll'}=4\pi W_{3j}^{2}(0,l,l')\cdot\{ i|\frac{\hbar^2l'(l'+1)(\mathfrak{R}-i\mathfrak{I})}{2m(\mathfrak{R}^2+\mathfrak{I}^2)}|j\}.
\end{equation}
Splitting real and imaginary parts,
\begin{equation}
L_{ij}^{ll'}=4\pi W_{3j}^{2}(0,l,l')\cdot\bigg[\{ i|\frac{\hbar^2l'(l'+1)(\mathfrak{R})}{2m(\mathfrak{R}^2+\mathfrak{I}^2)}|j\}-i\{ i|\frac{\hbar^2l'(l'+1)(\mathfrak{I})}{2m(\mathfrak{R}^2+\mathfrak{I}^2)}|j\}\bigg].
\end{equation}
For the DC electric field,
\begin{equation}
DC_{ij}^{ll'}=4\pi W_{3j}^{2}(1,l,l')\cdot\bigg[-\bigg\{ i~\bigg|eF_o \bigg(\bigg[r-r_o\bigg]\cos2\xi + r_o\cos\xi\bigg)\bigg|~j\bigg\} - i\bigg\{ i~\bigg|eF_o \bigg( \bigg[r-r_o\bigg]\sin2\xi+r_o\sin\xi \bigg)\bigg| ~j\bigg\}\bigg].
\end{equation}
Outside the scaling radius the total overlap matrix elements will be
\begin{equation}
S_{ij}^{ll'}=4\pi W_{3j}^{2}(0,l,l')\cdot e^{i\xi}\{ i|j\}.
\end{equation}
\subsection{Calculating Matrix Elements Inside The Scaling Radius}
We have the following definitions for the calculation of matrix elements inside the scaling radius. The kinetic energy is,
\begin{equation}
KE_{ij}^{ll'}=4\pi W_{3j}^{2}(0,l,l')\cdot\frac{\hbar^2}{2m_e}\{ \frac{d}{dr}i|\frac{d}{dr}j\}.
\end{equation}
For the potential energy we have,
\begin{equation}
PE_{ij}^{ll'}=4\pi W_{3j}^{2}(0,l,l')\cdot \{ i| \frac{-e^2}{4\pi \epsilon_or}|j\}.
\end{equation}
There is again an angular momentum piece,
\begin{equation}
L_{ij}^{ll'}=4\pi W_{3j}^{2}(0,l,l')\cdot\{ i|\frac{\hbar^2l'(l'+1)}{2m_er^2}|j\}.\,
\end{equation}
where $l'$ refers to the angular quantum number of $|j\}$.
\\~\\For the DC electric field we have,
\begin{equation}
DC_{ij}^{ll'}=4\pi W_{3j}^{2}(1,l,l')\cdot\{ i|-eF_o r| j\}.
\end{equation}
The overlap matrix elements are,
\begin{equation}
S_{ij}^{ll'}=4\pi W_{3j}^{2}(0,l,l')\cdot \{ i|j\}.
\end{equation}
\section{Separation of Variables and Coupled Equations for an External Field ($m \neq 0$)}\label{section:2point4}
Using spherical harmonics $Y_{ln}$ instead of Legendre Polynomials as a basis, the wave function can be expressed as,
\begin{equation}\label{eq:psi22}
\tilde{\Psi}(r,\theta,\phi) = \sum_{i,m,l}^{I,M,L}\frac{1}{r}c_{iml}f_{im}(r)Y_{ln}(\theta,\phi).
\end{equation}
Now, the matrix elements of $H$ can be expanded with the surface structure
\begin{align}
\begin{split}
H_{ij} = \langle u_l(r)Y_{ln}(\theta,\phi)|\frac{-\hbar^2}{2m}\frac{\partial^2}{\partial r^2}|u_{l'}(r)Y_{l'n}(\theta,\phi) \rangle+ \langle u_l(r)Y_{ln}(\theta,\phi)|\frac{\hbar^2l'(l'+1)}{2m_er^2}|u_{l'}(r)Y_{l'n}(\theta,\phi) \rangle\\+ \langle u_l(r)Y_{ln}(\theta,\phi)|\frac{-e^2}{4\pi \epsilon_o r}|u_{l'}(r)Y_{l'n}(\theta,\phi) \rangle+ \langle u_l(r)Y_{ln}(\theta,\phi)|-eF_or\cos\theta |u_{l'}(r)Y_{l'n}(\theta,\phi) \rangle.
\end{split}
\end{align}
Then just the DC field part where $F = -eF_or\cos\theta$,
\begin{align}
\begin{split}
DC_{ij} =  \langle u_l Y_{ln}(\theta,\phi)|F|u_{l'}Y_{l'n}(\theta,\phi) \rangle
 = \langle u_l Y_{ln}(\theta,\phi)|F_r(r)F_{\theta}(\theta)|u_{l'}Y_{l'n}(\theta,\phi) \rangle \\= \{ u_l |F_r(r)|u_{l'}\}[ Y_{ln} |F_r(\theta)|Y_{l'n}].
\end{split}
\end{align}
Explicitly,
\begin{align}\label{eq:DC}
\begin{split}
DC_{ij} =   \{ u_l |-eF_or|u_{l'}\}[ Y_{ln} |P_1|Y_{l'n}].
\end{split}
\end{align}
The general relationship between the integral of three spherical harmonics and the $3j$ coefficients is given by Eq. \ref{eq:legpoly1}.
So the integral in Eq. \ref{eq:DC} is,
\begin{align*}
\begin{split}
\sqrt{\frac{4\pi}{3}}\int_{0}^{2\pi}\int_{0}^{\pi}Y_{10}(\theta,\phi)Y_{l_2m_2}(\theta,\phi)Y_{l_3m_3}(\theta,\phi)\sin\theta d\theta d\phi\\=\sqrt{\frac{4\pi}{3}}\sqrt{\frac{(3)(2l_2+1)(2l_3+1)}{4\pi}}
\left(\begin{tabular}{c c c }
$1$ &$l_2$&$l_3$\\
$0$ &$m_2$&$m_3$\\
\end{tabular}\right)\left(\begin{tabular}{c c c }
$1$ &$l_2$&$l_3$\\
$0$ &$0$&$0$\\
\end{tabular}\right)
\\=\sqrt{(2l_2+1)(2l_3+1)}
\left(\begin{tabular}{c c c }
$1$ &$l_2$&$l_3$\\
$0$ &$m_2$&$m_3$\\
\end{tabular}\right)\left(\begin{tabular}{c c c }
$1$ &$l_2$&$l_3$\\
$0$ &$0$&$0$\\
\end{tabular}\right).
\end{split}
\end{align*}
The same integral with $P_1\rightarrow P_0$ (for the matrix elements other than the DC field) is,
\begin{align*}
\begin{split}
\sqrt{4\pi}\int_{0}^{2\pi}\int_{0}^{\pi}Y_{00}(\theta,\phi)Y_{l_2m_2}(\theta,\phi)Y_{l_3m_3}(\theta,\phi)\sin\theta d\theta d\phi\\=\sqrt{4\pi}\sqrt{\frac{(1)(2l_2+1)(2l_3+1)}{4\pi}}
\left(\begin{tabular}{c c c }
$0$ &$l_2$&$l_3$\\
$0$ &$m_2$&$m_3$\\
\end{tabular}\right)\left(\begin{tabular}{c c c }
$0$ &$l_2$&$l_3$\\
$0$ &$0$&$0$\\
\end{tabular}\right)
\\=\sqrt{(2l_2+1)(2l_3+1)}
\left(\begin{tabular}{c c c }
$0$ &$l_2$&$l_3$\\
$0$ &$m_2$&$m_3$\\
\end{tabular}\right)\left(\begin{tabular}{c c c }
$0$ &$l_2$&$l_3$\\
$0$ &$0$&$0$\\
\end{tabular}\right).
\end{split}
\end{align*}
\section{Stark Results for Hydrogen and Singly Ionized Helium}\label{section:2point5}
The following tables are data for ECS with a 3D hydrogen-like atom in a DC field represented by:
\begin{equation}
\bigg[\frac{-\hbar^2}{2m_e}\nabla^2 - \frac{Ze^2}{4\pi \epsilon_o r} - eF_or\cos\theta\bigg]\Psi = E\Psi.
\end{equation}
The tables are for hydrogen ($Z=1$) unless stated otherwise. One of the tables is for singly ionized helium ($Z=2$). 
The quantum numbers are designated by $n_1, n_2, m$ to describe the wave function in parabolic co-ordinates \cite{beth13,hey07,tel89} rather than $n,l,m$ since the addition of the electric field means that the eigenstates of the Hamiltonian will no longer generally be eigenstates of $L^2$ as they are for free hydrogen. The quantum number $m$ remains the same between the two systems. Of course, spherical harmonics can still be used as a basis since they, along with suitable basis functions for the radial part, span 3-space. It is interesting to note that the hydrogen system without a perturbation can be solved in the same co-ordinates with the same quantum numbers \cite{hey07}.\footnote{The parabolic co-ordinates $\xi$, $\eta$, $\phi$, are related to the spherical polar co-ordinates by
\begin{equation*}
r = \frac{1}{2}(\xi+\eta),
\end{equation*}
\begin{equation*}
\theta = \cos^{-1}\bigg(\frac{\xi-\eta}{\xi+\eta}\bigg),
\end{equation*}
\begin{equation*}
\phi =\phi.
\end{equation*}}
\\~\\However, the quantum numbers which designate a spherical harmonic cannot be used for the problem with a Stark perturbation.
\\~\\This becomes clear at the first order time-independent correction to the wave function, which has contributions for $L^2$ eigenstates which have a difference in angular momentum of $\Delta \ell =1$ (with the same quantum number $m$) measured against the original state for a perturbation of the form $\sim \cos\theta$. One can see a similar phenomenon occur in the time-dependent formalism in Fig. \ref{fig:figfit}, where states with $\Delta \ell =1$ initially become occupied, which later fuel occupation of states $\Delta \ell =2$, etc. The message is that the resultant state is no longer a single spherical harmonic.
\\~\\The radial variable is scaled into the complex plane by Eq. \ref{eq:scale4} to obtain the resonance parameters.
\\~\\In the following tables we explore the convergence and stability characteristics of our method. In Table 2.1 and Table 2.2 we explore the convergence characteristics with various sizes of radial and angular basis. The radial box is of size $r \in [0,100]$, with 100 subintervals for the FEM basis, leading to a $1~au$ box for every set of radial basis functions. The scaling radius is at $r = 10~au$. This box is actually larger than it needs to be, as around $20~au$ the probability density has settled to approximately zero ($\sim 3\cdot 10^{-10}$) which can be seen in Fig. $\ref{fig:ECSind}$ of the next chapter. As we increase the number of basis functions, the number of radial basis functions per subinterval is held equal to the number of total $L^2$ eigenstates.
\\~\\In Table 2.3 we explore the movement of the resonance with the scaling angle. Over a domain $\xi \in [0.2,1.0]$ in radians, the real part of the eigenvalue does not change. The imaginary part is fixed up to seven decimal places. However, changing the scaling angle to $1.2 ~rad$ breaks this consistency at the third decimal place in the real part, and the first decimal place in the imaginary part. This angle is rather close to $\pi/2$, which means that the radial co-ordinate has been scaled to the point it is almost purely imaginary. In the 1D hydrogen problem we treated in the previous chapter, this angle did not cause any problems. The cause of this is not straight forward, as the 1D hydrogen potential does go like $-1/x$ for large distances like the Coulomb potential. 
\\~\\However, one difference between the two problems is that we apply an electric field in the $z$ direction, so it is not entirely in the direction of the scaled variable, unlike the 1D case, where the field is in the $x$ direction. This may be the cause of the failure at the larger scaling angle and can be investigated in the future.
%\\~\\However, the radial momentum operator is not exactly like the linear momentum operator, since it only acts on functions over $r \in [0,\infty]$ as opposed to the linear momentum operator for $x$ which acts on functions over $x \in [-\infty, \infty]$. This means it is difficult to establish orthogonal radial momentum eigenstates since there is no symmetric cancellation of oscillations for the inner product between two eigenstates of the radial momentum operator. 
\begin{center}
Table 2.1: For the table below we use constants $F_o = 0.1 ~au$, and $\xi = 0.5 ~rad$. We show convergence of ECS with various parameters for the first resonant state ($n_1 = n_2 = m = 0$).\\~\\
\def\arraystretch{0.8}
\begin{tabular}{cccccc}

\toprule
 & (0, 100)$\times100$, $10.0~au$\\
 \midrule
(5,5) &  -0.527395257 -0.723420936(E-2)i
  &  &  \\
 \midrule
(6,6) &  -0.527427227 -0.72701379(E-2)i
 &  &  \\
 \midrule
(7,7) &  -0.527416500 -0.727033161(E-2)i
  &  &  \\
 \midrule
(8,8) & -0.527418278 -0.726857508(E-2)i
  &  &  \\
   \midrule
& -0.527418175 -0.726905676(E-2)$i$ \cite{tel89}  & \\
\bottomrule
\end{tabular}
\\~\\Intervals are shown on the top in the form $(start, end)\times subintervals,~r_o$, and number of basis functions (per subinterval for the radial basis functions; the angular basis functions are global) are shown on the left in the form $(\# ~radial, \# ~angular)$. Values of the first resonant eigenvalue are shown inside in atomic units.
\end{center}
\noindent 

\begin{center}
Table 2.2: Here $F_o = 0.5~au$, and $\xi = 0.5 ~rad$. We show convergence of ECS with various parameters for the first resonant state ($n_1 = n_2 = m = 0$).
\\~\\
\def\arraystretch{0.8}
\begin{tabular}{cccccc}
\toprule
 & (0, 100)$\times100$, $10.0~au$\\
 \midrule
(5,5) &  -0.623564039 -0.283481219i  
  &  &  \\
 \midrule
(6,6) &  -0.622693013 -0.277978947i   
 &  &  \\
 \midrule
(7,7) & -0.623355290     -0.280137628i  
  &  &  \\
 \midrule
(8,8) & -0.6229228448     -0.279640238i 
  &  &  \\
  \midrule
  & -0.623068026 -0.279744825$i$ \cite{tel89} & &
  \\
\bottomrule
\end{tabular}
\\~\\Intervals are shown on the top in the form $(start, end)\times subintervals,~r_o$, and number of basis functions (per subinterval for the radial basis functions; the angular basis functions are global) are shown on the left in the form $(\# ~radial, \# ~angular)$. Values of the first resonant eigenvalue are shown inside in atomic units.
\end{center}
\noindent
\begin{center}
Table 2.3: Here $F_o = 0.1~au$ again, and the number of basis functions is (5,5). We show the stability of ECS with respect to the scaling angle for the first resonant state ($n_1 = n_2 = m = 0$).
\\~\\
\def\arraystretch{0.8}
\begin{tabular}{cccc}
\toprule
 & (0, 100)$\times100$, $10.0 ~au$\\
\midrule
0.2 & -0.527395257 -0.723420920(E-2)i
\\
\midrule
0.5 & -0.527395257 -0.723420936(E-2)i &  & \\
\midrule
1.0 &  -0.527395257 -0.723420919(E-2)i &  & \\
\midrule
1.2 &  -0.526019077 -0.898024868(E-2)i &  & \\
\midrule
 &  -0.527418175 -0.726905676(E-2)$i$ \cite{tel89} &  & \\

\bottomrule
\end{tabular}
\\~\\Intervals are shown on the top in the form $(start, end)\times subintervals,~r_o$, and scaling parameter $\xi$ is on the left in $rad$. Values of the first resonant eigenvalue are shown inside in atomic units.
\end{center}
In Table 2.4 and Table 2.5 below, we study the resonances for the first excited state of singly ionized helium. More results are in our appendix. These results were not found in a literature search so they appear to be novel. 
\begin{center}
Table 2.4: Here we use $Z=2$ and thereby turn the above problem to the singly ionized helium problem. We also have $\xi=0.5$ $rad$ and $r_o = 10.0~au$, and for all eigenvalues, (0, 100)$\times100$, [7,7]. We explore ECS results for different field strengths for the first {\it excited} state of singly ionized helium (with $n_1 = n_2 = 0$, and $m = 1$). \\~\\
\begingroup
\def\arraystretch{0.8}
 \fontsize{9pt}{9pt}\selectfont
\begin{tabular}{cccc}
\toprule
 & $E$\\
 \midrule
  $F_o = 0.0005$ & -0.50000121 -0.64759586(E-15)i\\
\midrule
 $F_o = 0.005$ & -0.500122010 -0.618901281(E-15)i\\
\midrule
 $F_o = 0.01$ & -0.500489699  -0.131905200(E-14)i\\
\midrule
 $F_o = 0.018$ & -0.501603414 -0.106372922(E-12)i\\
\midrule
 $F_o = 0.02$ & -0.501986914  -0.533728998(E-11)i\\
\midrule
 $F_o = 0.022$ & -0.502414339      -0.123256709(E-9)i\\
\midrule
 $F_o = 0.024$ &  -0.502886979  -0.160512137(E-8)i\\
\midrule
 $F_o = 0.026$ & -0.503406404   -0.134859660(E-7)i\\
\midrule
 $F_o = 0.028$ & -0.503974558 -0.804693231(E-7)i\\
\midrule
 $F_o = 0.03$ & -0.504593894  -0.365771017(E-6)i\\
\midrule
 $F_o = 0.032$ & -0.505267516    -0.133462391(E-5)i\\
\midrule
 $F_o = 0.034$ & -0.505999257  -0.406797401(E-5)i\\
\midrule
 $F_o = 0.036$ & -0.506793607     -0.106818094(E-4)i\\
\midrule
 $F_o = 0.038$ & -0.507655401     -0.247550737(E-4)i\\
\midrule
 $F_o = 0.04$ & -0.508589263  -0.516194983(E-4)i\\
\midrule
 $F_o = 0.045$ & -0.511259479  -0.224738323(E-3)i\\
\midrule
$F_o = 0.05$ & -0.514405054  -0.663062430(E-3)i\\
\midrule
 $F_o = 0.06$ & -0.521788028     -0.278888570(E-2)i\\
 \midrule
 $F_o = 0.07$ & -0.529860127   -0.6786615104(E-2)i\\

 \midrule
 $F_o = 0.0$ &   $-Z/n^2=-2/n^2 = -0.5$\\
\bottomrule

\end{tabular}
\endgroup
 \\~\\Values of the first {\it excited} resonant eigenvalue are shown inside in atomic units.
\end{center}
\begin{center}
Table 2.5: Here we use $Z=2$ and thereby turn the above problem to the singly ionized helium problem. We also have $\xi=0.5$ $rad$ and $r_o = 10.0~au$, and for all eigenvalues, (0, 100)$\times100$, [7,7]. We explore ECS results for different field strengths for the first {\it excited} resonant state of singly ionized helium. The states are listed with subscripts $n_1,n_2,m$. \\~\\
\begingroup
\def\arraystretch{0.8}
 \fontsize{9pt}{9pt}\selectfont
\begin{tabular}{cccc}
\toprule
  & $E_{010}$&$E_{100}$\\
 \midrule
 $F_o = 0.0005$ & -0.500751313  -0.213476595(E-14)i&-0.499251310  -0.524848551(E-15)i\\
\midrule
 $F_o = 0.005$ & -0.507632936    -0.104345174(E-15)i&-0.492629878 -0.151457250(E-14)i\\
\midrule
 $F_o = 0.01$ & -0.515539926 -0.317310083(E-15)i&-0.485515194  -0.901263980(E-15)i\\
\midrule
 $F_o = 0.018$ &-0.528803660    -0.606870297(E-11)i&-0.474654296    -0.201288390(E-13)i\\
\midrule
 $F_o = 0.02$ & -0.532247036  -0.180149176(E-9)i&-0.472039514    -0.501988165(E-12)i\\
\midrule
 $F_o = 0.022$ &-0.535745614    -0.274875236(E-8)i&-0.469465289    -0.985656723(E-11)i\\
\midrule
 $F_o = 0.024$ &-0.539302237     -0.255448137(E-7)i&-0.466932002    -0.149537729(E-9)i\\
\midrule
 $F_o = 0.026$ &-0.542920427      -0.162525328(E-6)i&-0.464440172      -0.153398111(E-8)i\\
\midrule
 $F_o = 0.028$ &-0.546604648     -0.769153607(E-6)i&-0.461990484     -0.108857305(E-7)i\\
\midrule
 $F_o = 0.03$ & -0.550360580 -0.287579323(E-5)i&-0.459583833  -0.568965125(E-7)i\\
\midrule
 $F_o = 0.032$ & -0.554195212    -0.888324194(E-5)i&-0.457221383     -0.232105364(E-6)i\\
\midrule
 $F_o = 0.034$ & -0.558116536     -0.234488782(E-4)i&-0.454904617      -0.774269420(E-6)i\\
\midrule
 $F_o = 0.036$ & -0.562132686      -0.542874260(E-4)i&-0.452635370      -0.219064708(E-5)i\\
\midrule
 $F_o = 0.038$ & -0.566250567     -0.112507648(E-3)i&-0.450415790     -0.541013868(E-5)i\\
\midrule
 $F_o = 0.04$ & -0.570474331   -0.212191915(E-3)i&-0.448248251   -0.119325253(E-4)i\\
\midrule
 $F_o = 0.045$ & -0.581487242 -0.746731962(E-3)i&-0.443072637    -0.588090391(E-4)i\\
\midrule
$F_o = 0.05$ & -0.593042877     -0.185968860(E-2)i&-0.438267034     -0.195139214(E-3)i\\
\midrule
$F_o = 0.06$ & -0.616982341      -0.617046650(E-2)i&-0.429762048      -0.103352117(E-2)i\\
\midrule
$F_o = 0.07$ & -0.640971047   -0.130101241(E-1)i&-0.422432683  -0.307006557(E-2)i\\
 \midrule
 $F_o = 0.0$ &   $-Z/n^2=-2/n^2 = -0.5$&   $-Z/n^2=-2/n^2 = -0.5$\\
\bottomrule
\end{tabular}
\endgroup
 \\~\\Values of the first {\it excited} resonant eigenvalue are shown inside in atomic units. At low field strength we find the two eigenvalues converging, along with $E_{001}$ in the previous table.
\end{center}
We display the results from the previous tables for singly ionized helium graphically in the next figure. Depending on the strength of the field, the resonance can be classified as either over-the-barrier or under-the-barrier ionization. To find the dividing line for a state we start by interpolating between the real parts of the eigenvalues for the state, $Re(E)$, as a function of $F_o$. We then find the potential height at the local maximum of the barrier located at $r^*$, for $\cos\theta =1$ (so $-eF_o r^*\cos\theta \rightarrow -eF_o r^*$). That is, $H = V(r^*)-eF_o r^*$ for each $F_o$. We then find the interpolated eigenvalue, $Re(E)_{in}$, for which $Re(E)_{in} = H = V(r^*)-eF_o r^*$. The specific value of $F_o$ for this $Re(E)_{in}$ becomes $F_{crit}$, the dividing line between the tunnelling and over-the-barrier regimes for the state. The decay rates are in $au$ on the left, and in $1/s$ on the right. Right hand side values are calculate as follows: CODATA lists the atomic unit of time as $\hbar/E_h \approx  2.418 \times 10^{-17} ~s$ where $E_h$ is the atomic unit of energy. That makes the atomic unit of inverse time to be $E_h/\hbar \approx 4.134 \times 10^{16} ~s^{-1}$. So the width in inverse seconds is $E_h\Gamma/\hbar \approx \Gamma 4.134 \times 10^{16} ~s^{-1}$ where $\Gamma$ is the width in $au$.
\begin{figure*} [!h]
\centering
\includegraphics[width=350pt]{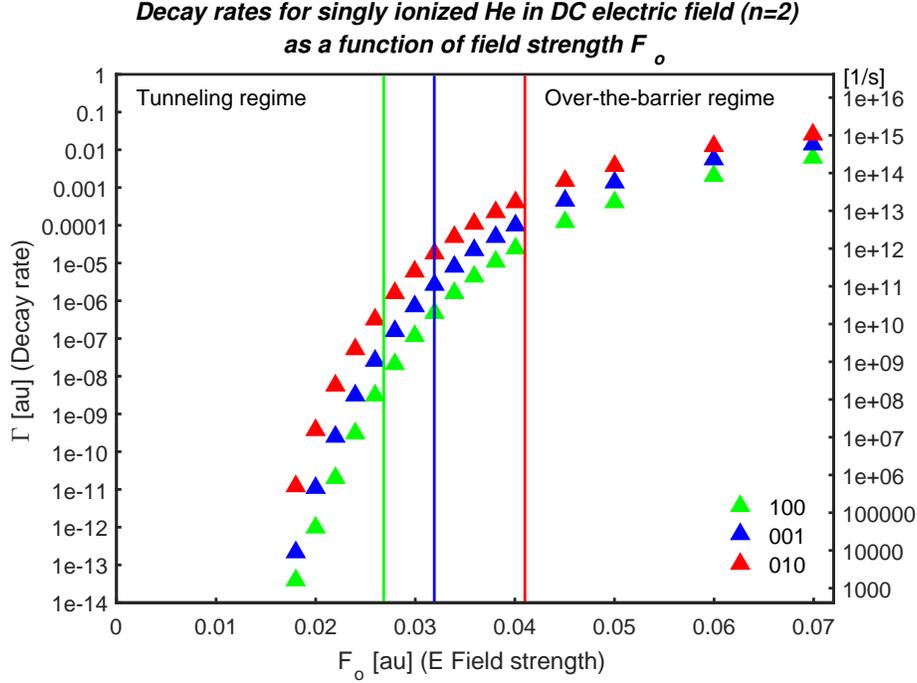}
\vspace{0.2cm}
\caption{\label{fig:decay}
 Plot of decay rate as a function of field strength for all three "first excited" states (for $n_1,n_2,m$ = 100, 010, 001) of singly ionized helium. The three vertical lines separate the tunnelling and over-the-barrier regimes for all three states. }
\end{figure*}

\section{Conclusion}\label{section:2point6}
We find that the 1D algorithm is not overly difficult to port to a 3D case. For weak external fields, a spherical basis converges quicker than for strong fields, because the mixing between spherical harmonics is stronger for stronger fields (therefore, more complicated, higher-order basis functions are needed). Furthermore, a high level of stability is still found with respect to the scaling angle. Our results for the Stark resonances of hydrogen match well with another methodology \cite{tel89} which is based on a parabolic co-ordinate system with complex rotated Coulomb-Sturmian basis functions. In that calculation the wave equation was split into two parts with no more than 50 basis functions for both solutions. We used on order of $\sim10$ basis functions for both the radial and angular parts. Our solutions generally match up to 3-4 decimal places for $\xi = 0.5$ for the most accurate runs that we did. Since it was not so difficult to port the code we also found an analogous result for singly ionized helium but for the first excited state. More results can be found in $\S \ref{section:5point01}$ of our appendix.
\newpage
\chapter{A 3D Time-dependent Model for Resonances}
\epigraph{Nature is crooked. I wanted right angles, straight lines.}{Allie Fox, {\it The Mosquito Coast}}
%\epigraph{The "paradox" is only a conflict between reality and your feeling of what reality "ought to be."}{Richard Feynman}
\section*{Overview}
In this chapter we discuss the implementation of a time-dependent solution of the Schr{\"o}dinger equation for hydrogen without and with ECS. In $\S \ref{section:3point1}$ we discuss our notation.
In $\S \ref{section:3point2}$ we discuss our solution of the time-dependent Schr{\"o}dinger equation using the Runge-Kutta 4 (RK4) method. In $\S \ref{section:3point3}$ we discuss the solution of the matrix problem for each time step and show our results. In $\S \ref{section:3point4}$ discuss the generalization to ECS and show our results. In $\S \ref{section:3point5}$ we discuss our conclusions.
\section{Notation} \label{section:3point1}
In this chapter we use the same notation as in the previous. For easy reference within the chapter we will repeat it here. The symbols $\langle~\rangle$ represent the volume integral with no radial Jacobian. We separate radial and angular parts as $\{ ~\}$ and $[~~]$ respectively,  such that $\{ ~\}[~~] =\int_0^{\infty}...dr \int_0^{2\pi} \int_0^{\pi}...~ \sin\theta d\theta d\phi$. When writing matrix elements with $\varphi_{iml}(r,\theta)$ or $\varphi_{iml}(r,\theta,\phi)$ (including $r\rightarrow \tilde{r}$), such as $\langle \varphi_{iml}|\mathcal{O}|\varphi_{i'm'l'}\rangle$, the brackets $\langle ~ \rangle = \int_0^{\infty}dr \int_0^{2\pi} \int_0^{\pi} \sin\theta d\theta d\phi$. The same integrals are later written with $f_{im}$ functions, and the angular parts are taken care of by Wigner $3j$ coefficients. These take the form $W_{3j}^{2}(...)\{ f_{im}|\mathcal{O}|f_{i'm'}\}=W_{3j}^{2}(...)\{ i|\mathcal{O}|j\}$. The definition $\{ ~ \} =\int_0^{\infty}dr$ is again used.
\\~\\Additionally, since we retain the notation from our previous chapter, where $m$ indicates the order of the local radial basis function, when $f_{im}$ appears together with spherical harmonics $Y_{lm}$, the conflict shall be solved by writing $Y_{ln}$. Everywhere else spherical harmonics shall be written $Y_{lm}$.
\\~\\Solutions are given in combined gaussian and atomic units. In particular $e^2/4\pi\epsilon_o \rightarrow e^2$, and further, $\hbar = m_e = e = 1$. For simplicity, we will just refer to this as atomic units.
\section{Use of the Runge-Kutta 4 Method}\label{section:3point2}
The time-dependent Schr{\"o}dinger equation (TDSE) in matrix form is,
\begin{equation}
i \hbar \underline{S}\frac{\partial }{\partial t}\vec{c}(t) = \underline{H}(t) \vec{c}(t),
\end{equation}
where $\underline{S}$ is the overlap matrix and $\underline{H}$ is the Hamiltonian matrix. 
\\~\\To solve this we will employ a fourth-order Runge-Kutta algorithm (RK4). However, as stated now this equation is not well-suited to be solved by an RK4 algorithm. First we must act on the left on both sides of the equation with $\underline{S}^{-1}$ and isolate the time derivative of $\vec{c}(t)$. This gives us,
\begin{equation}\label{eq:Sequation}
\frac{\partial }{\partial t}\vec{c}(t) = \frac{1}{i \hbar}\underline{S}^{-1}\underline{H}(t) \vec{c}(t),
\end{equation}
which can now be solved by an RK4 algorithm.
\section{Implementation Without ECS}\label{section:3point3}
We start with the time-dependent Schr{\"o}dinger equation (TDSE) for the hydrogen atom in an external DC electric field in spherical-polar co-ordinates. The mass of the proton is assumed to be infinite. This is,
\begin{equation}
H(t)\Psi(r,\theta,\phi,t) = \bigg[\frac{-\hbar^2}{2m_e}\nabla^2 - \frac{e^2}{4\pi \epsilon_o r} - eF_o(t)r\cos\theta\bigg]\Psi(r,\theta,\phi,t) = i\hbar\frac{\partial }{\partial t}\Psi(r,\theta,\phi,t).
\end{equation}
We will use a partial wave expansion (we enforce azimuthal symmetry of the problem) with basis vectors of the form
\begin{equation}
\Psi_l(r,\theta) = \frac{u_l(r)}{r}P_l(\cos\theta)= R_l(r)P_l(\cos\theta).
\end{equation}
Since the TDSE will be solved by an approximate wave function $\tilde{\Psi}(r,\theta,t) = \psi(r,\theta,t)/r$ expressed as an expansion with time-dependent coefficients $c_{iml}$, the working solution to the TDSE will be 
\begin{equation}\label{eq:psi2_2}
\tilde{\Psi}(r,\theta,t) = \sum_{i,m,l}\frac{1}{r}c_{iml}(t)f_{im}(r)P_l(\cos\theta)= \sum_{i,m,l}\frac{1}{r}c_{iml}(t)\varphi_{iml}(r,\theta),
\end{equation}
which we will refer to as $\Psi(r,\theta,t)$ with the understanding that the solution is approximate since the sums will be truncated.
\subsection{The Hamiltonian Matrix}\label{section:3point3point1}
The full matrix for this problem then looks like
\begin{equation}
H_{ij}(t)= \left(\begin{tabular}{c c c c}
$H^{00}(t)$ &$H^{01}(t)$&$H^{02}(t)$&\dots\\
$H^{10}(t)$ &$H^{11}(t)$&$H^{12}(t)$&\dots\\
$H^{20}(t)$ &$H^{21}(t)$&$H^{22}(t)$&\dots\\
\vdots & \vdots & \vdots &$\ddots$\\
\end{tabular}\right).
\end{equation}
Here $ll'$ in $H^{ll'}(t)$ refers to the $l$ quantum numbers of the two states involved in the matrix element $\langle \varphi_{iml}| H |\varphi_{i'm'l'}\rangle$. The subscripts $i$ and $j$ are $iml$ and $i'm'l'$ respectively.
\\~\\In a low order expansion, the sub-matrices have a block structure, e.g.,
\begin{equation}
H^{10}(t) = \left(\begin{tabular}{c c c c c}
$\langle \varphi_{111}|H|\varphi_{110}\rangle$&$\langle \varphi_{111}|H|\varphi_{130}\rangle$&$\langle \varphi_{111}|H|\varphi_{120}\rangle$&$0$&$0$\\
$\langle \varphi_{131}|H|\varphi_{110}\rangle$&$\langle \varphi_{131}|H|\varphi_{130}\rangle$&$\langle \varphi_{131}|H|\varphi_{120}\rangle$&$0$&$0$\\
$\langle \varphi_{121}|H|\varphi_{110}\rangle$&$\langle \varphi_{121}|H|\varphi_{130}\rangle$&$\langle \varphi_{121}|H|\varphi_{120}\rangle+\langle \varphi_{211}|H|\varphi_{210}\rangle$&$\langle \varphi_{211}|H|\varphi_{230}\rangle$&$\langle \varphi_{211}|H|\varphi_{220}\rangle$\\
$0$&$0$&$\langle \varphi_{231}|H|\varphi_{210}\rangle$&$\langle \varphi_{231}|H|\varphi_{230}\rangle$&$\langle \varphi_{231}|H|\varphi_{220}\rangle$\\
$0$&$0$&$\langle \varphi_{221}|H|\varphi_{210}\rangle$&$\langle \varphi_{221}|H|\varphi_{230}\rangle$&$\langle \varphi_{221}|H|\varphi_{220}\rangle$\\
\end{tabular}\right)
\end{equation}
for example. All the Hamiltonian matrix elements above are functions of time.
\\~\\The matrix element $H^{ll'}_{im,i'm'}(t)$ can be written as,
\begin{equation}\label{eq:mH2}
H^{ll'}_{im,i'm'}(t) =4\pi W_{3j}^{2}(0,l,l')\{ f_{im}|\tilde{H}_{l'}|f_{i'm'}\}-4\pi eF_o(t) W_{3j}^{2}(1,l,l')\{ f_{im}|r|f_{i'm'}\}.
\end{equation}
Here $W_{3j}^{2}(...)$ are the Wigner $3j$ coefficients squared and $\tilde{H}_{l'}$ is the Hamiltonian without the external electric field. It retains the subscript ${l'}$ from acting on $|\varphi_{i'm'l'}\rangle$.
\\~\\The eigenvectors will be of the same form as Eq. \ref{eq:longc}.
\\~\\We have the following definitions for the matrix elements. In the following definitions, $i,m,i',m'$ is also referred to as $i,j$, where $i = im$ and $j = i'm'$. Beginning with kinetic energy we have:
\begin{equation}
KE_{ij}^{ll'}=4\pi W_{3j}^{2}(0,l,l')\cdot\frac{\hbar^2}{2m_e}\{ \frac{d}{dr}i|\frac{d}{dr}j\}.
\end{equation}
For the potential energy, we have
\begin{equation}
PE_{ij}^{ll'}=4\pi W_{3j}^{2}(0,l,l')\cdot \{ i| \frac{-e^2}{4\pi \epsilon_or}|j\}.
\end{equation}
There is again an angular momentum piece, that is given by
\begin{equation}
L_{ij}^{ll'}=4\pi W_{3j}^{2}(0,l,l')\cdot\{ i|\frac{\hbar^2l'(l'+1)}{2m_er^2}|j\},
\end{equation}
where $l'$ refers to the angular quantum number of $|j\}$.
\\~\\For the DC electric field,
\begin{equation}
DC_{ij}^{ll'}=4\pi W_{3j}^{2}(1,l,l')\cdot\{ i|-eF_o(t) r| j\}.
\end{equation}
The overlap matrix elements are,
\begin{equation}
S_{ij}^{ll'}=4\pi W_{3j}^{2}(0,l,l')\cdot\{ i|j\}.
\end{equation}
\subsection{Solution}
To solve the above, we begin by solving the matrix problem for the Hamiltonian without the external field to find the initial eigenvector $\vec{c}_o$. This is used as the initial condition during solution of the time-dependent problem. The external DC field is then adiabatically switched on by the function $F_o(t)$ during the solution.
\\~\\The function $F_o(t)$ is described by
\begin{equation}
F_o(t) = \begin{cases}
    F_o \sin^2(\omega t), ~t\le t_o,\\
    F_o,~ t> t_o,\\

  \end{cases}
\end{equation}
where $\omega = \pi/2t_o$.
The external DC field is
\begin{equation}
F(t) = \begin{cases}
    eF_o \sin^2(\omega t)r\cos\theta,~t\le t_o,\\
    eF_or\cos\theta,~ t> t_o.\\

  \end{cases}
\end{equation}
The time evolution of the system of equations is defined by Eq. \ref{eq:Sequation}.
where $\underline{H}=H_{ij}$ defined in the equations of $\S \ref{section:3point3point1}$. This is solved using the RK4 algorithm.
When $\vec{c}(t)$ are found, we use
\begin{equation}\label{eq:psic}
\Psi(r,\theta,t)r = \psi(r,\theta,t) = \sum_{i,m,l}c_{iml}(t)f_{im}(r)P_l(\cos\theta)
\end{equation}
to find the adjusted wave function.
\\~\\For the choice $F_o$ = 0.1 $au$ and $t_o$ = 10 $au$, we plot $|\psi(r,\theta=0,t)|
^2=r^2|\Psi(r,\theta=0,t)|^2$ for an initial state of 1s at various time points in Fig. \ref{fig:fig2_2}.
\begin{figure*} [!h]
\centering
\includegraphics[width=350pt]{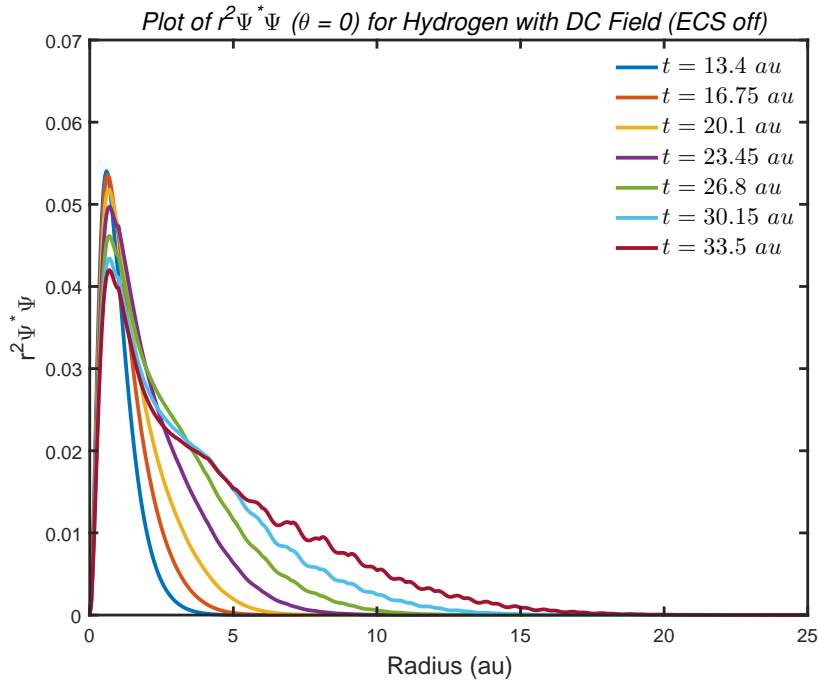}
\vspace{0.2cm}
\caption{\label{fig:fig2_2}
 Wave function profile for an initial state of 1s over an important range of time values. To be compared later to the case with ECS on with the same range.}
\end{figure*}
We note that the probability does not necessarily "appear" to be conserved here, although time evolution should always be unitary for a real Hamiltonian. However, we are plotting $|\tilde{\psi}(r,\theta=0,t)|^2$ so it is only the probability density at a single $\theta$ point and not the total integrated probability. In Fig. \ref{fig:fig_leg}, the initial state is the same as in Fig. \ref{fig:fig2_2}; the norm is unity for all times and $\psi(\vec{r},0)=\psi_{1s}(\vec{r})$. We denote the angular momentum eigenstate by $P_l$ since the quantum number $m=0$ in this example, so, the angular function associated with an angular momentum eigenstate is a Legendre polynomial. Notice the behaviour here matches what one would expect for an external field with one unit of angular momentum: the state picks up $l=1$ probability from the interaction between $P_0$ and $P_1$ achieved by the $\cos\theta$ term in the external field. Once $P_1$ begins to fill $P_2$ can as well by the same mechanism (population flows from from $P_0$ to $P_1$, then from $P_1$ to $P_2$). This shows that the external field causes eigenstates of the $L^2$ operator to mix.
%To check for norm-squared conservation we must check that 
%\begin{equation}
%\int r^2|\psi(r,\theta,t)|^2 dr\sin\theta d\theta d\phi= constant,~for~all~t
%\end{equation}
%However, to check that the norm-squared is really decreasing inside the potential well (that is, inside the atom), we suggest calculating 
%\begin{equation}
%\int_0^{2\pi} \int_0^{\pi} \int_{0}^{r_o}r^2|\psi(r,\theta,t)|^2 dr\sin\theta d\theta d\phi = P(t)
%\end{equation}
%Where $r_o$ should probably be around the scaling radius used for ECS. %ADD TO THIS THE ACTUAL RESULT.
\begin{figure*} [!h]
\centering
\includegraphics[width=350pt]{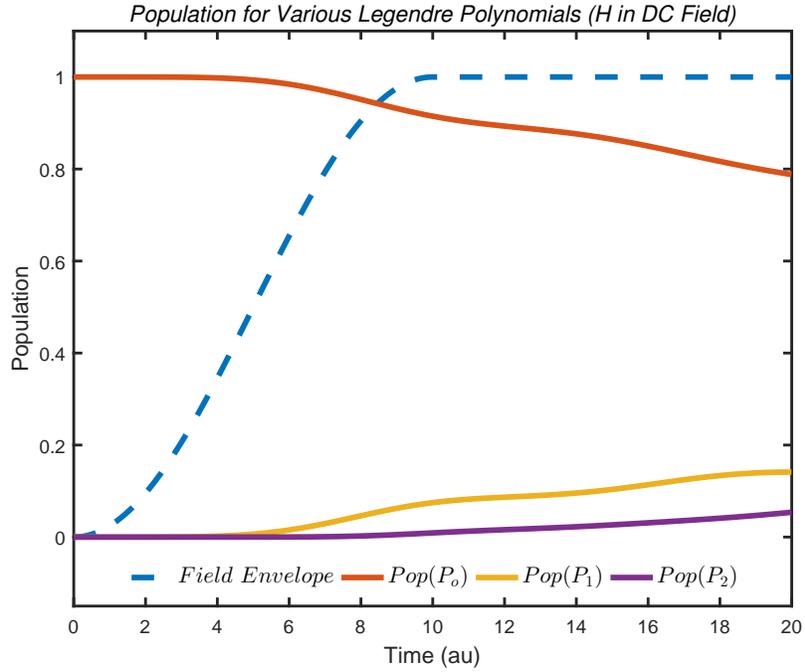}
\vspace{0.2cm}
\caption{\label{fig:fig_leg}
A plot of the total populations of being in an $l=0$, $l=1$, or $l=2$ eigenstate for an orbital initially in the 1s state. %That is, $Pop(P_l)=|a_l(t)\psi_l(\vec{r})|^2$ where $a_l(t)\psi_l(\vec{r},t)$ is the part of the wave function associated with a given $l$ value. 
The field envelope, $F_o(t)$ is plotted in blue along with the populations of Legendre polynomials.}
\end{figure*}
\section{A Time-dependent Model with Exterior Complex Scaling}\label{section:3point4}
\subsection{Instability Issues}
\begin{figure*} [!h]
\centering
\includegraphics[width=350pt]{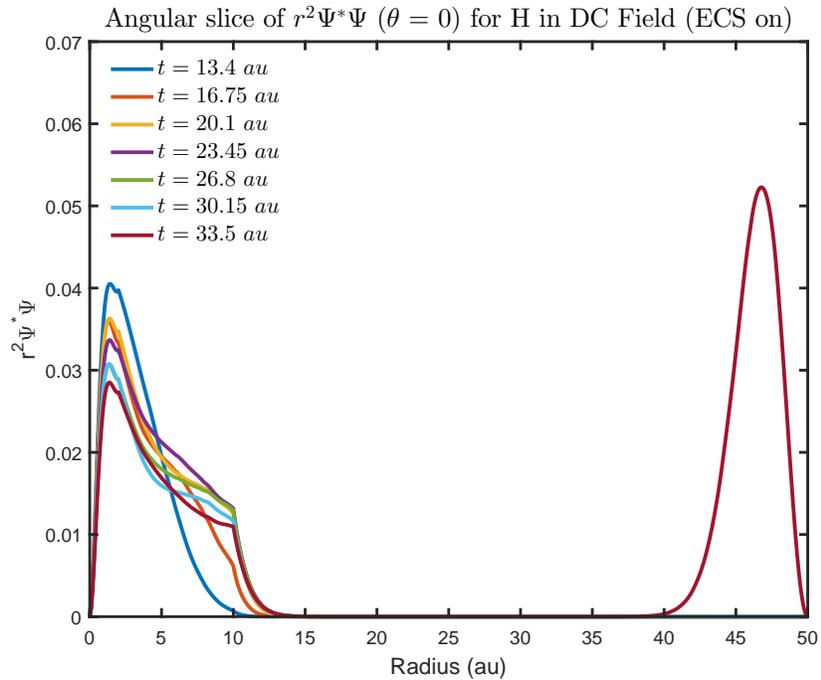}
\vspace{0.2cm}
\caption{\label{fig:psistarpsiECS}
A plot showing issues arising with the time-dependent ECS problem for the 1s state. %with an approximate Jacobian. 
}
\end{figure*}
\begin{figure*} [!h]
\centering
\includegraphics[width=350pt]{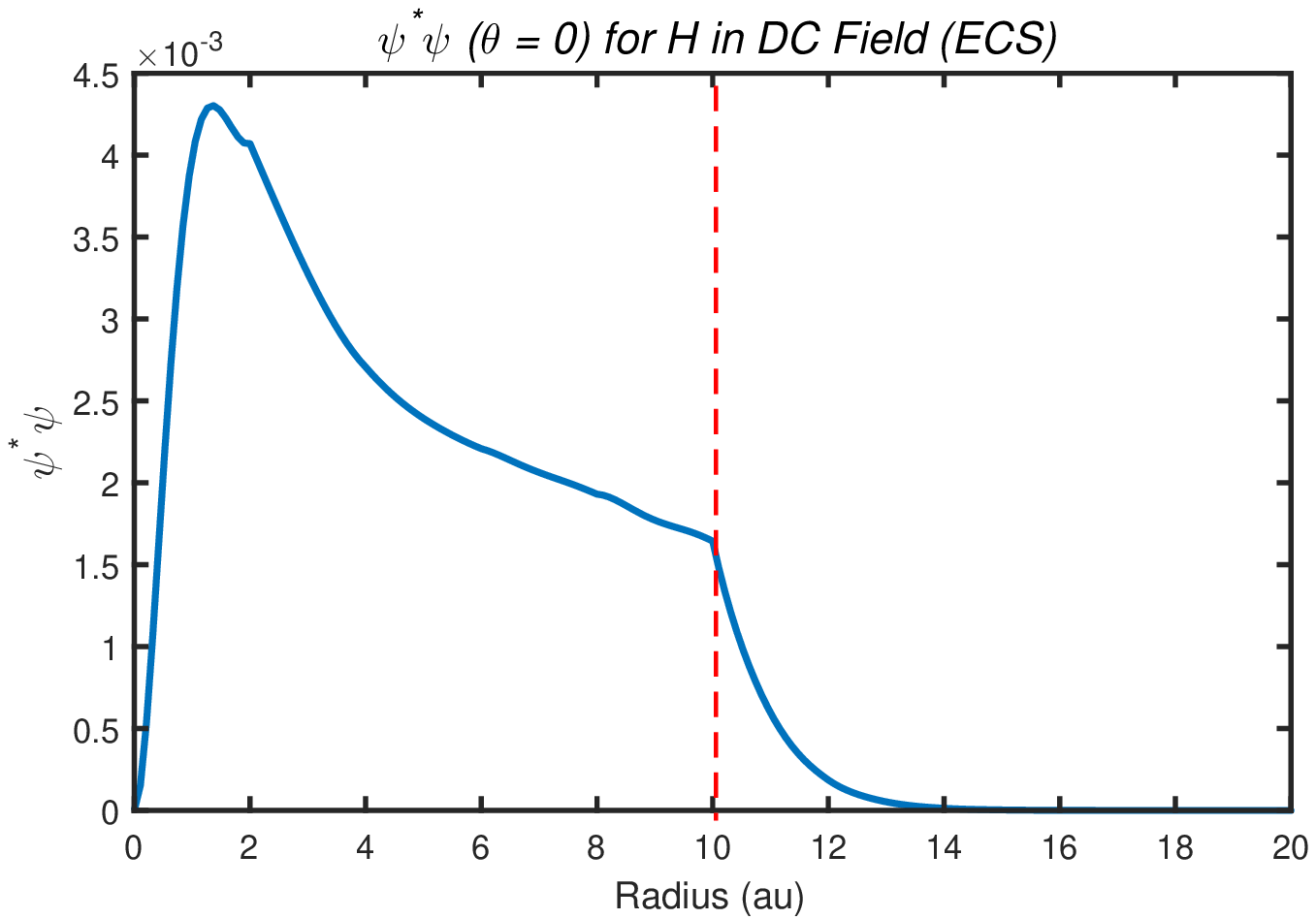}
\vspace{0.2cm}
\caption{\label{fig:ECSind}
A slice of $\psi^*\psi$ at $\theta=0$ for the 1s state of hydrogen with an external DC field (ECS on). Here the maximum external field strength is $F_o = 0.1 ~au$. This is calculated using a time-{\it independent} formalism, meaning that the shape of the profile must remain the same during the entire time-evolution. Usually, the amplitude will stay the same as well, since standard quantum mechanics requires real energy eigenvalues. However, since the eigenvalue is complex, the time-dependence is $\exp[-i(E-i\frac{\Gamma}{2})t/\hbar]$ so the profile will decrease in amplitude. The region outside of $r=25~au$ is totally flat so it is omitted. Compare this to to profile in Fig. \ref{fig:psistarpsiECS} around $t= 30.15~au$ before the instability forms. The dashed line indicates where the scaling radius is. Notice that there the probability density shows a first derivative discontinuity.}
\end{figure*}
To solve the problem with ECS turned on we use the same definitions for the matrix elements as in Ch. 2. In this approach applied one finds that an instability forms on the far right boundary of the radial box for some parameters. For example, if the scaling radius is at 10 $au$, and the radial box is $50~au$, with 25 subintervals and 4 basis functions per subinterval, after some time the instability grows indefinitely, apparently (see Fig. \ref{fig:psistarpsiECS}). In Fig. \ref{fig:psistarpsiECS} the time evolution of 1s goes from 0 $au$ to 40 $au$ but we have only shown a region of importance ($t\in[13.4~au,33.5~au]$) where one can see the formation of the profile affected by ECS (evidenced by the decay at $r_o=10~au$). Here the maximum external external field strength is $F_o = 0.1 ~au$. The issue we want to point out is that shortly after $t=30~au$ a large instability forms in the outer region (outside of $r=35~au$) and quickly grows to an amplitude much greater than that of the inner region feature. The evolution beyond $t=33.5~au$ is not shown since the instability quickly grows to overshadow the inner feature, making plotting difficult. One would like to note that the shape of the profile around $t\in[20.1~au,30.15~au]$ is in line with the time-independent profile for the analogous problem (see Fig. \ref{fig:ECSind}). It is only the formation of the instability which outright contradicts the time-independent results. To solve this issue we begin by applying the proper boundary condition at the far right of the radial box,
\begin{equation}
\psi(r=r_{max})=0,
\end{equation}
by removing the only basis function that is non-zero on the right boundary on the final subinterval of the box (see $\S\ref{section:5point1}$ for more on this matter). Before applying ECS for the time-dependent model, the solutions we considered did not require this boundary condition, and naturally fell towards zero at the right boundary.
\\~\\The instability still forms with this addition but now shifts towards the left in such a way that the solution is able to still go to zero at the end of the box.
\\~\\One way of dealing with this is to calculate all inner products using a truncated radial integral, that is, we calculate the expectation value of some arbitrary observable as,
\begin{equation*}
\langle \mathcal{O} \rangle = \int_{0}^{2\pi}\int_{0}^{\pi}\int_{0}^{\mathcal{R}} \psi^*(\tilde{r},\theta,\phi,t) \mathcal{O}\psi(\tilde{r},\theta,\phi,t)dr \sin \theta d\theta d\phi.
\end{equation*}
Here $\mathcal{R}$ is a chosen radius that is inside the instability, so as to not include it in a calculation.
\\~\\In the case that $\mathcal{O}=1$, that is, when one is calculating the total probability $P(t)$, one can calculate the resonance parameter $\Gamma$ by fitting an exponential to the dying part of $P(t)$. In other words,
\begin{equation}
P(t) = A\exp[-\Gamma (t-t_{fall})], ~ t > t_{fall},
\end{equation}
where $t_{fall}$ is the time at which the total probability begins falling as the electron begins exiting the potential well of the atom. An example of this approach is in Fig. \ref{fig:figfit} for the first resonant state. Note that a good fit of an exponential to the dying part of the total probability means that the "time-independent" ansatz of exponential time-dependence is justified. It also means that our definition of expectation values of observables is useful. In Fig 3.5 we have chosen the radius $\mathcal{R} = 30~au$ to truncate the radial integral. The field $F$ has maximum field strength $max[F(t)]=F_o=0.1 ~au$. In this case the external field looks like $\sim eF(t)r\cos\theta$ and turns on like $F(t) = F_o \sin^2 \omega t$. At $t=10~au$ the field reaches the max, $0.1 ~au$, and stays at that level for 30 more $au$ (the full run time of this simulation). The time grid is more dense than displayed; some data points have been omitted for visual clarity. In the region where the total probability dies we have fit an exponential (red line). The decay rate is $\Gamma = 0.0157975~au$ with a 0.95 confidence level interval of $\Gamma\in[0.015789~au,0.0158059~au]$. The value $t_{fall}$ is marked by a dashed line at $18.75~au$ and denotes the beginning of the fitting region. As a small note the location of $t_{fall}$ was chosen somewhat arbitrarily, but its location is meant to represent a happy medium between the turn over region where decay begins and the start of the fully exponential region (which is roughly a region after 15 $au$ and before 20 $au$). The total probability has been normalized such that the initial value at $t=0$ is unity. So looking at the initial norm of the fit (0.979445), one can see that the value of $t_{fall}$ in this case doesn't line up with an initial norm of unity. The issue here is with the turn-over region. Depending on where one sets $t_{fall}$ one can get closer to an initial probability of unity. More importantly, one can get a $\Gamma$ that is closer to the time-independent result for the analogous problem. In fact, one can get a $\Gamma$ that is smaller or larger than the time-dependent result depending on the location of $t_{fall}$, showing the general weakness of this approach (as far as it comes to comparisons with the time-independent approach). In this plotted case the $\Gamma$ fit is larger than the time-independent result. Note that the best result we got for $F_o = 0.1$ in the time-dependent approach was $\Gamma / 2 \sim 0.00727$, giving $\Gamma \sim 0.0146$. That means that (1) truncating the radial integral works decently, (2) there is decent agreement between the time-dependent and time-independent approaches to ECS, and (3) our definition of observables is useful. The confidence interval for the fit can be largely ignored due to the issues with choosing $t_{fall}$. If $t_{fall}$ is chosen to be deeper in the exponential curve one can even get a $\Gamma$ $\sim 0.013$.  Suffice to say there is decent agreement between the exponential decay region of the time-dependent solution and the time-independent solution, even if there isn't a systematic way of determining the actual confidence interval. It is probably fair to say that the range is something like 0.016 to 0.013 which gives an average of 0.0145, close to the desired result. It isn't clear what level of confidence this interval represents.
\begin{figure*} [!h]
\centering
\includegraphics[width=400pt]{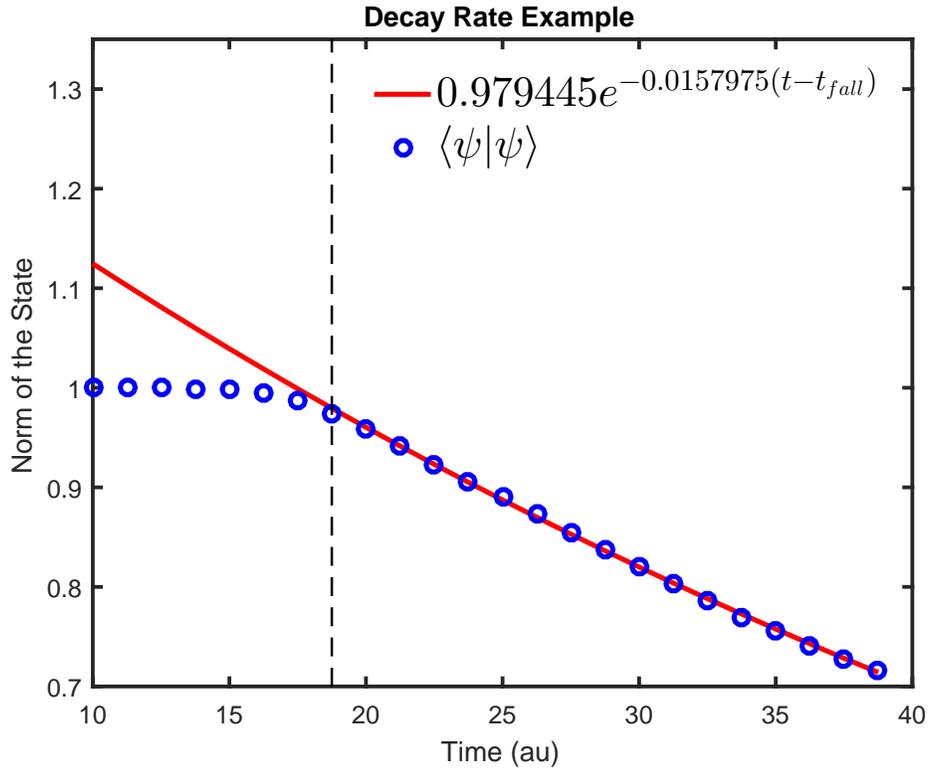}
\vspace{0.2cm}
\caption{\label{fig:figfit}
A plot of the norm of the state (in blue circles) from $t = 10~au$ to $t=40~au$ of the first resonance of hydrogen with ECS on for a time-dependent external field. A fit line is plotted in solid red. The vertical dashed line indicates our choice for $t_{fall}$.}
\end{figure*}
\clearpage
\noindent
The problem is ultimately solved at the numerical level by making the radial box $(0~au,15~au)\times 15$, that is, by using a radial box 15 $au$ wide with 15 subintervals. The scaling radius is kept the same as before (10 $au$). These parameters were sufficient over a solution time 80 atomic time units in length (with turn on of the external field at 10 $au$). We also do not see the instability with the radial box $(0~au,25~au)\times 25$. The scaling radius is again kept the same as before (10 $au$). These parameters were sufficient over a solution time 40 atomic time units in length (with turn on of external field at 10 $au$). However the instability returns if the time interval is increased back up to 80 $au$. Thus we find that there is some direction toward convergence, that is to say, the instability does not show up if one 
\begin{itemize}
\item increases the density of subintervals, and
\item makes the radial box smaller.
\end{itemize}
However we did not conduct an in depth study of the effects of these parameters. Both points seem to be required more as one increases the total time interval one is solving over. Based on this evidence it appears that higher accuracy is required to capture the correct wave function in the outer regions. This may be because a smaller basis does not capture the correct expectation value of the potential in that region and therefore "sees" the wrong field. This is not altogether satisfying as there doesn't appear to be the same trouble with the time-dependent approach without ECS, and the field is the same in that region aside from the complex scaling. 
\section{Conclusion}\label{section:3point5}
We find that our time-dependent 3D algorithm without ECS produces expected results for the hydrogen problem with the external electric field. That is, the probability slowly moves out of the potential well region. When ECS is turned on with a large radial box, an instability appears in the outer region which grows as a function of time. The instability is cured by increasing the density of radial subintervals and making the radial box smaller. Therefore it likely appears since the algorithm allows for it when the approximation is too drastic. This either means that the implementation of the ECS algorithm doesn't fully absorb the outer part of the wave function, or, it only does when the the approximation is sufficiently high level, especially for the outer region of the box. 
\chapter{A Time-independent Model of Water Molecules in Strong DC Fields}
\epigraph{You must never give in to despair. Allow yourself to slip down that road and you surrender to your lowest instincts. In the darkest times, hope is something you give yourself. That is the meaning of inner strength.} {Uncle Iroh, {\it Avatar: The Last Airbender}}

\section*{Overview}
Density functional theory (or DFT, as discussed in our appendix in $\S \ref{section:5point1}$) is a methodology used to calculate the effects of multi-electron interactions in quantum mechanics. %In short, one calculates a basic nuclear potential derived from an initial charge distribution, and then derives an electronic charge distribution from this, recalculates an effective potential from this new information, re-derives an electronic charge distribution, and so forth and so on until one finds a suitable level of accuracy. Density functional theory (DFT) thus is used to find effective potentials for multi-electron systems. This frees one from the tedious calculations involved in describing all electron-electron interactions. 
One maps the many-electron problem onto an effective single-particle problem which yields the electron density to establish the Kohn-Sham equations in second generation DFT. Density functional models have also been proposed to include electron-electron correlation effects.
\\~\\One can find a single potential energy that includes the nuclear potential, plus what is called a "screening" effect from the electrons which balances out the positive charge of the nucleus as one moves from short to larger electron-nucleus distances. The screening part replaces the repulsion an outer electron would feel from an inner one. This theory can be used to motivate effective single electron models of molecules in which all interactions that electron would feel (including $e^-$ to $e^-$ interactions) are approximated by a single potential energy. One can therefore solve the single-electron Schr{\"o}dinger equation and find acceptable solutions for the energy levels of the molecule. 
\\~\\Using a single effective potential energy, it is not difficult to port the techniques we have developed in the previous chapters to solving the problem of ionization of water by strong electric fields. We present a solution involving the use of ECS for the Stark resonances in water using an effective potential which we have borrowed from another work \cite{ice11}.
\\~\\In $\S \ref{section:4point1}$ we briefly discuss some of the types of DFT and our notation.
In $\S \ref{section:4point2}$ we discuss the form of the Hamiltonian we will use. In $\S \ref{section:4point3}$ we discuss the technical details of our solution. In $\S \ref{section:4point4}$ we show our results. In $\S \ref{section:4point5}$ we discuss our conclusions.
\section{Introduction}\label{section:4point1}
When attacking the problem of DC Stark ionization of water molecules one is met with great technical difficulties. Why? First it is a multi-center system so to fully solve the problem one would have to solve for the motion of all three nuclei. However, used Hartree-Fock methods or experiment one can find the ground-state geometric configuration of the nuclei and then use the Born-Oppenheimer approximation to solve for only the electronic part of the system with the nuclei fixed. It is not only a multi-center system but also has multiple electrons. One knows from the helium problem that unless the system is ionized, meaning one electron is removed, the problem cannot be solved straightforwardly with the single-electron Schr{\"o}dinger equation. For DC Stark ionization of water molecules, there are multiple electrons {\it and} the problem cannot be solved without somehow making the Schr{\"o}dinger equation non-Hermitian, as we mentioned before. Therefore, it is a technically detailed problem that goes beyond the level of the fundamental theory of the single-electron Schr{\"o}dinger equation. 
\\~\\Density functional theory (DFT) however, is a well known technique for dealing with multi-electron systems. One starts with the basic premise that the total energy can be calculated as a functional of the total electronic charge distribution. In mathematical terms, $E= E[\rho]$ where $E$ is the total energy and $\rho$ is the total electronic charge distribution. The energy can also be written as a functional of gradients of the density and other combinations involving the density. One can minimize this energy to find the ground-state energy and thereby find the ground-state density (see $\S \ref{section:HKT}$).
\\~\\It is also possible to come up with an effective potential model for the orbitals of a multi-electron system within the context of DFT. These orbitals do not in any straight-forward way combine to the many-electron wave function. However, they can be used to find a total electron density. This equation, or rather, set of equations, are known as the Kohn-Sham equations which are discussed in the appendix ($\S \ref{section:5point2point3}$).
\\~\\However, this effective equation is hard to find because it requires an effective potential which is not exactly known. Thus, one must come up with more density functional models to find an effective and approximate potential which can be used for the Kohn-Sham equations.
\\~\\What makes the multi-electron problem so difficult is that one finds that on top of the simply motivated Hartree energy\footnote{This is given in atomic units as \begin{equation}\label{eq:rhoo}
W_H[\rho] =\frac{1}{2} \int d^3x' \int d^3x\frac{\rho(\vec{r'})\rho(\vec{r})}{|\vec{r'}-\vec{r}|}.
\end{equation}} in the system there are non-obvious effects which are captured in what is called the exchange-correlation energy. Since the exchange-correlation energy is a number and not a function it is not straight-forward to motivate a simple scalar potential from this that can be added to the nuclear Coulomb attraction. This means that to capture the system's dynamics exactly, one could not use a single-electron Schr{\"o}dinger type equation. However, in the study of the Kohn-Sham equations, an argument is made for a single effective potential that tries to capture the average dynamics for every electron.
\\~\\The idea of an effective potential is an interesting and useful one even outside of the context of DFT. The Kohn-Sham equations suggest that one could suppose the existence of a single effective potential energy that captures an averaged effect of electron-electron interactions for every electron in such a way that the orbitals one solves for can actually give the measured ionization energies in terms of the negative of the modelled valence orbital energies. With this effective potential energy in hand it should not be overly difficult to use the ECS techniques we have developed so far to find the Stark resonance parameters for the water molecule's valence orbitals as so defined.
\\~\\We retain some of the notation used in previous chapters. In this chapter $\langle~\rangle$ represents the volume integral with no radial Jacobian. When we can, we separate the radial part as $\{ ~ \} =\int_0^{\infty}dr$. Since we no longer use $3j$ coefficients, we make the following modification to the notation of spherical integrals: we write the full form of the 3D integral as $Y(...)\{ f_{im}|\mathcal{O}|f_{i'm'}\} = Y(...)\{ i|\mathcal{O}|j\}$ where $Y(...)$ contains the angular part. 
\\~\\Additionally, since we retain the notation from our earlier chapters, where $m$ indicates the order of the basis function, when $f_{im}$ appears together with spherical harmonics $Y_{lm}$, the conflict shall be solved by writing $Y_{ln}$. 
\\~\\Solutions are given in combined gaussian and atomic units. In particular $e^2/4\pi\epsilon_o \rightarrow e^2$, and further, $\hbar = m_e = e = 1$. For simplicity, we will just refer to this as atomic units.
\section{A Hamiltonian For Water}\label{section:4point2}
Since the effective potential energy of water cannot be split into a purely angular and purely radial part, we must briefly consider the form of the matrix element when split into real an imaginary parts. Here one should not confuse the local interval $i$ of the basis functions $f_{im}$ as defined earlier in this work, with the imaginary unit we have denoted here with ${\rm i}$.
\\~\\One of the main differences when comparing water to hydrogen is that a technical difficulty arises when calculating a general potential $V(r,\theta,\phi)$ which cannot be separated into radial and angular parts. Given a matrix element of the form,
\begin{equation}
V_{ij} = \langle f_{im}(r)N_{ln}P_{ln}(\cos\theta)e^{{\rm i} n\phi}|V(r,\theta,\phi)|  f_{i'm'}(r)N_{l'n'}P_{l'n'}(\cos\theta)e^{{\rm i}n'\phi}\rangle,
\end{equation}
where $N_{ln}$ is the normalization factor for spherical harmonics, we cannot go back and use Wigner $3j$ coefficients, but must integrate the whole expression explicitly. Note that here $V_{ij}$ represents the $ith$ row with $imln$ numbers indicating the desired basis functions to the left, and the $jth$ column with $i'm'l'n'$ numbers indicating the desired basis functions to the right.
\\~\\We can split real and imaginary parts to get,
\begin{align}
\begin{split}
V_{ij} = \int f_{im}(r)N_{ln}P_{ln}(\cos\theta)\cos\bigg(\bigg[n'-n\bigg]\phi\bigg)V(r,\theta,\phi)f_{i'm'}(r)N_{l'n'}P_{l'n'}(\cos\theta)d^3x
\\+ {\rm i}\int f_{im}(r)N_{ln}P_{ln}(\cos\theta)\sin\bigg(\bigg[n'-n\bigg]\phi\bigg)V(r,\theta,\phi)  f_{i'm'}(r)N_{l'n'}P_{l'n'}(\cos\theta)d^3x.
\end{split}
\end{align}
Here the general form of the potential energy is given by the potential energies of three nuclei, or,
\begin{align}\label{eq:waterp1}
\begin{split}
V(r,\theta,\phi) = -\frac{Z_1(r,r_1)}{\sqrt{(r\sin\theta\cos\phi-r_1\sin\theta_1\cos\phi_1)^2+(r\sin\theta\sin\phi-r_1\sin\theta_1\sin\phi_1)^2+(r\cos\theta-r_1\cos\theta_1)^2}}
\\-\frac{Z_2(r,r_{2})}{\sqrt{(r\sin\theta\cos\phi-r_2\sin\theta_2\cos\phi_2)^2+(r\sin\theta\sin\phi-r_2\sin\theta_2\sin\phi_2)^2+(r\cos\theta-r_2\cos\theta_2)^2}}
\\-\frac{Z_3(r,r_{3})}{\sqrt{(r\sin\theta\cos\phi-r_3\sin\theta_3\cos\phi_3)^2+(r\sin\theta\sin\phi-r_3\sin\theta_3\sin\phi_3)^2+(r\cos\theta-r_3\cos\theta_3)^2}}.
\end{split}
\end{align}
Here $Z_k(r,r_k)$ are the effective atomic numbers and $(r_k,\theta_k,\phi_k)$ are the locations of the three nuclei of the water molecule.
\\~\\This is the general approach of an effective potential for a molecule made up of three atoms. We take a specific form from Illescas et al. \cite{ice11} where
\begin{equation}
V(\vec{r}) = V_O(r_O)+V_{H1}(r_{H1})+V_{H2}(r_{H2}),
\end{equation}
with,
\begin{equation}
V_O(\vec{r}_O) = - \frac{8-N_O}{r_O}-\frac{N_O}{r_O}(1+\alpha_Or_O)\exp(-2\alpha_Or_O),
\end{equation}
and,
\begin{equation}\label{eq:waterp4}
V_{H}(\vec{r}_{H})- \frac{1-N_H}{r_H}-\frac{N_H}{r_H}(1+\alpha_Hr_H)\exp(-2\alpha_Hr_H).
\end{equation}
This is in atomic units $\hbar=m_e=e=4\pi\epsilon_o =1$.
A plot of this potential is given in Fig. \ref{fig:fig_water}. The parameters $N_O$, $N_H$, $\alpha_O$, and $\alpha_H$ are given below.
\begin{figure*} [!h]
\centering
\includegraphics[width=420pt]{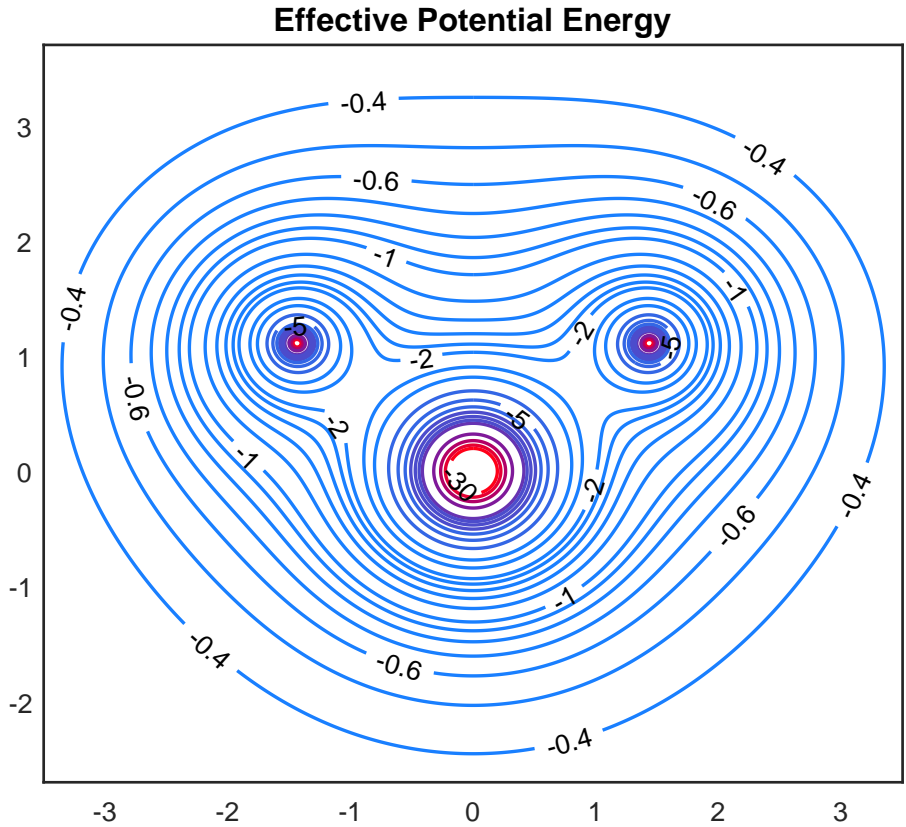}
\vspace{0.2cm}
\caption{\label{fig:fig_water}
A plot of the effective potential energy given by Eqs. \ref{eq:waterp1}-\ref{eq:waterp4}. The horizontal axis is $y$, and the vertical axis is $x$; both are in atomic units. The molecule is sliced at $\theta = \pi/2$. The centres of each individual atom are cut off at $PE = -30 ~au$ to make the plot more readable. The energies of the contour lines from outer to inner are: $[-0.4, -0.5,  -0.6, -0.7, -0.8, -0.9, -1, -1.2, -1.4, -1.6, -1.8, -2, -2.5, -3, -4, -5, -6, -7, -8, -9, -10, -15,\\ -20, -25, -30]$. These values are in atomic units.}
\end{figure*}
\section{Technical Details}\label{section:4point3}
Practically all of the code that was developed for the time-independent hydrogen problem is still relevant, but here we lay out some of the distinct properties of the approach for the water molecule.
\subsection{Co-ordinates}
We choose spherical polar co-ordinates and center the origin on the Oxygen nucleus. Thus, the above potential is re-written as,
\begin{align}
\begin{split}
V(r,\theta,\phi) = -\frac{Z_O(r_O)}{r_O}
-\frac{Z_{H1}(r_{H1})}{r_{H1}}-\frac{Z_{H2}(r_{H2})}{r_{H2}},
\end{split}
\end{align}
where,
\begin{align}
\begin{split}
r_O= \sqrt{(r\sin\theta\cos\phi)^2+(r\sin\theta\sin\phi)^2+(r\cos\theta)^2},
\\r_{H1}=\sqrt{(r\sin\theta\cos\phi-r_2\sin\theta_2\cos\phi_2)^2+(r\sin\theta\sin\phi-r_2\sin\theta_2\sin\phi_2)^2+(r\cos\theta-r_2\cos\theta_2)^2},
\\r_{H2}=\sqrt{(r\sin\theta\cos\phi-r_3\sin\theta_3\cos\phi_3)^2+(r\sin\theta\sin\phi-r_3\sin\theta_3\sin\phi_3)^2+(r\cos\theta-r_3\cos\theta_3)^2},
\end{split}
\end{align}
and where, $\theta_1=\theta_2 = \theta_3 = \pi/2$, $\phi_2=1.8238691/2$, $\phi_3=-1.8238691/2$, and $r_2 = r_3 = 1.8140$ (all of these are in radians and atomic units respectively).
The effective screening functions are,
\begin{equation}
Z_{O}(r_O) = (8-N_O)+N_O(1+\alpha_Or_O)\exp(-2\alpha_Or_O),
\end{equation}
\begin{equation}
Z_{H1}(r_{H1}) = (1-N_H)+N_H(1+\alpha_Hr_{H1})\exp(-2\alpha_Hr_{H1}),
\end{equation}
\begin{equation}
Z_{H2}(r_{H2}) = (1-N_H)+N_H(1+\alpha_Hr_{H2})\exp(-2\alpha_Hr_{H2}),
\end{equation}
with $\alpha_O= 1.6025$,
$\alpha_H = 0.617$,
$N_O = 7.185$, and
$N_H = 0.9075$. The molecular potential energy is thus modelled as a superposition of three radially symmetric functions.
\\~\\Since the model water potential we use goes like $-1/r$ at large distances these results can be compared to hydrogen and helium. At large distances, the sum of the potential energies of the three radially symmetric functions is $\approx -(Z_1+Z_2+Z_3)/r$. $Z_n$ are functions of the radius in our model but reduce to constants for large $r$. It can be shown that $\lim \limits_{r\rightarrow \infty} (Z_1+Z_2+Z_3) = 1.027$.
\\~\\That means the model potential has Rydberg states like hydrogen even though is much stronger than hydrogen near the nuclei. The strength of the water potential near the nuclei makes it somewhat like helium, although the deepest water eigenvalue is much deeper than that for helium. One source lists the water eigenvalues as $E_{1b1}
= -0.5187~au$, $E_{3a1}
= -0.5772~au$, $E_{1b2}
= -0.7363~au$, $E_{2a1}
= -1.194~au$, and $E_{1a1}
= -20.25~au$ \cite{alba19}. There is clearly some overlap between the eigenvalues of water, hydrogen and helium since they all have some eigenvalues near $-0.5~au$.
\\~\\We are solving the time-independent Schr{\"o}dinger equation, given in atomic units as,
\begin{equation}
H\Psi(r,\theta,\phi) = \bigg[\frac{-1}{2}\nabla^2 + V(r,\theta,\phi) - F_or\cos\theta\bigg]\Psi(r,\theta,\phi) = E\Psi(r,\theta,\phi)
\end{equation}
where $V(r,\theta,\phi)$ is taken from Eqs. \ref{eq:waterp1}-\ref{eq:waterp4}. We will use an expansion with spherical harmonics, where the working solution will be 
\begin{equation}\label{eq:psi2_3}
\tilde{\Psi}(r,\theta,\phi) = \sum_{i,m,l,n}\frac{1}{r}c_{imln}f_{im}(r)Y_{ln}(\theta,\phi)= \sum_{i,m,l,n}\frac{1}{r}c_{imln}\varphi_{imln}(r,\theta,\phi),
\end{equation}
which we will refer to as $\Psi(r,\theta,\phi)$ with the understanding that the solution is approximate. The summation here is over all local basis functions listed by local interval $i$ and order $m$, and all spherical harmonics listed by angular number $l$ ($0 \le l \le l_{max}$) and magnetic number $n$ ($-l,...0,...l$).
The full matrix for this problem then looks like
\begin{equation}
H_{ij}= \left(\begin{tabular}{c c c c}
$H^{0000}$ &$H^{001-1}$&$H^{0010}$&\dots\\
$H^{1-100}$ &$H^{1-11-1}$&$H^{1-110}$&\dots\\
$H^{1000}$ &$H^{101-1}$&$H^{1010}$&\dots\\
\vdots & \vdots & \vdots &$\ddots$\\
\end{tabular}\right),
\end{equation}
where $H^{lnl'n'}$ is a sub-matrix where $l,n,l',n'$ are fixed and $i,m,i',m'$ are not (the matrix elements associated with these numbers are arranged as before).
\\~\\The matrix element $H^{lnl'n'}_{im,i'm'}$ before scaling can also be written as,
\begin{equation}\label{eq:mH3}
H^{lnl'n'}_{im,i'm'} = \langle \varphi_{imln}| H |\varphi_{i'm'l'n'}\rangle.
\end{equation}
From now on we will refer to the radial bras and kets as $\{ i|$ and $|j\}$. In problems where the radial and angular parts can be separated, this notation will be used. In those cases we write the angular part of the integral as $Y(l,n,l',n')$. In the case of the DC field, we write $Y(l,n,l',n',1)$ to indicate the factor of $\cos\theta$ contributed by the field. Otherwise angular bras and kets with $\varphi_{imln}$ will be used. All the definitions above are also transformed by exterior complex scaling.
\subsection{Calculating Matrix Elements Outside The Scaling Radius}
In the following definitions, the radial variable is scaled the same as before, by Eq. \ref{eq:scale4}.
For simplicity, in this section we will simply refer to the radial variable as $\tilde{r}$ since we are outside $r_o$. In the next section, where the calculations are inside $r_o$, we will refer to the radial variable as $r$ since there is no scaling applied.
The value for $r_o$ is $10~au$ and the value for $\xi$ is $0.5~rad$. In the following definitions, $i,m,i',m'$ is also referred to as $i,j$, where $i = im$ and $j = i'm'$.
The kinetic energy piece is,
\begin{equation}
KE^{lnl'n'}_{imi'm'}=KE^{lnl'n'}_{ij}=Y(l,n,l',n')\cdot\frac{\hbar^2}{2m_e}e^{-i\xi}\{ \frac{d}{dr}i|\frac{d}{dr}j\}.
\end{equation}
One can see the appendix for the derivation of the symmetric form of the kinetic energy matrix element.
\\~\\For the potential energy,
\begin{equation}
PE^{lnl'n'}_{imi'm'}=e^{i\xi}\langle \varphi_{imln}| V(\tilde{r},\theta,\phi)|\varphi_{i'm'l'n'}\rangle.
\end{equation}
Following Arias Laso \cite{laso16}, we set $V(\tilde{r},\theta,\phi)$ equal to $-1/\tilde{r}$, since Eqs. \ref{eq:waterp1}-\ref{eq:waterp4} give a potential energy $\approx -1/r$ at large distances.
\\~\\There is an angular momentum piece, given by,
\begin{equation}
L^{lnl'n'}_{ij}=Y(l,n,l',n')\cdot e^{i\xi}\{ i|\frac{\hbar^2l'(l'+1)}{2m_e\tilde{r}^2}|j\},
\end{equation}
For the DC electric field,
\begin{equation}
DC^{lnl'n'}_{ij}=Y(l,n,l',n',1)\cdot  e^{i\xi}\{ i|-eF_o \tilde{r}|j\}
\end{equation}
Outside the scaling radius the total overlap matrix elements will be
\begin{equation}
S^{lnl'n'}_{ij}=Y(l,n,l',n')\cdot e^{i\xi}\{ i|j\}.
\end{equation}
\subsection{Calculating Matrix Elements Inside The Scaling Radius}
We have the following definitions for the calculation of matrix elements inside the scaling radius. In the following definitions, $i,m,i',m'$ is also referred to as $i,j$, where $i = im$ and $j = i'm'$. The kinetic energy piece is,
\begin{equation}
KE^{lnl'n'}_{ij}=Y(l,n,l',n')\cdot\frac{\hbar^2}{2m_e}\{ \frac{d}{dr}i|\frac{d}{dr}j\}.
\end{equation}
For the potential energy,
\begin{equation}
PE^{lnl'n'}_{imi'm'}=\langle \varphi_{imln}| V(r,\theta,\phi)|\varphi_{i'm'l'n'}\rangle.
\end{equation}
There is again an angular momentum piece,
\begin{equation}
L^{lnl'n'}_{ij}=Y(l,n,l',n')\cdot\{ i|\frac{\hbar^2l'(l'+1)}{2m_er^2}|j\},
\end{equation}
where $l'$ refers to the angular quantum number of $|j\}$.
\\~\\For the DC electric field,
\begin{equation}
DC^{lnl'n'}_{ij}=Y(l,n,l',n',1)\cdot\{ i|-eF_o r| j\}. 
\end{equation}
The overlap matrix elements are,
\begin{equation}
S^{lnl'n'}_{ij}=Y(l,n,l',n')\cdot \{ i|j\}.
\end{equation}
\section{Results}\label{section:4point4}
Table 4.1 contains our orbital energy results for a free water molecule, which in atomic units obeys,
\begin{equation}
\bigg[\frac{-1}{2}\nabla^2 - \sum_{i=1}^{3}\frac{Z_i(|\vec{r}_i|)}{|\vec{r}_i|} \bigg]\Psi = E\Psi,
\end{equation}
where $Z_i$ and $\vec{r}_i$ are defined in $\S\ref{section:4point3}$. Here $\hbar=m_e=e=4\pi\epsilon_o =1$. We display them along with the results for two other sources using the same form of potential energy but different basis sets. We use almost exactly the same parameters for the potential energy as Jorge et al. \cite{alba19,ice11}, except our $\alpha_O$ differs at the fourth decimal place.  We use exactly the same potential energy as Errea et al. \cite{er15}.
\\~\\In Table 4.1, the results listed by Jorge et al. use very nearly the same potential energy as us but are calculated using gaussian type orbitals (GTO) \protect \cite{ice11}. The results of Errea et al. use the same potential energy and are listed for a lattice calculation with grid density $25/a_o$ (G25) \protect \cite{er15}. A method using one-center basis set self consistent field (OCBS/SCF) molecular orbital (MO) wavefunctions \protect \cite{moccia} is listed. We also list a method that used SCF to derive the potential energy and slater type orbitals \protect \cite{laso17,laso16}. This is listed as SCF/STO. The results for four Hartree-Fock wave functions (successively meant to be increasing in accuracy) are listed along with experimental comparison from Aung, Pitzer, $\&$ Chan \protect \cite{pitzer}.
 \setlength{\tabcolsep}{2.0pt}
\bgroup
\def\arraystretch{1.2}
\begin{table}[!h]\label{table:first}
\begin{center}
 \begin{tabular}{ c c c c c c } 
 \hline
\hline
Molecular Orbital (MO) &$1a_1$&$2a_1$&$1b_2$&$3a_1$&$1b_1$\\
\hline
 $l_{max}=2$ (Current Work) & $ -20.29 $& $-1.067$&  $-0.706$&$-0.471$ &$-0.436$ \\ 
\hline
GTO \cite{alba19} & $ -20.25 $& $-1.194$&  $-0.737$&$-0.578$ &$-0.519$ \\ 
\hline
 G25 \cite{er15} &  NA & NA&  $-0.737$&$-0.568$ &$-0.518$ \\ 
\hline
 %GTO \cite{er15} &  NA& NA&  $-0.737$&$-0.578$ &$-0.519$ \\ 

%\hline
OCBS/SCF \cite{moccia}&$-20.5249$ &$-1.3261$&$-0.6814$ &$-0.5561$  &$-0.4954$\\
\hline
 SCF/STO\footnotemark &  NA& NA&  $-0.682$&$-0.557$
 &$-0.497$ \\ 
\hline

WF I \cite{pitzer} &  $-20.5559$& $-1.2850$&  $-0.6242$&$-0.4661$ &$-0.4026$ \\ 
\hline
 WF II \cite{pitzer} & $-20.5421$ &$-1.3534$ &$-0.7099$ &$-0.5638$ &$-0.5077$ \\ 
\hline
 WF III \cite{pitzer} &  $-20.5541$& $-1.3356$&  $-0.7153$&$-0.5840$& $-0.5130$ \\ 
\hline
 WF IV \cite{pitzer} & $-20.5654$& $-1.3392$& $-0.7283$& $-0.5950$ & $-0.5211$\\ 
\hline
 Experiment \cite{pitzer} &  NA& NA&  $-0.595(11)$&$-0.533(11)$ &$-0.463(4)$ \\ 
\hline
\hline
\end{tabular}

\caption{Energy eigenvalues of free water in atomic units. The results of the current work uses a radial box from 0 to 20 $au$ with 20 subintervals and 10 basis functions per subinterval. It also uses the first and second spherical harmonics with all $m$ values. References are listed as well for other works.}
\end{center}
\end{table}
\footnotetext{These values are digitally extracted from plots from the works of Arias Laso \cite{laso17,laso16}.}
\noindent
\\Table 4.2 and Table 4.3 contain our results for water in a DC electric field, which in atomic units obeys,
\begin{equation}\label{eq:waterfield}
\bigg[\frac{-1}{2}\nabla^2 - \sum_{i=1}^{3}\frac{Z_i(|\vec{r}_i|)}{|\vec{r}_i|} -F_or\cos\theta \bigg]\Psi = E\Psi,
\end{equation}
where the radial variable is understood to be scaled.
 \setlength{\tabcolsep}{14pt}
\bgroup
\def\arraystretch{1.2}
\begin{table}[!h]
\begin{center}
 \begin{tabular}{| c c c c|} 
 \hline
\hline
   Field Strength ($au$) & MO: $1b_1$ & MO: $3a_1$& MO: $1b_2$\\ [0.2ex] 
 \hline\hline

  0.00 &-0.4360&-0.4705&-0.7054\\

    0.06 &-0.4411 -9.086E-004 &-0.4757 -3.059E-005 &-0.7071  -1.129E-012\\
    0.08 &-0.4456 -5.548E-003 &-0.4800 -5.323E-004&-0.7083 -2.982E-011\\
   
     0.10 &-0.4482 -1.348E-002& -0.4849 -2.124E-003&-0.7100 -9.930E-009\\
   
       0.14 &-0.4445 -3.388E-002&-0.4951 -5.794E-003&-0.7145 -5.205E-006 \\

          0.20 &-0.4212 -5.624E-002& -0.5145 -1.384E-002&-0.7250 -3.951E-004 \\
           
\hline
\hline
\end{tabular}

\caption{Energy eigenvalues for the water molecule for different field strengths (Eq. \ref{eq:waterfield}) from the current work with $l_{max}=2$. Here $E = E_r -i\frac{\Gamma}{2}$.}
\end{center}
\end{table}
\\Table 4.3 contains the energy shifts and widths calculated from Table 4.2 table using the value at $F_o = 0.0$ as a reference. The MO $1b_1$ is the valence orbital, and the widths for this have good consistency with the valence orbital widths for the "coupled-cluster singles and doubles with perturbative triples excitations" or CCSD (T) method calculations by Jagau in Table 4.4.
 \setlength{\tabcolsep}{6pt}
\bgroup
\def\arraystretch{1.2}
\begin{table}[!h]
\begin{center}
 \begin{tabular}{| c c c c c c c|} 
\hline
\hline
 & MO: $1b_1$ && MO: $3a_1$& &MO: $1b_2$&\\ [0.2ex] 

   Field Strength ($au$) & $\Delta E$ &$\Gamma$ & $\Delta E$ &$\Gamma$  & $\Delta E$ &$\Gamma$\\ [0.2ex] 
\hline
\hline

    0.06 &-5.093E-003&1.817E-003&-5.155E-003&6.117E-005&-1.592E-003&2.258E-012\\
      0.08&-9.579E-003&1.110E-002&-9.467E-003&1.064E-003&-2.847E-003&5.963E-011 \\
        0.10 &-1.218E-002&2.697E-002&-1.442E-002&4.247E-003&-4.481E-003&1.986E-008 \\
         
            0.14 &-8.474E-003&6.777E-002&-2.456E-002&1.159E-002&-8.990E-003&1.041E-005\\
             0.20 &+1.479E-002&1.125E-001&-4.445E-002&2.768E-002&-1.952E-002&7.902E-004\\

\hline
\hline

\end{tabular}

\caption{Energy shifts and widths for the water molecule for different field strengths (Eq. \ref{eq:waterfield}) from the current work with $l_{max}=2$. Here $\Delta E = E_r(F_o) - E_o$ where $E_o$ is the energy at $F_o = 0.0~au$. The width is $\Gamma = -2\mathfrak{Im}(E)$.}
\end{center}
\end{table}
 \setlength{\tabcolsep}{6pt}
\bgroup
\def\arraystretch{1.2}
\begin{table}[!h]
\begin{center}
 \begin{tabular}{|c c c|} 
 \hline

&\cite{jag18}&\cite{jag18}\\
\hline
\hline
 Field Strength ($au$)& Hartree-Fock & Coupled-Cluster Singles and Doubles (T) \\ [0.2ex] 
 \hline\hline

  0.06&$\Delta E$ = -1.5381E-002, $\Gamma$ = 5.23E-004 &   $\Delta E$ = -1.8097E-002, $\Gamma$ = 1.885E-003

 \\0.08&$\Delta E$ = -2.8639E-002, $\Gamma$ = 4.346E-003  & $\Delta E$ = -3.3293E-002, $\Gamma$ =  1.0475E-002
  \\0.10&$\Delta E$ = -4.5285E-002, $\Gamma$ = 1.4036E-002  & $\Delta E$ = -5.1238E-002, $\Gamma$ = 2.7403E-002
\\0.12&$\Delta E$ = -6.3967E-002, $\Gamma$ = 3.0614E-002 & $\Delta E$ = -7.0468E-002, $\Gamma$ = 5.0967E-002

\\0.14&$\Delta E$ = -8.3259E-002, $\Gamma$ = 5.2159E-002  &  $\Delta E$ = -9.0255E-002, $\Gamma$ = 8.0214E-002
\\
\hline
\hline

\end{tabular}

\caption{Energy shifts for the valence orbital of water for two different methods from Jagau \protect \cite{jag18}. These values were taken from the supplement to the paper. Here $E = E_o + \Delta E - i\frac{\Gamma}{2}$ where $E_o$ is the valence energy for free water. The external electric field is perpendicular to the plane of the molecule as in our work.}
\end{center}
\end{table}
\clearpage
\noindent The figure below contain a plot of the data for the results for $\Delta E$ (shifts) and $\Gamma$ (widths) for Jagau's work and ours.
\\
\begin{figure}[htbp]
  \centering
  \setlength{\unitlength}{\textwidth} 
    \begin{picture}(1,0.5)%in case your image is twice as wide as it is high
                          %(otherwise change the 0.5 to your file's height/width).
       \put(-0.1,0){\includegraphics[width=550pt]{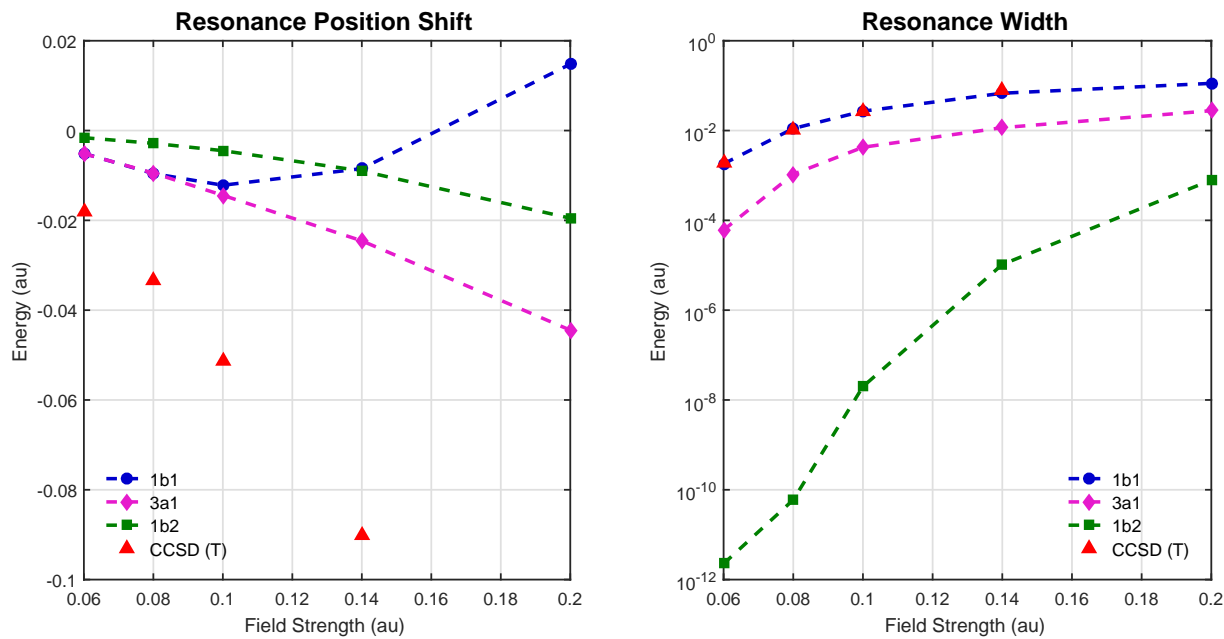}}
       
    \end{picture}
    \caption{Plots of resonance parameters for the valence orbitals of water in an external electric field at various field strengths. The left panel shows the shift of the real part of the energy eigenvalue relative to $F_o = 0.0$, while the right panel shows $\Gamma = -2\mathfrak{Im}(E)$.}
\end{figure}

\noindent 
Table 4.5 contains Arias Laso's results for water in a DC electric field pointed from the oxygen atom towards the hydrogen atoms in the plane of the molecule. Therefore it is not directly comparable to our results. Arias Laso uses an effective potential derived from the single-centre calculations of Moccia \cite{laso16,moccia}.
 \setlength{\tabcolsep}{14pt}
\bgroup
\def\arraystretch{1.2}
\begin{table}[!h]
\begin{center}
 \begin{tabular}{| c c c c c|} 
 \hline
\hline
    & MO: $1b_1$ & & MO: $1b_2$ &\\
   \hline
   \hline
   Field Strength (au) & $E_r$ & $\Gamma$& $E_r$ &$\Gamma$\\ [0.2ex] 
 \hline\hline
     0.1 & -0.506&1.14E-003&-0.689&4.04E-005\\
         0.2 &-0.525&2.28E-002 &-0.718&1.23E-002\\
          0.3 & -0.546&6.74E-002&-0.760&7.51E-002\\
           0.4 &-0.564 &1.24E-001&-0.790&1.91E-001\\
            0.5 &-0.580 &1.90E-001&-0.796&3.11E-001\\
             0.6 &-0.593 &2.61E-001&-0.797&4.11E-001\\
           
\hline
\hline

\end{tabular}

\caption{Energy eigenvalues of water with various field strengths from Arias Laso \protect\cite{laso16}. Here $E = E_r -i\frac{\Gamma}{2}$.}
\end{center}
\end{table}

\section{Conclusion}\label{section:4point5}
From the above results it is clear that the potential energy we use in common with Jorge et al., Errea et al., and Illescas et al. \cite{alba19,er15,ice11} produces reasonably close eigenvalues for free water for three solution methods. The lowest three eigenvalues, listed by Jorge et al. as: -20.25, -1.194, 0.737 \cite{alba19} match up to $0.04$, $0.127$, and $0.031$, or $\sim 0.2\%$, $\sim 11\%$, and $\sim 4\%$ error taking the values listed by Jorge et al. as a reference. However the next two, -0.578, and -0.519 \cite{alba19} match up to $0.107$, and $0.083$, or $\sim 19\%$ and $\sim 16\%$. An earlier version of the code produced better results for $1b1$ and $3a1$ even with $l_{max}=1$. However, the code and results was left out of this work so that we could focus on $l_{max}=2$ results, which allowed for a response from $1b2$ under the influence of the external electric field. That earlier version was modified to the one used in this work by setting certain matrix element integrals to zero which were by definition zero and did not need to be calculated. This was done to speed up calculation time for the $l_{max}=2$ calculation. The likely cause of the skew for the top two eigenvalues is that the consistency of the calculation was artificially broken by a pre-determined result for the null integrals. The basis/Hamiltonian system is misrepresented if all integrals are not calculated the same way. Only a study of the convergence of the matrix element integrator would elucidate the possible remedy of the skew, as a higher accuracy in the non-zero integrals should mean that they come closer to representing the same system as the exactly calculated null integrals. That is a matter for future studies along with convergence with respect to $l_{max}$.
\\~\\None of these modelled eigenvalues agree within the experimental error for the first three negative ionization energies found by experiment, listed by Aung et al. \cite{pitzer}. However, the experimental values $-0.595(11)$, $-0.533(11)$, and $-0.463(4)$ agree with the listed results from Jorge et al. \cite{alba19} to roughly $24\%$, $8\%$, and $12\%$ error respectively, using the experimental values as a comparison. 
\\~\\Using the exterior complex scaling method we have good agreement with the widths in Jagau's results for CCSD (T) at four different field strengths (0.06, 0.08, 0.10, 0.14), with absolute differences of roughly $7\cdot 10^{-5}$, $6\cdot 10^{-4}$, $4\cdot 10^{-4}$, and $2\cdot 10^{-2}$ respectively corresponding to percentage differences of $\sim 4\%$, $\sim 6\%$, $\sim 2\%$ and $\sim 33\%$. Comparing Jagau's widths for HF and CCSD (T), one can see that correlation effects contribute $\sim$ 72, 59, 49, 40, and 35$\%$ differences at field strengths $F_o$ = 0.06, 0.08, 0.1, 0.12, and 0.14 $au$, respectively, using the CCSD (T) width as a comparison. This would imply correlation effects are less important at higher field strengths. Using exterior complex scaling we also have rough agreement with Arias Laso \cite{laso16} for the low field strengths but since the external field is oriented differently in Arias Laso's work the comparison is not really possible between the eigenvalues. 
\\~\\The movement in the real part of the eigenvalue of the MO $1b1$ with respect to external field strength is unsettling since it initially becomes more negative but then turns around and becomes more positive. This is likely due to an issue with a parameter of the solution protocol (such as the number of basis functions) which is purely an artefact of the numerical scheme and has nothing to do with the physics of the problem or the mathematics of the exact solution. In the future a resolution of this problem can be addressed by studying the change in the eigenvalue with respect to these parameters.
\\~\\In conclusion we have presented preliminary results for the problem of a water molecule ionizing in a strong DC electric field. Agreement with literature in terms of the structure of the eigenvalues for a free water molecule is found. The bottom three eigenvalues agree within $11\%$ of the results listed by Jorge et al. \cite{alba19}. The top two agree within $19\%$. This seems to be evidence of the expectation that the accuracy of the higher eigenvalues depends on the accuracy of the lower ones. However, one must consider that the labels $a1$, $b1$, and $b2$ of the orbitals account for different symmetry properties, and therefore an orbital labelled $a1$ cannot depend on an orbital labelled $b1$, etc. Therefore $3a1$ depends on $2a1$ and $1a1$ but $1b1$ does not depend on any of the lower eigenvalues, and its discrepancy cannot be accounted for by this explanation. One should note that since the potential energy is fixed for all orbitals, in an exact representation all orbitals are really determined independently. However, in a finite sized matrix representation the orthogonality considerations of the upper to the lower states implies that if the lower states are inaccurate then the upper states are affected. The agreement with Jagau's results for the widths of the top orbital indicate that the calculations are on the right track to an accurate representation of the ionization of water. Further study must be made to refine the calculations.

 \chapter{Appendix}
\epigraph{Albert Einstein: I was a clerk in a patent office. Faraday was a carpenter. Isaac Newton was an insurance salesman.
\\Catherine Boyd: Isaac Newton was *not* an insurance salesman!} {{\it I.Q.}}
 \section{Testing the Code}\label{section:5point0}
In this section we outline the specifics of implementing the previously discussed finite element method for the case of $N=2$ where $N$ is the number of subintervals. We will use the 1D quantum harmonic oscillator as the test case for a Hermitian problem. For the purposes of checking the results of our code, we will produce a matrix truncated at third order for the $f_{im}(x)$. Subsequently the $h_m(x)$ will be truncated at third order as well. This should give a global matrix with $5\times 5 = 25$ elements, 17 being non-zero. This is just for demonstrating the structure of the matrix. The fidelity of the approach will be tested by using a larger matrix.
\\~\\The fundamental functions we use throughout this thesis are given by monomials defined over $x \in [0,1]$:
\begin{align}
h_m(x) = 1,~m=1,
\\
h_m(x) = \frac{x^{m-1}}{m-1},~m>1.
\end{align}
The corresponding boundary matrix is,
\begin{equation}
B_h= \left(\begin{tabular}{c c c }
$h_1(0)$ &$h_2(0)$&$h_3(0)$\\
$h_1(1)$ &$h_2(1)$&$h_3(1)$\\

\end{tabular}\right)=\left(\begin{tabular}{c c c }
$1$ &$0$&$0$\\
$1$ &$1$&$1/2$\\

\end{tabular}\right).
\end{equation}
The matrix $B_h$ can be split as $(P,Q)$ where $P$ contains the first two columns of $B_h$.
Subsequently, the $W$ matrix is,
\begin{equation}
W = \left(\begin{tabular}{c c}
$P^{-1}$ &$-P^{-1}\cdot Q$\\
$0$ &$\bold{1}$\\
\end{tabular}\right)=\left(\begin{tabular}{c c c}
$1$ &$0$&$0$\\
$-1$ &$1$&$-1/2$\\
$0$ &$0$&$1$\\
\end{tabular}\right).
\end{equation}
We can then construct the $f_{im}(x)$, given by,
\begin{equation}
f_{im}(x) = \sum_{m'}W_{m'm}h_{m'}\left(\frac{x-x_{i-1}}{\delta}\right).
\end{equation}
We note we have substituted $\delta$ in the place of $x_{i}-x_{i-1}$.
We can explicitly write this out for the three basis functions, which are,
\begin{equation}
f_{i,1}(x) = 1 -\left(\frac{x-x_{i-1}}{\delta}\right),
\end{equation}
\begin{equation}
f_{i,2}(x) = \left(\frac{x-x_{i-1}}{\delta}\right),
\end{equation}
\begin{equation}
f_{i,3}(x) = -\frac{1}{2}\left(\frac{x-x_{i-1}}{\delta}\right) + \frac{1}{2}\left(\frac{x-x_{i-1}}{\delta}\right)^2.
\end{equation}
The global matrix is stitched together from the individual sub-matrices.
The sub-matrix elements are generated by evaluating $\langle i|H|j\rangle= \langle f_{im}|H|f_{i'm'}\rangle$ over the $i^{th}$ domain, where
\begin{equation}
H = -\frac{\hbar^2}{2m}\frac{d^2}{dx^2} + \frac{1}{2}kx^2.
\end{equation}
Additionally we add the continuity constraint by replacing the highest-order index ($m$ or $m'$) with the $2nd$ index and then having the last element of the $ith$ sub-matrix be added to the first element of the $(i+1st)$ sub-matrix. The second index is then replaced with the third, and the interval between the third order and the second order is $[4,M]$. In other words, the indices $m$ or $m'$ as seen in the sub-matrix go $[1~ 3 ~4...M~ 2]$, where $M$ is the highest order. If $M < 4$, then the indices go $[1~ 3 ~2]$.
\\~\\ This results in element overlaps between each sub-matrix.
Since we are not yet using ECS, we do not implement the discontinuity at the scaling radius.
\\~\\To make manifest the symmetry of the problem, the Hamiltonian is split into its kinetic and potential energy parts, each evaluated separately, then added together. This looks like,
\begin{equation}\label{eq:ke}
H_{ij}=\frac{\hbar^2}{2m_e}\langle \frac{d}{dx}i|\frac{d}{dx}j\rangle + \frac{1}{2}\langle i|kx^2|j\rangle.
\end{equation}
One can see the appendix for a derivation of the symmetric form of the kinetic energy part ($\S \ref{section:5point1}$).
\\~\\The final matrix looks like,
 \setlength{\tabcolsep}{6pt}
\begin{equation}\label{referencemat}
H_{global} = \left(\begin{tabular}{c c c c c}
$\langle f_{11}|H|f_{11}\rangle$&$\langle f_{11}|H|f_{13}\rangle$&$\langle f_{11}|H|f_{12}\rangle$&$0$&$0$\\
$\langle f_{13}|H|f_{11}\rangle$&$\langle f_{13}|H|f_{13}\rangle$&$\langle f_{13}|H|f_{12}\rangle$&$0$&$0$\\
$\langle f_{12}|H|f_{11}\rangle$&$\langle f_{12}|H|f_{13}\rangle$&$\langle f_{12}|H|f_{12}\rangle+\langle f_{21}|H|f_{21}\rangle$&$\langle f_{21}|H|f_{23}\rangle$&$\langle f_{21}|H|f_{22}\rangle$\\
$0$&$0$&$\langle f_{23}|H|f_{21}\rangle$&$\langle f_{23}|H|f_{23}\rangle$&$\langle f_{23}|H|f_{22}\rangle$\\
$0$&$0$&$\langle f_{22}|H|f_{21}\rangle$&$\langle f_{22}|H|f_{23}\rangle$&$\langle f_{22}|H|f_{22}\rangle$\\
\end{tabular}\right).
\end{equation}
The overlap matrix looks like,
\begin{equation}
S_{global} = \left(\begin{tabular}{c c c c c}
$\langle f_{11}|f_{11}\rangle$&$\langle f_{11}|f_{13}\rangle$&$\langle f_{11}|f_{12}\rangle$&$0$&$0$\\
$\langle f_{13}|f_{11}\rangle$&$\langle f_{13}|f_{13}\rangle$&$\langle f_{13}|f_{12}\rangle$&$0$&$0$\\
$\langle f_{12}|f_{11}\rangle$&$\langle f_{12}|f_{13}\rangle$&$\langle f_{12}|f_{12}\rangle+\langle f_{21}|f_{21}\rangle$&$\langle f_{21}|f_{23}\rangle$&$\langle f_{21}|f_{22}\rangle$\\
$0$&$0$&$\langle f_{23}|f_{21}\rangle$&$\langle f_{23}|f_{23}\rangle$&$\langle f_{23}|f_{22}\rangle$\\
$0$&$0$&$\langle f_{22}|f_{21}\rangle$&$\langle f_{22}|f_{23}\rangle$&$\langle f_{22}|f_{22}\rangle$\\
\end{tabular}\right).
\end{equation}
%** I did not put this in the code ** Additionally the Dirichlet boundary conditions are implemented by... **
The problem to solve is then,
\begin{equation}
H_{global}\vec{c} =E S_{global}\vec{c}.
\end{equation}
We found our code to give expected results for this case and for larger matrices (with a larger basis/more subintervals) as well.

 \section{Smoothness of Transformed Integrands}\label{sec:smooth}
 \subsection{Smoothness of the Radial Basis Functions}
 This section shows the smoothness of functions mentioned in $\S \ref{sec:smoothper}$.
 \\~\\To prove smoothness (in particular C-infinity level smoothness, but we will just call it {\it smoothness}), we must show that the function $f_M(\rho)$ is differentiable everywhere at all orders \cite{rowland}. In other words, every derivative of $f_M(\rho)$ with respect to $\rho$ must be smooth itself. This requires two proofs: (P1) the smoothness of a product of smooth functions, and (P2) the smoothness of a sum of smooth functions. 
\\~\\If some basis set of functions $h_m(\rho)$ which are used to represent $f_M(\rho)$ are smooth then $f_M(\rho)$ is smooth if a sum of smooth functions is smooth. 
\\~\\We first must show $h_m(\rho)$ are smooth. If the functions $h_m(x)$ are all monomials of the form $a_m x^{m}$, then $h_m(\rho)$ must be of the form 
\begin{equation}
h_m(\rho)=a_m\bigg(\frac{b-a}{2}\cos \rho + \frac{b+a}{2}\bigg)^m,
\end{equation}
which is just a product of smooth functions in the $\rho$ variable since $a_m$, $b$ and $a$ are constants, and $\cos\rho$ is smooth. Therefore we show to start that a product of smooth functions is smooth.
The continued use of the product rule ensures that all derivatives of a product of smooth functions will be sums of products of functions of $\rho$ which are all smooth. To show this, suppose set $S_1$ of functions $f(\rho), g(\rho)...h(\rho)$ are all smooth functions. Then the product rule gives, for the derivative of a function which is the product of the functions of this set $S_1$,
\begin{equation}
\frac{\partial }{\partial \rho} f(\rho)g(\rho)...h(\rho) = \frac{\partial }{\partial \rho} f(\rho)[g(\rho)...h(\rho)]= \frac{\partial }{\partial \rho} f(\rho)[p(\rho)], 
\end{equation}
or,
\begin{equation}
\frac{\partial }{\partial \rho} f(\rho)[p(\rho)]= f'(\rho)[p(\rho)]+f(\rho)[p(\rho)]'= f'(\rho)[p(\rho)]+f(\rho)\frac{\partial }{\partial \rho} [g(\rho)...h(\rho)].
\end{equation}
So as long as functions $g(\rho)...h(\rho)$ are smooth then the final term on the RHS will be a sum of products of smooth functions since every product rule applied on a product of smooth functions gives a sum of products of smooth functions (we will show this). Similarly, the first term on the RHS is just a product of smooth functions. Therefore as long as a sum of products of smooth functions is smooth, then $h_m(\cos\rho)$ is smooth in the $\rho$ variable. Here we can take care of (P2) since a sum of smooth functions is smooth because of the linearity of the derivative operator. So the only thing we need to know is if the product of smooth functions is always smooth. 
\\~\\One can always use the product rule for all orders of derivatives on a product of two smooth functions, since it will always produce sums of products of two smooth functions. 
\\~\\Furthermore, the product of $N$ smooth functions can always be re-written as a product of two functions, where one, call it, $f$, is a lone function in that product ($f$ is smooth by definition), and the other, call it, $p$ is the product of the other $N-1$ functions in that product (which we must show is smooth). By (P2) if $f'p$ and $fp'$ have analytic derivatives at all orders then $fp$ is smooth. We know $f'p$ is smooth if $p$ is smooth, since the smoothness of $p$ proves the smoothness of the product of smooth functions, and $f'$ is smooth by definition. It follows since $f$ is smooth, that if $p$ is smooth then $fp'$ is also smooth since $p'$ must be smooth by definition and again, the smoothness of $p$ proves the smoothness of a product of smooth functions. The question here at this stage is whether $p$ (and therefore the product of smooth functions) is smooth. First, let us deal with the question of whether the product of $N-1$ functions $p$ is composed of is smooth. The $N-1$ functions can be re-written the same way as $f$ and $p$, call the decomposition $g$ and $k$, where $k$ is the product of $N-2$ functions. So on and so forth one can keep grouping lesser products of functions. Because of this one can always use the product rule (as defined for a product of two smooth functions) on $p$ to obtain derivatives of $p$ at all orders. Thus $p$ is smooth. Then too is $f'p + fp'$. So too is $fp$ and more specifically $h_m(\cos\rho)$.
\\~\\By (P1) and (P2) we find that $f_M(\rho)$ is smooth as well since it is a sum of products of smooth functions.
\\~\\Therefore, since $f_M(\rho)$ is smooth and periodic in $\rho$, and additionally even, a Fourier cosine series representation is optimal.
 \subsection{Smoothness of the Radial Integrands}
  This section shows the smoothness of functions mentioned in $\S \ref{sec:loose}$.
  \\~\\We must admit that $f_M(x)$ as we have defined them are not the full integrand which we use our radial integration method for. 
\\~\\$f_{im}'(r)f_{i'm'}'(r)$, $f_{im}(r)f_{i'm'}(r)/\tilde{r}$, and $f_{im}(r)f_{i'm'}(r)/\tilde{r}^{2}$ are all integrated using this method, and are smooth. We will soon show this. 
\\~\\For ECS non-smoothness of $\tilde{r}$ at $r_o$ is not a problem since we employ a finite element method (FEM), and so it only appears during a transition from one finite interval to another, and every integral is calculated on a finite interval. 
\\~\\However, $\tilde{r}$ is smooth everywhere else. 
\\~\\The function $f_{im}'\left(\frac{b-a}{2}\cos \rho + \frac{b+a}{2}\right)$ is a derivative of a sum of smooth functions so it is smooth. 
\\$f_{im}'\left(\frac{b-a}{2}\cos \rho + \frac{b+a}{2}\right)f_{i'm'}'\left(\frac{b-a}{2}\cos \rho + \frac{b+a}{2}\right)$ is a product of smooth functions so it is smooth. \\Likewise $f_{im}\left(\frac{b-a}{2}\cos \rho + \frac{b+a}{2}\right)f_{i'm'}\left(\frac{b-a}{2}\cos \rho + \frac{b+a}{2}\right)$ is smooth.
\\~\\We just have to show the last two integrands are smooth in the $\rho$ variable, or more specifically, we only have to show $1/\tilde{r}^n$ is smooth.
\\~\\Consider the derivative of $V_n(\rho) = 1/\tilde{r}^n$ outside of the scaling radius under the same transformation of variables as $h_M(x)$,
\begin{equation}
V_n'(\rho)=\frac{\partial }{\partial \rho}\bigg[ e^{i\xi}\bigg(\left[\frac{b-a}{2}\cos \rho + \frac{b+a}{2}\right]- r_o\bigg)+ r_o\bigg]^{-n}.
\end{equation}
By the chain rule,
\begin{equation}
V_n'(\rho)=-n\bigg[ e^{i\xi}\bigg(\left[\frac{b-a}{2}\cos \rho + \frac{b+a}{2}\right]- r_o\bigg)+ r_o\bigg]^{-n-1}\bigg[\bigg(e^{i\xi}\bigg) \bigg( \frac{a-b}{2}\sin \rho\bigg)\bigg].
\end{equation}
Consider a sketch of this proof. The term inside $[~]^{-n-1}$ is $\Phi(\rho)$, and we can see that $\Phi'(\rho)$ inside the second set of square brackets is clearly smooth.
Now note the first term is still of the form of $V_n(\rho)$ and its derivative will be of the form of $V_n'(\rho)$ (and so on and so forth for the first term of every consecutive derivative) so it is smooth. The product of two smooth functions is smooth so $V_n(\rho)$ is smooth. More clearly,
\begin{equation}
V_n'(\rho)=-n\bigg[\Phi(\rho)\bigg]^{-n-1} \Phi'(\rho),
\end{equation}
is smooth if,
\begin{equation}
H(\rho) = \bigg[\Phi(\rho)\bigg]^{-n-1}
\end{equation}
is smooth.
But, 
\begin{equation}
H'(\rho) = \frac{\partial }{\partial \rho}\bigg[\Phi(\rho)\bigg]^{-n-1}=\bigg(-n-1\bigg)\bigg[\Phi(\rho)\bigg]^{-n-2}\Phi'(\rho),
\end{equation}
which is smooth if,
\begin{equation}
G(\rho) = \bigg[\Phi(\rho)\bigg]^{-n-2}
\end{equation}
is smooth. Its derivative is,
\begin{equation}
G'(\rho) = \bigg(-n-2\bigg)\bigg[\Phi(\rho)\bigg]^{-n-3}\Phi'(\rho),
\end{equation}
and so on one can argue for smoothness in this way.
Thus the only condition left that is required to satisfy smoothness of $H(\rho)$ and thus $V_n(\rho)$ is the smoothness of $x^{-m}$ where $m$ is a positive integer (this allows us to continue this argument {\it ad infinitum}). The $Jth$ derivative of $x^{-m}$ is,
\begin{equation}
\frac{\partial ^J}{\partial x^J}x^{-m}=\left[\prod_{j=1}^{J}\bigg(-m-j+1\bigg)\right]x^{-m-J},
\end{equation}
which exists for all positive $J$. Thus $x^{-m}$ is smooth. Inside the scaling radius the proof goes the same way.
So since the product of smooth functions is smooth, $V_n(\rho)$ is smooth inside and outside the scaling radius. By the same principle the last two integrands we mentioned are all smooth in the $\rho$ variable. 
\\~\\One more consideration must be made since we split real and imaginary parts before integration. However, because of the linearity of a derivative operator, a smooth function cannot be decomposed into the sum of two functions unless they are both smooth. Therefore our integrands are smooth, even-symmetric and periodic (as mentioned before even symmetry and periodicity is strictly enforced by the change of variables).
 \newpage
\section{More Stark Results For Hydrogen And Singly Ionized Helium}\label{section:5point01}
\begin{center}
Table 5.1: Here $F_o = 0.5~au$ and the number of basis functions is (5,5). We show the stability of ECS with respect to the scaling angle for the first resonant state ($n_1 = n_2 = m = 0$). \\~\\
\def\arraystretch{0.8}
\begin{tabular}{cccc}
\toprule
 & (0, 100)$\times100$, $10.0 ~au$\\
\midrule
0.2 & -0.626735751     -0.281667480i 
\\
\midrule
0.5 &  -0.623564039     -0.283481219i  &  & \\
\midrule
1.0 & -0.624224716     -0.282596082i    &  & \\
\midrule
1.2 & -0.624136364     -0.282943183i     &  & \\
\midrule
 & -0.623068026 -0.279744825$i$ \cite{tel89}  &  & \\

\bottomrule
\end{tabular}
\\~\\Intervals are shown on the top in the form $(start, end)\times subintervals,~r_o$, and scaling parameter $\xi$ is on the left in $rad$. Values of the first resonant eigenvalue are shown inside in atomic units.
\end{center}

\begin{center}

Table 5.2: Here $F_o = 0.1~au$, $\xi = 0.5 ~rad$, and the number of basis functions is (5,5). We show the stability of ECS with respect to the scaling radius for the first resonant state ($n_1 = n_2 = m = 0$).\\~\\
\def\arraystretch{0.8}
\begin{tabular}{cccc}
\toprule
 & (0, 100)$\times100$\\
\midrule
8 $au$ & -0.527395257 -0.723420953(E-2)i
 
\\
\midrule
9 $au$&  -0.527395256 -0.723420936(E-2)i
 &  & \\
\midrule
10 $au$& -0.527395257 -0.723420936(E-2)i  &  & \\
\midrule
11 $au$& -0.527395256 -0.723420999(E-2)i
  &  & \\
\midrule
12 $au$ & -0.527395256 -0.723420862(E-2)i
 &  & \\
\midrule
 & -0.527418175 -0.726905676(E-2)$i$ \cite{tel89}
 &  & \\
\bottomrule
\end{tabular}
\\~\\Intervals are shown on the top in the form $(start, end)\times subintervals$, and scaling the scaling radius is on the left. Values of the first resonant eigenvalue are shown inside in atomic units.
\end{center}
\newpage
\begin{center}

Table 5.3: Here $F_o = 0.5~au$ again, $\xi = 0.5 ~rad$, and the number of basis functions is (5,5). We show the stability of ECS with respect to scaling radius for the first resonant state ($n_1 = n_2 = m = 0$).\\~\\
\def\arraystretch{0.8}
\begin{tabular}{cccc}
\toprule
 & (0, 100)$\times100$\\
\midrule
8 $au$ & -0.622714896     -0.284486217i 
 
\\
\midrule
9 $au$& -0.623188534     -0.284444458i   
 &  & \\
\midrule
10 $au$& -0.623564039     -0.283481219i   &  & \\
\midrule
11 $au$& -0.622820143     -0.2813606844i  
  &  & \\
\midrule
12 $au$ & -0.620124345     -0.277875440i   &  & \\\midrule
& -0.623068026 -0.279744825$i$ \cite{tel89}  &  & \\

\bottomrule
\end{tabular}
\\~\\Intervals are shown on the top in the form $(start, end)\times subintervals$, and scaling the scaling radius is on the left. Values of the first resonant eigenvalue are shown inside in atomic units.
\end{center}
\begin{center}
Table 5.4: Here $F_o = 0.005 ~au$, $\xi = 0.5 ~rad$, and $r_o = 10.0~au$. We explore the convergence of ECS results with various parameters for the first {\it excited} resonant state. The states are listed with subscripts $n_1,n_2,m$.  \\~\\
\begingroup
\def\arraystretch{0.8}
 \fontsize{9pt}{12pt}\selectfont
\begin{tabular}{cccc}
\toprule
 & $E_{100}$ & $E_{010}$ \\
 \midrule
(0, 100)$\times100$, [5,5] & -0.112060042 -0.330738212(E-6)i&-0.142620039     -0.504520381(E-4)i\\
\midrule
(0, 100)$\times100$, [6,6] & -0.112062422   -0.227726822(E-5)i&-0.142619051    -0.531180793(E-4)i\\
\midrule
(0, 100)$\times100$, [7,7] & -0.112062062 -0.298313131(E-5)i&-0.142618582     -0.530479782(E-4)i\\
\midrule
(0, 100)$\times100$, [8,8] & -0.112061889 -0.289201267(E-5)i&-0.142618594     -0.529658319(E-4)i\\
\midrule
(0, 100)$\times100$, [9,9] &  -0.112061920 -0.285462323(E-5)i & -0.142618609 -0.529702603(E-4)i & \\
\midrule
 &  -0.112061924 -0.2864684(E-5)$i$ \cite{tel89}& -0.142618608 -0.52972232(E-4)$i$  \cite{tel89}& \\

\bottomrule
\end{tabular}
\endgroup
 \\~\\Intervals are shown on the left in the form $(start, end)\times subintervals$, and number of basis functions (per subinterval for the radial basis functions; the angular basis functions are global) are shown next to them in the form $\left[\# ~radial, \# ~angular\right]$. Values of the first {\it excited} resonant eigenvalue are shown inside in atomic units.
\end{center}
\newpage
\begin{center}
Table 5.5: Here $F_o = 0.005 ~au$, $\xi = 0.5 ~rad$, and $r_o = 10.0~au$. We explore the convergence of ECS results with various parameters for the first {\it excited} resonant state ($n_1 = n_2 = 0, m = 1$). \\~\\
\begingroup
\def\arraystretch{0.8}
 \fontsize{9pt}{12pt}\selectfont
\begin{tabular}{cccc}
\toprule
 & $E$\\
 \midrule
(0, 100)$\times100$, [5,5] & -0.127127654 -0.165250916(E-5)i\\

\midrule
(0, 100)$\times100$, [7,7] & -0.127147315 -0.129048745(E-4)i\\
\midrule
(0, 100)$\times100$, [8,8] & -0.127146646 -0.132155098(E-4)i\\
\midrule
(0, 100)$\times100$, [9,9] & -0.127146583     -0.130782907(E-4)i \\
\midrule
(0, 100)$\times100$, [10,10] & -0.127146614 -0.130705073(E-4)i\\
\midrule
& -0.127146612 -0.13076427(E-4)$i$ \cite{tel89}\\
\bottomrule
\end{tabular}
\endgroup
 \\~\\Intervals are shown on the left in the form $(start, end)\times subintervals$, and number of basis functions (per subinterval for the radial basis functions; the angular basis functions are global) are shown next to them in the form $\left[\# ~radial, \# ~angular\right]$. Values of the first {\it excited} resonant eigenvalue are shown inside in atomic units.
\end{center}
\begin{center}
Table 5.6: Here $F_o = 0.01 ~au$, $\xi = 0.5 ~rad$, and $r_o = 10.0~au$. We explore the convergence of ECS results with various parameters for the first {\it excited} resonant state ($n_1 = n_2 = 0, m = 1$). \\~\\
\begingroup
\def\arraystretch{0.8}
 \fontsize{9pt}{12pt}\selectfont
\begin{tabular}{cccc}
\toprule
 & $E$\\
 \midrule
(0, 100)$\times100$, [5,5] & -0.135275988  -0.308809820(E-2)i\\
\midrule

(0, 100)$\times100$, [7,7] & -0.134490707     -0.310651813(E-2)i\\
\midrule
(0, 100)$\times100$, [8,8] & -0.134535964  -0.313643312(E-2)i\\
\midrule
(0, 100)$\times100$, [9,9] & -0.134523503  -0.314114866(E-2)i \\
\midrule
 & -0.134524888 -0.313865388(E-2)$i$ \cite{tel89}\\
\bottomrule
\end{tabular}
\endgroup
 \\~\\Intervals are shown on the left in the form $(start, end)\times subintervals$, and number of basis functions (per subinterval for the radial basis functions; the angular basis functions are global) are shown next to them in the form $\left[\# ~radial, \# ~angular\right]$. Values of the first {\it excited} resonant eigenvalue are shown inside in atomic units.
\end{center}
\newpage
\begin{center}
Table 5.7: Here we use $Z=2$ and thereby turn the hydrogen problem into the singly ionized helium problem. We also have $\xi=0.5$ $rad$ and $r_o = 10.0~au$, and for all results, (0, 100)$\times100$, [5,5]. We explore ECS results for different field strengths for the first resonant state of singly ionized helium (with $n_1 = n_2 = m = 0$). \\~\\
\begingroup
\def\arraystretch{0.8}
 \fontsize{9pt}{12pt}\selectfont
\begin{tabular}{cccc}
\toprule
 & $E$\\
 \midrule
 $F_o = 0.0005$ & -1.999628691  -0.395286371(E-14)i\\
\midrule
 $F_o = 0.005$ & -1.999632173 -0.274745739(E-15)i\\
\midrule
 $F_o = 0.01$ & -1.999642724  -0.134510948(E-13)i\\
\midrule
 $F_o = 0.02$ &  -1.999684935   -0.934649409(E-14)i\\
\midrule
 $F_o = 0.03$ &  -1.999755308   -0.247446670(E-14)i\\
 \midrule
 $F_o = 0.0$ &   $-Z/n^2=-2/n^2 = -2.0$\\
\bottomrule
\end{tabular}
\endgroup
 \\~\\Values of the first resonant eigenvalue are shown inside in atomic units.
\end{center}

\begin{center}
Table 5.8: Here we use $Z=2$ and thereby turn the hydrogen problem into the singly ionized helium problem. We also have $\xi=0.5$ $rad$ and $r_o = 10.0~au$, and for all results, (0, 100)$\times100$, [5,5]. We explore ECS results for different field strengths for the first {\it excited} resonant state of singly ionized helium. The states are listed with subscripts $n_1,n_2,m$. \\~\\
\begingroup
\def\arraystretch{0.8}
 \fontsize{9pt}{12pt}\selectfont
\begin{tabular}{cccc}
\toprule
 & $E_{010}$&$E_{100}$\\
 \midrule
 $F_o = 0.0005$ & -0.500744032 -0.187613294(E-13)i&-0.499244006 -0.207735065(E-13)i\\
\midrule
 $F_o = 0.005$ & -0.507625670 -0.191289879(E-13)i&-0.492622573 -0.194754788(E-13)i\\
\midrule
 $F_o = 0.01$ & -0.515532700 -0.187543364(E-13)i&-0.485507891-0.223133786(E-13)i\\
\midrule
 $F_o = 0.02$ & -0.532239935 -0.843372764(E-10)i&-0.472032244  -0.857866648(E-11)i\\
\midrule
 $F_o = 0.03$ & -0.550352998  -0.227243594(E-5)i&-0.459576095  -0.265274683(E-7)i\\
\midrule
 $F_o = 0.04$ & -0.570473793 -0.201878513(E-3)i&-0.448233210 -0.132193685(E-5)i\\
\midrule
$F_o = 0.05$ & -0.593083190 -0.186621982(E-2)i&-0.438315916  -0.107351991(E-3)i\\
\midrule
$F_o = 0.06$ & -0.617037114 -0.625834894(E-2)i&-0.430119151  -0.108046990(E-2)i\\
 \midrule
 $F_o = 0.0$ &   $-Z/n^2=-2/n^2 = -0.5$&   $-Z/n^2=-2/n^2 = -0.5$\\
\bottomrule
\end{tabular}
\endgroup
 \\~\\Values of the first {\it excited} resonant eigenvalue are shown inside in atomic units. At low field strength we find two eigenvalues converging.
\end{center}
 \section{Symmetric Form of Kinetic Energy Matrix Elements}\label{section:5point1}
The kinetic energy operator in a matrix element can be re-framed in terms of first derivatives rather than second derivatives. Consider our goal representation:
 \begin{equation*}
KE_{ij}=\frac{\hbar^2}{2m}\langle \frac{d}{dx}i|\frac{d}{dx}j\rangle.
\end{equation*}
This can be expressed with an identity operator as,
 \begin{equation}
KE_{ij}=-\frac{\hbar^2}{2m}\langle i|\frac{d^2}{dx^2}|j\rangle =-\frac{\hbar^2}{2m}\langle i|\hat{\mathcal{I}}|\frac{d^2}{dx^2}j\rangle = -\frac{\hbar^2}{2m}\int_{-\infty}^{\infty} \langle x|i\rangle^*\frac{d^2}{dx^2}\langle x|j\rangle dx
\end{equation}
where $\hat{\mathcal{I}} = \int_{-\infty}^{\infty}|x\rangle \langle x|dx$ and $\langle x|i\rangle^* = \langle i|x\rangle$.
\\~\\Then by integration by parts,
 \begin{equation}
-\frac{\hbar^2}{2m}\int_{-\infty}^{\infty} \langle x|i\rangle^*\frac{d^2}{dx^2}\langle x|j\rangle dx = -\frac{\hbar^2}{2m}\langle x|i\rangle^*\frac{d}{dx}\langle x|j\rangle\bigg\rvert_{-\infty}^{\infty} + \frac{\hbar^2}{2m}\int_{-\infty}^{\infty} \frac{d}{dx}\langle x|i\rangle^*\frac{d}{dx}\langle x|j\rangle dx.
\end{equation}
For a normalizable wave function derived from a Hamiltonian with a symmetric potential well (like the 1D hydrogen potential), one expects that tails of the wave function drop to zero symmetrically,
meaning that the nature of the solution should set the first term on the RHS to zero. One can think of this in the following way: a bound state has an energy level which sits at some point in the potential well of the atom. The potential well of 1D hydrogen has sides that go to zero symmetrically. Therefore the barrier is higher than the energy of the state and the wave function gets eaten up symmetrically just like a particle tunnelling out of an square 1D well with infinitely long barriers. Then we get,
 \begin{equation}
-\frac{\hbar^2}{2m}\int_{-\infty}^{\infty} \langle x|i\rangle^*\frac{d^2}{dx^2}\langle x|j\rangle dx = \frac{\hbar^2}{2m}\int_{-\infty}^{\infty} \frac{d}{dx}\langle x|i\rangle^*\frac{d}{dx}\langle x|j\rangle dx = \frac{\hbar^2}{2m}\langle \frac{d}{dx}i|\frac{d}{dx}j\rangle,
\end{equation}
We use a similar method for the radial part of the kinetic energy operator acting on the $R(r)$ function. We can show why this is possible.
\\~\\Consider the expectation value of the second radial derivative on the $u(r)=rR(r)$ function,
\begin{equation}
\int_{0}^{\infty}u^*(r)\frac{d^2}{dr^2}u(r)dr, 
\end{equation}
which we can expand in the following way,
\begin{equation}
 \int_{0}^{\infty}u^*(r)\frac{d^2}{dr^2}u(r)dr =\int_{0}^{\infty}rR^*(r)\frac{d}{dr}(rR'(r)+R(r))dr =\int_{0}^{\infty}rR^*(r)(2R'(r)+rR''(r))dr.
\end{equation}
This is just
\begin{equation}
\int_{0}^{\infty}rR^*(r)(2R'(r)+rR''(r))dr  =  \int_{0}^{\infty}R^*(r)(\frac{2}{r}R'(r) +R''(r)) r^2dr.
\end{equation}
Which is the right form for the expectation value of the radial part of the kinetic energy operator aside from the constants \cite{scherrer} and the angular part of the integral. So one can continue as before for the 1D case.
\\~\\Since we can use the second derivative on the $u(r)$ functions to attain the right form for the derivatives of the $R(r)$ function, we can proceed pretty much the same as before. The question for the radial integral is then whether or not, 
\begin{equation}\label{eq:rdr}
-\frac{\hbar^2}{2m}\langle r|i\rangle^*\frac{d}{dr}\langle r|j\rangle\bigg\rvert_{0}^{\infty},
\end{equation}
goes to zero for all $i,j$. More pragmatically, in reference to the discretized space where we solve the problem, we could say that
\begin{equation}\label{eq:rdr1}
-\frac{\hbar^2}{2m}\langle r|i\rangle^*\frac{d}{dr}\langle r|j\rangle\bigg\rvert_{0}^{\mathcal{R}}
\end{equation}
should go to zero, where $\mathcal{R}$ is the outer limit of the radial box.
However, $\langle r|i\rangle$ and $\langle r|j\rangle$ are part of a basis to expand $rR(r)$ in our case, which means that $\langle r|i\rangle^*\frac{d}{dr}\langle r|j\rangle$ effectively has an $r$ in both $\langle r|i\rangle$ and $\langle r|j\rangle$. To set Eq. $\ref{eq:rdr}$ to zero using FEM techniques one can delete all basis functions on the first and last interval that are non-zero on the far left and far right boundaries. Then one can use the symmetric form of the kinetic energy matrix element. In our work, the boundary condition on the far right boundary was only introduced for the time-dependent model of hydrogen with ECS due to instability issues. In the time-independent model of hydrogen with ECS and the time-dependent model of hydrogen without ECS, no explicit care was taken to enforce this condition, and we found that the wave function naturally went to zero at the right boundary of the radial box.
\section{Time-independent Density Functional Theory}\label{section:5point2}
One can take two approaches inspired from DFT: use the full theory or just use an effective potential. We will describe the three generations of DFT so that we may motivate the existence of a useful effective potential. Above we introduced an effective potential which we solved as a kind of warm-up for the full DFT, but will leave the solution of DFT as it is outside the scope of this work.
 \\~\\Before we discuss this we quickly introduce the functional derivative which is necessary for DFT.
\subsection{The Functional Derivative}
Calculus arose to describe the evolution of physical systems through time. A major problem in calculus, regardless of the type, is finding minima or maxima (generally called an ex·tre·mum) of some problem. Suppose one has solved Newton's second law for a projectile in earth's gravity (a second order differential equation in time). In other words, consider a system defined by the ODE
\begin{equation}
-g = \frac{d^2x}{dt^2},
\end{equation}
for which the solution is,
\begin{equation}
x = -\frac{1}{2}gt^2 + v_ot.
\end{equation}
Then one may ask, what point in time is the $x$ value at a maximum or minimum (in this case a maximum since the gravitational acceleration is downwards)?
One takes a time derivative and sets this to zero, giving,
\begin{equation}
\frac{dx}{dt} = -gt + v_o = 0.
\end{equation}
Thus at $t = v_o/g$ one finds the peak of the particle's arc. 
Similarly, instead of having a function $x(t)$ one attempts to find the extremum for, one can find the extremum of a {\it functional} $F[h(\vec{y})]$, which is a function of a function of a variable. One defines the functional derivative in the following way:
\begin{equation}\label{eq:fder}
\frac{\delta F[h]}{\delta h(\vec{y})} = \lim\limits_{\epsilon \rightarrow 0}\frac{1}{\epsilon}  \{F[h(\vec{x})+\epsilon \delta(\vec{x}-\vec{y})]-F[h(\vec{x})]\},
\end{equation}
where the similarity with the limit definition of a standard derivative should be noted. We warn the reader that here when it comes to Dirac brackets, only standard angled Dirac brackets are used in the section, so every other bracket, including the ones above, are just used to separate parts of an expression. 
\\~\\ In functional calculus, finding an extremum involves finding a special function rather than a special point. To do this, one sets the functional derivative to zero.
\\~\\The chain rule can also be extended to functional derivatives.
The chain rule for functional derivatives is defined as the continuum limit of the regular chain rule. Thus an integral would survive, which takes the place of the sum over discrete variables that depend on the original variable which the derivative is taken with respect to. In other words, since the discrete set of variables are replaced with at least one function when going from calculus to functional calculus, which has a value for every point on the domain, we get a continuous set which must be integrated over. This can be written as \cite{engel}
\begin{equation}\label{eq:chainfunc}
\frac{\delta F[h]}{\delta h(\vec{y})} = \int d^3x \frac{\delta F[h]}{\delta D(\vec{x})}\frac{\delta D(\vec{x}) }{\delta h(\vec{y})},
\end{equation}
where the functional $F$ depends on the function $D$.
For multiple functions that depend on $ h(\vec{y})$ we get \cite{krieg93}
\begin{equation}\label{eq:chainfunc2}
\frac{\delta F[h]}{\delta h(\vec{y})} = \sum_i\int d^3x \frac{\delta F[h]}{\delta D_i(\vec{x})}\frac{\delta D_i(\vec{x}) }{\delta h(\vec{y})}.
\end{equation}
\subsection{Hohenberg-Kohn Theory (first generation)}\label{section:HKT}
We will study all three generations of density functional theory to see how intuition and accuracy are built up in successive approximations of the guts of the the theory. In sections where we discuss the orbitals associated with the second generation Kohn-sham equation (which the third generation theories rely on as well), the spin of the orbital is generally left out for simplicity. All the equations are in atomic units $\hbar=m_e=e=4\pi \epsilon_o=1$.
\\~\\We begin with Hohenberg-Kohn theory.
\\~\\Consider first the Schr{\"o}dinger equation for a time-independent system with internal interactions,
\begin{equation}
\hat{H}\Psi = E\Psi,
\end{equation}
or,
\begin{equation}
[\hat{T}+\hat{W}+\hat{V}]\Psi = E\Psi.
\end{equation}
Here $\hat{T}$ is the kinetic energy operator, $\hat{W}$ is the internal interaction energy operator, and $\hat{V}$ is the external potential.
\\~\\The kinetic energy operator as usual is an operator independent of the form of the final wave function. Suppose $\hat{W}$ is fixed as well \cite{engel}.
\\~\\Then $\Psi$ cannot vary by them. Let 
\begin{equation}\label{eq:bigV}
\hat{V} = \int d^3x v(\vec{r})\hat{\rho}(\vec{r}),
\end{equation}
where $\hat{\rho}(\vec{r})$ is the operator that yields the total density, given by,
\begin{equation}\label{eq:bigV2}
\hat{\rho}(\vec{r}) = \sum_i \delta^{3}(\vec{r}-\vec{r}_i),
\end{equation}
where $i$ enumerates the electron in the system.
Then let $\Psi = \Psi[v(\vec{r})]$, that is, $\Psi$ is a functional of only $v(\vec{r})$. 
\\~\\We can then write,
\begin{equation}\label{eq:rho0}
[\hat{T}+\hat{W}+\hat{V}]\Psi[v(\vec{r})] = E\Psi[v(\vec{r})].
\end{equation}
The next trick is to say that
\begin{equation}\label{eq:rho}
[\hat{T}+\hat{W}+\hat{V}[\rho]]\Psi[\rho] = E\Psi[\rho],
\end{equation}
where $\rho = \langle \Psi|\hat{\rho}(\vec{r})| \Psi \rangle$. For Eq. \ref{eq:rho} to be correct we require that, first the density is a functional of the wave function, which is a functional of $v(\vec{r})$,
\begin{equation}
\rho = \rho[\Psi[v]],
\end{equation}
which is evidently correct from the above consideration in Eq. \ref{eq:rho0} and the fact that $\rho = \langle \Psi|\hat{\rho}(\vec{r})| \Psi \rangle$. Then, we require this is invertible, or
\begin{equation}
v = v[\Psi[\rho]].
\end{equation}
In which case one can write Eq. \ref{eq:rho}.
\\~\\If this turns out to be the case, which it is, we can then write an expectation value of the total Hamiltonian as \cite{Tue11,engel}
\begin{equation}
E[\rho(\vec{r})]=\langle \hat{T} \rangle + \langle \hat{W} \rangle + \langle \hat{V}[\rho] \rangle,
\end{equation}
which can be written as
\begin{equation}
E[\rho]=F_{HK}+ \langle \hat{V}[\rho] \rangle.
\end{equation}
Using the functional derivative, we can find a ground state density using 
\begin{equation}
\frac{\delta E[\rho]}{\delta \rho(\vec{r})} = 0,
\end{equation}
or,
\begin{equation}
\frac{\delta F_{HK}}{\delta \rho(\vec{r})}=- \frac{\delta }{\delta \rho(\vec{r})}\langle \hat{V}[\rho] \rangle.
\end{equation}
Due to the variational principle the wave function that corresponds to the density for which the energy is a minimum has an energy which is larger or equal to the actual ground state \cite{Tue11,engel,scherrer}.
\\~\\Using Eqs. \ref{eq:bigV}-\ref{eq:bigV2} we get,
\begin{equation}
\bigg \langle \Psi \bigg|\hat{V}\bigg| \Psi \bigg\rangle = \bigg\langle \Psi\bigg|\int d^3x v(\vec{r})\bigg(\sum_i \delta^{3}(\vec{r}-\vec{r}_i)\bigg)\bigg|\Psi \bigg\rangle,
\end{equation}
or,
\begin{equation}
\bigg \langle \Psi \bigg|\hat{V}\bigg| \Psi \bigg\rangle = \bigg\langle \Psi\bigg|\sum_iv(\vec{r_i})\bigg|\Psi \bigg\rangle.
\end{equation}
To return a number rather than a function, this means that the integral defined by the Dirac brackets must be over the $i$ variables. Therefore we can re-write this expectation value as,
\begin{equation}
\bigg \langle \Psi \bigg|\hat{V}\bigg| \Psi \bigg\rangle = \int d^3x v(\vec{r})\bigg\langle \Psi\bigg|\bigg(\sum_i \delta^{3}(\vec{r}-\vec{r}_i)\bigg)\bigg|\Psi \bigg\rangle.
\end{equation}
Using the knowledge that the sum over Dirac delta functions is the density operator,
\begin{equation}
\bigg \langle \Psi \bigg|\hat{V}\bigg| \Psi \bigg\rangle = \int d^3x v(\vec{r})\rho(\vec{r}).
\end{equation}
Using this result in Eq. \ref{eq:fder} we get
\begin{equation}
\frac{\delta F_{HK}}{\delta \rho(\vec{r})}=- \lim\limits_{\epsilon \rightarrow 0}\frac{1}{\epsilon}\bigg[\int \bigg\{\rho(\vec{w})+\epsilon\delta(\vec{w}-\vec{r})\bigg\}v(\vec{w})d^3w-\int \rho(\vec{w})v(\vec{w})d^3w\bigg],
\end{equation}
so,
\begin{equation}
\frac{\delta F_{HK}}{\delta \rho(\vec{r})}=-\int \delta^3(\vec{w}-\vec{r})v(\vec{w})d^3w,
\end{equation}
or,
\begin{equation}
\frac{\delta F_{HK}}{\delta \rho(\vec{r})}=-v(\vec{r}),
\end{equation}
where $v(\vec{r})$ is known. As before $\rho = \langle \Psi|\hat{\rho}(\vec{r})| \Psi \rangle$ but $\Psi$ is the ground state wave function so $\rho$ is the ground state density. The is the essential equation of the Hohenberg-Kohn theory. Solving this equation gives the ground state density \cite{Tue11}.
\subsection{Kohn-Sham Theory (second generation)}\label{section:5point2point3}
For the work of our previous section to stand we require what is known as {\it v-representability}. This is to say,
\begin{equation}
\rho = \rho[\Psi_{AS}[v]],
\end{equation}
where $\Psi_{AS}$ denotes an anti-symmetric ground state wave function \cite{Tue11}.
\\~\\This is not always the case. 
\\~\\However, one can find a ground state density which has what is known as {\it N-representability}. 
\\~\\That is to say, any non-negative differentiable function for which
\begin{equation}
\int \rho(\vec{r}) d^3x = N,
\end{equation}
and,
\begin{equation}
\int |\nabla\rho(\vec{r})^{1/2}|^2 d^3x < \infty,
\end{equation}
where $N$ is the number of particles.
\\~\\Such a density is larger than the class of {\it v-representable} densities \cite{Tue11, gilbert75}.
\\~\\The approach for such a density now is a general two-step minimization procedure:
\begin{equation}
E_o = \underset{\rho}{\text{min}} \{ \underset{\Psi \rightarrow \rho}{\text{min}}\{\langle \Psi | \hat{T} + \hat{W}|\Psi\rangle \}+\int v(\vec{r})\rho(\vec{r})d^3x \},
\end{equation}
where the inner minimization searches all $\Psi$ that produce a fixed $\rho$ for the local minimal energy solution, then searches all $\rho$ for the global minimal energy solution. The functional $\underset{\Psi \rightarrow \rho}{\text{min}}\{\langle \Psi | \hat{T} + \hat{W}|\Psi\rangle \}$ is called $F_{LL}$, the Levy-Lieb functional \cite{engel}.
\\~\\To find $F_{LL}$ we now consider the Kohn-Sham method.
\\~\\Let us back-track a little. We know that in general,
\begin{equation}\label{eq:rho2}
E[\rho]=\langle \hat{T} \rangle + \langle \hat{W} \rangle + \langle \hat{V} \rangle=T[\rho] + W[\rho] + V[\rho].
\end{equation}
Suppose then we write
\begin{equation}\label{eq:rho3}
E[\rho]\approx \tilde{E}[\rho]= T_{\Sigma s}[\rho] + W_H[\rho] + V[\rho],
\end{equation}
where we define
\begin{equation}\label{eq:rho4}
T_{\Sigma s}[\rho] = \sum_{i}f_i\langle \phi_i[\rho]|\hat{T_i}|\phi_i[\rho] \rangle,
\end{equation}
where $\Sigma s$ denotes the sum over non-interacting single particle expectation values of kinetic energy, $T_i$ is the single particle kinetic energy operator and $\phi_i$ are single particle orbitals, and, in atomic units,
\begin{equation}\label{eq:rho5}
W_H[\rho] =\frac{1}{2} \int d^3x' \int d^3x\frac{\rho(\vec{r'})\rho(\vec{r})}{|\vec{r'}-\vec{r}|},
\end{equation}
also known as the Hartree energy. 
\\~\\Then we can write,
\begin{equation}\label{eq:rho52}
E_{xc}[\rho]\equiv E[\rho]- \tilde{E}[\rho]= T[\rho] - T_{\Sigma s}[\rho] + W[\rho] - W_H[\rho],
\end{equation}
where the quantity $E_{xc}$ is known as the exchange correlation energy \cite{cap06}.
\\~\\Suppose then we find the minimum energy again using,
\begin{equation}
\frac{\delta E[\rho]}{\delta \rho} = 0 = \frac{\delta}{\delta \rho} \bigg(T_{\Sigma s}[\rho] + W_H[\rho] + V[\rho] + E_{xc}[\rho]\bigg).
\end{equation}
As before the variational derivative of a potential energy functional is the potential energy, so, treating everything but the total single particle kinetic energy as a potential energy functional,
\begin{equation}
\frac{\delta E[\rho]}{\delta \rho} = 0 = \frac{\delta T_{\Sigma s}[\rho]}{\delta \rho}  + w_H(\vec{r}) + v(\vec{r}) + v_{xc}(\vec{r}),
\end{equation}
which can be written as 
\begin{equation}
\frac{\delta E[\rho]}{\delta \rho} = 0 = \frac{\delta T_{\Sigma s}[\rho]}{\delta \rho}  + v_s(\vec{r}),
\end{equation}
where all the potentials have been collected into $v_s(\vec{r})$.
\\~\\The trick is then to write a single-particle equation,\footnote{This allows us to use a kind of average exchange-correlation potential that applies for all orbitals (however fictitious they may be), rather than an orbital dependent exchange-correlation potential. This is always the case for the Kohn-Sham equations since they allow only one potential for all the orbitals. The orbital dependent exchange-correlation potential will be covered later, given by $v_{xci}=\delta E_{xc}/\delta \phi^*_i$.}
\begin{equation}
\bigg[-\frac{1}{2}\nabla^2 + v_s(\vec{r})\bigg]\phi_i(\vec{r})=\varepsilon_i\phi_i(\vec{r}),
\end{equation}
with, 
\begin{equation}
\rho(\vec{r}) = \sum_{i}^{N}f_i|\phi_i(\vec{r})|^2,
\end{equation}
where $f_i$ is the occupation function, given by,
\begin{equation}
f_i = \Theta (\varepsilon_F - \varepsilon_i)
\end{equation}
where $\Theta$ is the step function and $\varepsilon_F$ is the Fermi energy \cite{engel,cap06}. In words, this function makes sure only the orbitals below or equal to the Fermi energy are occupied when forming the total density of the ground state.
\\~\\Then one proceeds by starting with a guess $\rho_o(\vec{r})$.
\\~\\In principle, from this we can calculate $w_H(\vec{r})$ and $v_{xc}(\vec{r})$. As before, $v(\vec{r})$ is an external potential that is known. Thus one can calculate $v_s = w_h + v + v_{xc}$, which allows one to compute the eigenvectors from the single particle Schr{\"o}dinger equation with $v_s$, then compute the new density, and repeat. Thus the trick here is to fictitiously treat the system as non-interacting.
\\~\\So first, we tackle $w_H(\vec{r})$,
\begin{equation}
w_H(\vec{r})=\frac{\delta W_{H}}{\delta \rho(\vec{r})}=
 \int d^3x'\frac{\rho(\vec{r'})}{|\vec{r'}-\vec{r}|}.
\end{equation}
Then we have to face,
\begin{equation}
v_{xc}(\vec{r})=\frac{\delta E_{xc}}{\delta \rho(\vec{r})},
\end{equation}
which is more tricky to calculate.
\subsection{Calculating $E_{xc}$}
The quantity $E_{xc}$ is usually split as 
\begin{equation}
E_{xc}[\rho] = E_x[\rho] + E_c[\rho],
\end{equation}
where  the correlation part is \cite{cap06}
\begin{equation}
E_{c}[\rho] = C[\rho] + T[\rho] - T_{\Sigma s}[\rho],
\end{equation}
and the exchange part is due to the Pauli principle, given by
\begin{equation}
E_{x} = E_{xc}[\rho] - E_{c}[\rho] = W[\rho] - W_{H}[\rho]  -C[\rho],
\end{equation}
where $C[\rho]$ denotes the part of the correlation energy that is taken from the difference in internal interaction energy $W-W_H$. The other part is just the difference between the kinetic energy $T$ of the total electron wave function $\Psi[\rho]$ and the total kinetic energy $T_{\Sigma s}$ of all non-interacting single particle orbitals which can be solved for by the Kohn-Sham equation. This can be called $T_c$ \cite{cap06}.
\\~\\The exchange energy can be derived from the Hartree-Fock theory, and explicitly including both spin states $\sigma$ is given as \cite{Tue11}
\begin{equation}\label{eq:HFEQ}
E_x^{HF}[\rho] = -\frac{1}{2}\sum_{\sigma}\sum^{N_{\sigma}}_{i=1}\sum^{N_{\sigma}}_{j=1}\int d^3x'\int d^3x\frac{\phi_{i\sigma}^*(\vec{r})\phi_{j\sigma}^*(\vec{r'})\phi_{i\sigma}(\vec{r'})\phi_{j\sigma}(\vec{r})}{|\vec{r'}-\vec{r}|}.
\end{equation}
The correlation energy can be given as \cite{Tue11}
\begin{equation}
E_c[\rho]=
 \int d^3x' \int d^3x\frac{\rho(\vec{r})f(\vec{r'},\vec{r})}{|\vec{r'}-\vec{r}|}.
\end{equation}
%Where $f(\vec{r'},\vec{r})$ is an integral kernel. If we think of it as a Green's function, then,
%\begin{equation}
%\hat{L}f(\vec{r'},\vec{r}) =\delta^3(\vec{r'},\vec{r})
%\end{equation}
%Then since for,
%\begin{equation}
%\hat{L}S(\vec{r}) =D(\vec{r})
%\end{equation}
%The solution is,
%\begin{equation}
%S(\vec{r})= \int  f(\vec{r'},\vec{r}) D(\vec{r'})d^3x'
%\end{equation}
%Clearly for the correlation energy,
%\begin{equation}
%D(\vec{r'})=
%\int d^3x\frac{\rho(\vec{r})}{|\vec{r'}-%\vec{r}|}
%\end{equation}
%Which is just $w_H(\vec{r})$.
%Which is to say that the correlation energy arises out of a system that reacts to a Coloumb-like potential $w_H(\vec{r})$ derived from the total density, or,
%\begin{equation}
%\hat{L}E_c[\rho] =w_H(\vec{r})
%\end{equation}
%In which case, $f(\vec{r'},\vec{r})$ the potential energy arising from a point potential and $\int  f(\vec{r'},\vec{r})w_H(\vec{r'})d^3x'$ is the potential energy arising from a Hartree potential.
%In other words, the correlation energy is a back reaction to the Hartree potential. All that needs to be done is to quantify $\hat{L}$ to find $f(\vec{r'},\vec{r})$ and ultimately $E_c[\rho]$.
Various approximations to the $E_{xc}$ term exist.
\\~\\One is called the local density approximation or LDA where one assumes the exchange and correlation energies takes the form \cite{cap06},
\begin{equation}
E_{xc}^{LDA} =\int \rho(\vec{r})\mathcal{E}_x(\vec{r})d^3x + \int \rho(\vec{r})\mathcal{E}_c(\vec{r})d^3x.
\end{equation}
The homogeneous electron gas local density approximation or HEG-LDA is given as \cite{Tue11},
\begin{equation}
E_{xc}^{LDA} =-\frac{3}{4}\bigg(\frac{3}{\pi}\bigg)^{1/3}\int \rho(\vec{r})^{4/3}d^3x + \int \rho(\vec{r})\mathcal{E}_c(\vec{r})d^3x.
\end{equation}
The Becke hybrid is another approach, given as \cite{Tue11},
\begin{equation}
E_{xc}^{Becke}[\rho] = c_oE_{x}^{HF}[\rho]+c_1E_{xc}^{LDA},
\end{equation}
or more fully as,
\begin{align}
\begin{split}
E_{xc}^{Becke}[\rho]
=c_o\left\{-\frac{1}{2}\sum_{\sigma}\sum^{N_{\sigma}}_{i=1}\sum^{N_{\sigma}}_{j=1}\int d^3x'\int d^3x\frac{\phi_{i\sigma}^*(\vec{r})\phi_{j\sigma}^*(\vec{r'})\phi_{i\sigma}(\vec{r'})\phi_{j\sigma}(\vec{r})}{|\vec{r'}-\vec{r}|}\right\} \\+c_1\left\{-\frac{3}{4}\bigg(\frac{3}{\pi}\bigg)^{1/3}\int \rho(\vec{r})^{4/3}d^3x + \int \rho(\vec{r})\mathcal{E}_c(\vec{r})d^3x\right\}.
\end{split}
\end{align}
\subsection{Optimized Effective Potential Theory (third generation)}
LDA theory fails to account for the existence of negative ions since $v_s^{LDA}$ decays faster than $-1/r$, implying the neutral atom does not have a Rydberg series, preventing binding of an additional electron \cite{engel}. It also fails to account for the London dispersion force since in LDA theory correlation energy between atoms or molecules only appears when their electron densities overlap \cite{engel}. LDA also runs into problems when dealing with strongly correlated systems. The generalized gradient expansion, or GGA, which involves considering derivatives of the density in the determination of $E_{xc}$ also fails in these three regards \cite{engel}.
\\~\\Therefore a better approximation for $E_{xc}$ must be introduced.
Instead of solving for $\rho$ by minimizing the energy functional $E[\rho]$, one can solve for $v_{xc}$ by minimizing $E[v_{xc}]$.
\\~\\Consider using the Kohn-Sham method.
\\~\\Using the former KS method, one begins with a guess $\rho_o(\vec{r})$ to find $w_H$ and finds an expression for $E_{xc}[\rho]$, calculates $v_{xc}(\vec{r})= \delta E_{xc}[\rho]/ \delta \rho$ which is a necessary ingredient for the potential $v_s(\vec{r})$, then solves for the orbitals from the single particle Schr{\"o}dinger equation, computes the new $\rho$ and then iterates. However, unless one makes an approximation for $E_c[\rho]$, the correlation energy, such as $E_c[\rho]=0$, one cannot compute $v_s(\vec{r})$ since the exact form of $E_c[\rho]$ is unknown. 
\\~\\As mentioned, one can modify the KS method by solving for $v_{xc}$. Here one solves the KS equation simultaneously with equations which determine the optimal $v_{xc}$ at each step of iteration.
\subsubsection{The OEP Equations} 
%One first considers that
%\begin{equation}
%v_{xc}^{\sigma}(\vec{r}) = \frac{\delta E_{xc}}{\delta \rho_{\sigma}(\vec{r})} = \sum_{\sigma'}\int d^3x' \frac{\delta v_s^{\sigma'}(\vec{r'})}{\delta \rho_{\sigma}(\vec{r})}\frac{\delta E_{xc}}{\delta v_s^{\sigma'}(\vec{r'})}
%\end{equation}
%For now we will consider the orbitals $\phi_i$ as bi-spinors, which is to say, they include both up and down spin. We can write these as,
%\begin{equation}
%\phi_i(\vec{r}) = \left(   
%\begin{tabular}{c}
%$\phi_i^{+}(\vec{r})$  \\
%$\phi_i^{-}(\vec{r})$  \\
%\end{tabular}\right).
%\end{equation}
%The KS potential will be,
%\begin{equation}
%v_s(\vec{r}) = \left(   
%\begin{tabular}{c c}
%$v_s^{+}(\vec{r})$ & $0$ \\
%$0$ & $v_s^{-}(\vec{r})$ \\
%\end{tabular}\right).
%\end{equation}
There are at least three ways to find the optimized effective potential (OEP) equations, the first is by a direct functional derivative, the second is by total energy minimization, and the third is by using the equality of the Kohn-Sham total density and the density of the interacting system \cite{engel}. We will follow the second way. Since we are driving towards the Krieger-Li-Iafrate (KLI) approximation of the OEP equations, we shall ignore the $\delta E_{xc} /\delta \varepsilon_i$ term (where $\varepsilon_i$ are eigenvalues of the KS equation) that arises in the equations generally. This is because the KLI approximation is ambiguous if it is not set to zero \cite{engel}. 
Applying first the minimization condition to the energy functional, with respect to $v_s$, we get \cite{krieg93}
\begin{equation}\label{eq:vs}
\frac{\delta E}{\delta v_{s}(\vec{r})} = \sum_{i}\int d^3x'\frac{\delta E}{\delta\phi^*_{i}(\vec{r'})} \frac{\delta\phi^*_{i}(\vec{r'})}{\delta v_{s}(\vec{r})} +c.c.= 0.
\end{equation}
To calculate the functional derivative $\delta\phi^*_{i}(\vec{r'})/\delta v_{s}(\vec{r})$, consider the first order perturbation correction to the wave function, where one gets,
\begin{equation}
\delta \phi_i(\vec{r'}) = \sum_{j\neq i }\frac{\langle \phi_j| \delta v_s|\phi_i\rangle }{\varepsilon_i-\varepsilon_j}\phi_j(\vec{r'}),
\end{equation}
which is,
\begin{equation}
\delta \phi_i(\vec{r'}) = \sum_{j\neq i }\frac{\int  \phi^*_j(\vec{r''}) \delta v_s(\vec{r''})\phi_i(\vec{r''})d^3x'' }{\varepsilon_i-\varepsilon_j}\phi_j(\vec{r'}),
\end{equation}
or,
\begin{equation}
\delta \phi_i(\vec{r'}) = -\int d^3x'' \sum_{j\neq i }\frac{\phi_j(\vec{r'})\phi^*_j(\vec{r''})}{\varepsilon_j-\varepsilon_i}\delta v_s(\vec{r''})\phi_i(\vec{r''}).
\end{equation}
Then taking the complex conjugate, and assuming $\delta v_s(\vec{r''})$ is real, we get,
\begin{equation}\label{eq:firstorder}
\delta \phi^*_i(\vec{r'}) = -\int d^3x'' \sum_{j\neq i }\frac{\phi^*_j(\vec{r'})\phi_j(\vec{r''})}{\varepsilon_j-\varepsilon_i}\delta v_s(\vec{r''})\phi^*_i(\vec{r''}).
\end{equation}
Dividing through by $\delta v_s(\vec{r})$,
\begin{equation}
\frac{\delta \phi^*_i(\vec{r'})}{\delta v_s(\vec{r})} = -\int d^3x'' \sum_{j\neq i }\frac{\phi^*_j(\vec{r'})\phi_j(\vec{r''})}{\varepsilon_j-\varepsilon_i}\frac{\delta v_s(\vec{r''})}{\delta v_s(\vec{r})}\phi^*_i(\vec{r''}),
\end{equation}
which is,\footnote{Consider the functional derivative
\begin{equation*}
\frac{\delta F[g]}{\delta g(\vec{y})} = \lim\limits_{\epsilon \rightarrow 0}\frac{1}{\epsilon}  \{F[g(\vec{x})+\epsilon \delta(\vec{x}-\vec{y})]-F[g(\vec{x})]\}
\end{equation*}
with $F[g(\vec{x})] = g(\vec{x})$, 
\\~\\so,
\begin{equation*}
\frac{\delta g(\vec{x})}{\delta g(\vec{y})} = \lim\limits_{\epsilon \rightarrow 0}\frac{1}{\epsilon}  \{g(\vec{x})+\epsilon \delta(\vec{x}-\vec{y})-g(\vec{x})\} = \delta(\vec{x}-\vec{y}).
\end{equation*}}
\begin{equation}
\frac{\delta \phi^*_i(\vec{r'})}{\delta v_s(\vec{r})} = -\int d^3x'' \sum_{j\neq i }\frac{\phi^*_j(\vec{r'})\phi_j(\vec{r''})}{\varepsilon_j-\varepsilon_i}\delta^{(3)}(\vec{r''},\vec{r})\phi^*_i(\vec{r''}),
\end{equation}
so,
\begin{equation}\label{eq:gphi}
\frac{\delta\phi^*_{i}(\vec{r'})}{\delta v_{s}(\vec{r})} = - \bar{G}_{i}(\vec{r'},\vec{r})\phi^*_{i}(\vec{r})
\end{equation}
where,
\begin{equation}\label{eq:greens}
\bar{G}_{i}(\vec{r'},\vec{r}) = \sum_{j\neq i} \frac{\phi_{j}^*(\vec{r'})\phi_{j}(\vec{r})}{\varepsilon_{j}-\varepsilon_{i}}.
\end{equation}
This is a modified Green's function for the equation
\begin{equation}\label{eq:eqgreens}
[\hat{h}-\varepsilon_i]\phi(\vec{r}) = y(\vec{r}),
\end{equation}
which has a standard Green's function,
\begin{equation}
[\hat{h}-\varepsilon_i]G_i(\vec{r}) = \delta^{(3)}(\vec{r'},\vec{r}).
\end{equation}
To find the Green's function one can write the eigenvalue problem for Eq. \ref{eq:eqgreens}, which is, 
\begin{equation}\label{eq:eqgreensev}
[\hat{h}-\varepsilon_i]\phi_j(\vec{r}) = \lambda \phi_j(\vec{r}).
\end{equation}
Then if $\phi_j(\vec{r})$ is an eigenvector of $\hat{h}$,
\begin{equation}\label{eq:eqgreensev2}
[\hat{h}-\varepsilon_i]\phi_j(\vec{r}) = [\epsilon_j - \epsilon_i] \phi_j(\vec{r}).
\end{equation}
Then applying the standard definition of a Green's function, one finds that,
\begin{equation}
G_{i}(\vec{r'},\vec{r}) = \sum_{j} \frac{\phi_{j}^*(\vec{r'})\phi_{j}(\vec{r})}{\varepsilon_{j}-\varepsilon_{i}}.
\end{equation}
Then to find the modified Green's function,  we separate the differential equation defining Green's function into two parts, giving,
\begin{equation}
\bigg[\hat{h}-\varepsilon_i\bigg]\sum_{j\neq i } \frac{\phi_{j}^*(\vec{r'})\phi_{j}(\vec{r})}{\varepsilon_{j}-\varepsilon_{i}} + \bigg[\hat{h}-\varepsilon_i\bigg]\frac{\phi_{i}^*(\vec{r'})\phi_{i}(\vec{r})}{\varepsilon_{i}-\varepsilon_{i}} = \delta^{(3)}(\vec{r'},\vec{r}),
\end{equation}
or,
\begin{equation}
\bigg[\hat{h}-\varepsilon_i\bigg]\bar{G}_{i}(\vec{r'},\vec{r}) = \delta^{(3)}(\vec{r'},\vec{r}) -  \bigg[\hat{h}-\varepsilon_i\bigg]\frac{\phi_{i}^*(\vec{r'})\phi_{i}(\vec{r})}{\varepsilon_{i}-\varepsilon_{i}},
\end{equation}
since $\hat{h}$ only acts on the $\vec{r}$ variable,
\begin{equation}
\bigg[\hat{h}-\varepsilon_i\bigg]\bar{G}_{i}(\vec{r'},\vec{r}) = \delta^{(3)}(\vec{r'},\vec{r}) -  \bigg[\varepsilon_{i}-\varepsilon_i\bigg]\frac{\phi_{i}^*(\vec{r'})\phi_{i}(\vec{r})}{\varepsilon_{i}-\varepsilon_{i}},
\end{equation}
so,
\begin{equation}
[\hat{h}-\varepsilon_i]\bar{G}_{i}(\vec{r'},\vec{r}) = \delta^{(3)}(\vec{r'},\vec{r}) -  \phi_{i}^*(\vec{r'})\phi_{i}(\vec{r}).
\end{equation}
This is the differential equation defining the modified Green's function \cite{krieg93}.
The derivatives $\delta E/\delta \phi^*_{i}(\vec{r'})$ can be calculated with the definition for the functional derivative, but one expects they need an explicit expression for $E[\phi^*_{i}]$.
%\\~\\This Green's function solves the equation
%\begin{equation}\label{eq:ksg}
%[-\frac{1}{2}\nabla^2 + v_{s,\sigma}(\vec{r})]G_{i\sigma}(\vec{r'},\vec{r})=\delta^3(\vec{r'},\vec{r})-\phi_{i\sigma}(\vec{r})\phi_{i\sigma}^*(\vec{r'})
%\end{equation}
\\~\\Suppose $E[\phi^*_{i}]$, in similar fashion to $\rho$ as defined earlier, is the sum over all expectation values of the single particle Hamiltonians (evaluated for the eigenfunctions of the KS equation) weighted by $f_{i}$ plus $E_{xc}$,
\begin{equation}
E[\phi_{i},\phi^*_{i}] = \sum_{j} f_{j}\int   \phi^*_{j}(\vec{r'})\hat{\underline{h}}'\phi_{j}(\vec{r'})d^3x' + E_{xc}.
\end{equation}
The functional derivative is,
\begin{equation}
\frac{\delta E[\phi_i,\phi^*_i]}{\delta \phi^*_{i}(\vec{r'})} = \lim\limits_{\epsilon \rightarrow 0}\frac{1}{\epsilon}  \{E[\phi_i,\phi^*_i(\vec{r})+\epsilon \delta(\vec{r}-\vec{r'})]-E[\phi_i,\phi^*_i(\vec{r})]\},
\end{equation}
or,
\begin{align}
\begin{split}
\frac{\delta E[\phi_i,\phi^*_i]}{\delta \phi^*_{i}(\vec{r'})} = \lim\limits_{\epsilon \rightarrow 0}\frac{1}{\epsilon}\bigg[\sum_{j\neq i} f_{j}\int \phi^*_j(\vec{r})\hat{\underline{h}}\phi_{j}(\vec{r})d^3x + f_i\int \bigg\{\phi^*_i(\vec{r})+\epsilon \delta(\vec{r}-\vec{r'})\bigg\}\hat{\underline{h}}\phi_{i}(\vec{r})d^3x \\+E_{xc}[\phi_i,\phi^*_i(\vec{r})+\epsilon \delta(\vec{r}-\vec{r'})]-\sum_{j\neq i} f_{j}\int  \phi^*_j(\vec{r})\hat{\underline{h}}\phi_{j}(\vec{r})d^3x \\-f_i\int \bigg\{\phi^*_i(\vec{r})\bigg\}\hat{\underline{h}}\phi_{i}(\vec{r})d^3x -E_{xc}[\phi_i,\phi^*_i(\vec{r})]\bigg],
\end{split}
\end{align}
or,
\begin{align}
\begin{split}
\frac{\delta E[\phi_i,\phi^*_i]}{\delta \phi^*_{i}(\vec{r'})} = \lim\limits_{\epsilon \rightarrow 0}\frac{1}{\epsilon}\bigg[ f_{i}\int  \epsilon \delta(\vec{r}-\vec{r'})\hat{\underline{h}}\phi_{i}(\vec{r})d^3x + E_{xc}[\phi_i,\phi^*_i(\vec{r})+\epsilon \delta(\vec{r}-\vec{r'})]- E_{xc}[\phi_i,\phi^*_i(\vec{r})]\bigg],
\end{split}
\end{align}
or,
\begin{align}
\begin{split}
\frac{\delta E[\phi_i,\phi^*_i]}{\delta \phi^*_{i}(\vec{r'})} = \lim\limits_{\epsilon \rightarrow 0}\frac{1}{\epsilon}\bigg[f_{i}\int  \epsilon \delta(\vec{r}-\vec{r'})\hat{\underline{h}}\phi_{i}(\vec{r})d^3x \bigg]+\frac{\delta E_{xc}[\phi_{i},\phi^*_{i}]}{\delta \phi^*_{i}(\vec{r'})},
\end{split}
\end{align}
or,
\begin{equation}
\frac{\delta E[\phi_{i},\phi^*_{i}]}{\delta \phi^*_{i}(\vec{r'}) } = f_{i}\hat{\underline{h}}'\phi_{i}(\vec{r'}) +\frac{\delta E_{xc}[\phi_{i},\phi^*_{i}]}{\delta \phi^*_{i}(\vec{r'})},
\end{equation}
with,
\begin{equation}\label{eq:ksn1}
\hat{\underline{h}}'=-\frac{1}{2}\nabla'^2 + w_{H}(\vec{r'}) + v(\vec{r'}).
\end{equation}
We can then write,
\begin{equation}
\frac{\delta E[\phi_{i},\phi^*_{i}]}{\delta \phi^*_{i}(\vec{r'})} = f_{i}\bigg[-\frac{1}{2}\nabla'^2 + w_{H}(\vec{r'}) + v(\vec{r'})\bigg]\phi_{i}(\vec{r'}) + \frac{\delta E_{xc}[\phi_{i},\phi^*_{i}]}{\delta \phi^*_{i}(\vec{r'}) } .
\end{equation}
Remember that,
\begin{equation*}
\hat{h}\phi_{i}(\vec{r})=\bigg[-\frac{1}{2}\nabla^2 + w_{H}(\vec{r}) + v(\vec{r}) + v_{xc}(\vec{r})\bigg]\phi_{i}(\vec{r})=\varepsilon_{i}\phi_{i}(\vec{r}),
\end{equation*}
so,
\begin{equation}
\frac{\delta E[\phi_{i},\phi^*_{i}]}{\delta \phi^*_{i}(\vec{r'})} = f_{i}\bigg[\varepsilon_{i} - v_{xc}(\vec{r'})\bigg]\phi_{i}(\vec{r'}) + \frac{\delta E_{xc}[\phi_{i},\phi^*_{i}]}{\delta \phi^*_{i}(\vec{r'}) }.
\end{equation}
Therefore we can take the total energy minimization,
\begin{equation*}\label{eq:vs0}
\frac{\delta E}{\delta v_{s}(\vec{r})} = \sum_{i}\int d^3x'\frac{\delta E}{\delta\phi^*_{i}(\vec{r'})} \frac{\delta\phi^*_{i}(\vec{r'})}{\delta v_{s}(\vec{r})} +c.c.= 0,
\end{equation*}
and use the definition of $\delta \phi^*_i(\vec{r'})/\delta v_s(\vec{r})$ arrived at above, to get,
\begin{equation}\label{eq:vs02}
\frac{\delta E}{\delta v_{s}(\vec{r})} = \sum_{i}\int d^3x'\frac{\delta E}{\delta\phi^*_{i}(\vec{r'})}\bigg\{ - \bar{G}_{i}(\vec{r'},\vec{r})\phi^*_{i}(\vec{r})\bigg\} +c.c.= 0,
\end{equation}
or, adding the functional derivative of $E$ with respect to the $c.c.$ of an orbital we just derived,
\begin{equation}\label{eq:vs2}
\frac{\delta E}{\delta v_{s}(\vec{r})} = \sum_{i}\int d^3x' \Bigg [\Bigg \{-\bar{G}_{i}(\vec{r'},\vec{r})\phi^*_{i}(\vec{r})\Bigg \}\Bigg \{f_{i}\Bigg(\varepsilon_{i} - v_{xc}(\vec{r'})\Bigg)\phi_{i}(\vec{r'})+ \frac{\delta E_{xc}[\phi_{i},\phi^*_{i}]}{\delta \phi^*_{i}(\vec{r'}) } \Bigg \} +c.c.\Bigg ]= 0.
\end{equation}
Splitting this up,
\begin{align}\label{eq:vs3}
\begin{split}
 \sum_{i}\int d^3x' \Bigg [\Bigg \{\bar{G}_{i}(\vec{r'},\vec{r})\phi^*_{i}(\vec{r})\Bigg \}\Bigg \{f_{i}v_{xc}(\vec{r'})\phi_{i}(\vec{r'})\Bigg \} +c.c.\Bigg ]\\
 +
 \sum_{i}\int d^3x' \Bigg [\Bigg \{-\bar{G}_{i}(\vec{r'},\vec{r})\phi^*_{i}(\vec{r})\Bigg \}\Bigg \{f_{i}\varepsilon_{i} \phi_{i}(\vec{r'})\Bigg \} +c.c.\Bigg ]
 \\-  \sum_{i}\int d^3x' \Bigg [\Bigg \{-\bar{G}_{i}(\vec{r'},\vec{r})\phi^*_{i}(\vec{r})\Bigg \}\Bigg \{ \frac{\delta E_{xc}[\phi_{i},\phi^*_{i}]}{\delta \phi^*_{i}(\vec{r'}) } \Bigg \} +c.c.\Bigg ] = 0.
 \end{split}
\end{align}
Due to the orthogonality relation,
\begin{equation}\label{eq:orth}
\int d^3x' \bar{G}_{i}(\vec{r'},\vec{r})\phi_{i}(\vec{r'})=0,
\end{equation}
the second line in Eq. \ref{eq:vs3} is zero, so we have,
\begin{align}\label{eq:vs4}
 \sum_{i}\int d^3x' \Bigg [\Bigg \{\bar{G}_{i}(\vec{r'},\vec{r})\phi^*_{i}(\vec{r})\Bigg \}\Bigg \{f_{i}\Bigg(v_{xc}(\vec{r'})-v_{xci}(\vec{r'})\Bigg)\phi_{i}(\vec{r'})\Bigg \} +c.c.\Bigg ] = 0.
\end{align}
We make a note about the definition of $v_{xci}$. One can find $v_{xci}$ by the standard definition,
\begin{align}
E_{xc}[\phi_{j},\phi^*_{j}]= \sum_j \int d^3x f_j \phi^*_{j}(\vec{r})v_{xcj}(\vec{r})\phi_{j}(\vec{r}),
\end{align}
so,
\begin{align}\label{eq:vxcis}
\begin{split}
 \frac{\delta E_{xc}[\phi_i,\phi^*_i]}{\delta \phi^*_i(\vec{r})} = \lim\limits_{\epsilon \to 0}\frac{1}{\epsilon}\bigg[\sum_{j\neq i}\int d^3x'f_{j}\phi^*_j(\vec{r'})v_{xcj}(\vec{r'})\phi_{j}(\vec{r'}) + \int d^3x' f_i \bigg\{\phi^*_i(\vec{r'})+\epsilon \delta(\vec{r'}-\vec{r})\bigg\}v_{xcj}(\vec{r'})\phi_i(\vec{r'}) \\- \sum_{j}\int d^3x'f_{j}\phi^*_j(\vec{r'})v_{xcj}(\vec{r'})\phi_{j}(\vec{r'})\bigg],
 \end{split}
\end{align}
so,
\begin{equation}\label{eq:vxcis2}
 \frac{\delta E_{xc}[\phi_i,\phi^*_i]}{\delta \phi^*_i(\vec{r})} = \lim\limits_{\epsilon \to 0}\frac{1}{\epsilon}\bigg[\int d^3x'f_i \bigg\{\epsilon \delta(\vec{r'}-\vec{r})\bigg\}v_{xcj}(\vec{r'})\phi_i(\vec{r'})\bigg].
\end{equation}
Then we can write (here we switch the primed co-ordinate to be consistent with Eq. \ref{eq:vs3})
\begin{align}
\frac{\delta E_{xc}[\phi_{i},\phi^*_{i}]}{\delta \phi^*_{i}(\vec{r'}) } = f_i v_{xci}(\vec{r'})\phi_{j}(\vec{r'}), \frac{\delta E_{xc}[\phi_{i},\phi^*_{i}]}{\delta \phi_{i}(\vec{r'}) } = f_i \phi^*_{i}(\vec{r'}) v_{xci}(\vec{r'}).
\end{align}
Eq. \ref{eq:vs4} is often written instead as,
\begin{equation}\label{eq:vs5}
 \sum_{i}\int d^3x' \Bigg [\Bigg \{\bar{G}_{i}(\vec{r},\vec{r'})\phi_{i}(\vec{r})\Bigg \}\Bigg \{f_{i}\Bigg(v_{xc}(\vec{r'})-v_{xci}(\vec{r'})\Bigg)\phi^*_{i}(\vec{r'})\Bigg \} +c.c.\Bigg ] = 0,
\end{equation}
where have just flipped which part is contained in $c.c.$\footnote{Note that 
\begin{equation}\label{eq:g}
\bar{G}_{i}^*(\vec{r'},\vec{r}) =\bar{G}_{i}(\vec{r},\vec{r'}).
\end{equation}}
\\~\\If we multiple through by $-1$, which does not change the form of the equation but flips the order of $v_{xc}$ and $v_{xci}$, this can be written in compact form as \cite{k03}
\begin{equation}\label{eq:vs6}
 \sum_{i}f_{i}\psi^*_{i}(\vec{r})\phi_{i}(\vec{r}) +c.c.= 0,
\end{equation}
where,
\begin{equation}\label{eq:vs7}
\psi^*_{i}(\vec{r}) =  \int d^3x' \Bigg [\Bigg \{\bar{G}_{i}(\vec{r},\vec{r'})\Bigg \}\Bigg \{[v_{xci}(\vec{r'})-v_{xc}(\vec{r'})]\phi^*_{i}(\vec{r'})\Bigg \}\Bigg ].
\end{equation}
Note that first order perturbation theory gives the result (re-arranged from Eq. \ref{eq:gphi})
\begin{equation}\label{eq:gphi2}
\frac{\delta\phi^*_{i}(\vec{r})}{\delta v_{s}(\vec{r'})} = - \bar{G}_{i}(\vec{r},\vec{r'})\phi^*_{i}(\vec{r'}),
\end{equation}
or by Eq. \ref{eq:firstorder},
\begin{equation}\label{eq:gphi3}
\delta\phi^*_{i}(\vec{r}) = \int d^3x' \Bigg [\bar{G}_{i}(\vec{r},\vec{r'})\phi^*_{i}(\vec{r'}) \bigg\{-\delta v_{s}(\vec{r'})\bigg\}\bigg],
\end{equation}
where $\delta v_s = v_{xc}-v_{xci}$ and
\begin{equation}\label{eq:greensstar}
\bar{G}_{i}(\vec{r},\vec{r'}) = \sum_{j\neq i} \frac{\phi_{j}^*(\vec{r})\phi_{j}(\vec{r'})}{\varepsilon_{j}-\varepsilon_{i}}.
\end{equation}
So $\psi^*_i(\vec{r})$ is just $\delta \phi^*_{i}(\vec{r})$ at first order for a change $v_{xc} \rightarrow v_{xci}$ according to Eq. \ref{eq:firstorder} and Eq. \ref{eq:greens}. 
Note that the density can be defined as,
\begin{equation}\label{eq:vs6change}
 \rho = \sum_{j}f_{j}\phi^*_j(\vec{r})\phi_{j}(\vec{r}),
\end{equation}
so the minimization condition for $\rho$ is
\begin{equation}\label{eq:vs6change2}
 \frac{\delta \rho[\phi_i,\phi^*_i]}{\delta \phi^*_i(\vec{r})} = \lim\limits_{\epsilon \to 0}\frac{1}{\epsilon}\bigg[\sum_{j\neq i}f_{j}\phi^*_j(\vec{r'})\phi_{j}(\vec{r'}) + f_i \bigg\{\phi^*_i(\vec{r'})+\epsilon \delta(\vec{r'}-\vec{r})\bigg\}\phi_i(\vec{r'}) - \sum_{j}f_{j}\phi^*_j(\vec{r'})\phi_{j}(\vec{r'})\bigg]= 0,
\end{equation}
or,
\begin{equation}\label{eq:vs6change22}
 \frac{\delta \rho[\phi_i,\phi^*_i]}{\delta \phi^*_i(\vec{r})} = \lim\limits_{\epsilon \to 0}\frac{1}{\epsilon}\bigg[ f_i \bigg\{\epsilon \delta(\vec{r'}-\vec{r})\bigg\}\phi_i(\vec{r'})\bigg]= 0.
\end{equation}
That makes,
\begin{equation}\label{eq:vs6change4}
 \delta \rho_{\phi^*_i}(\vec{r'}) = \int d^3x\bigg[ f_i \bigg\{ \delta(\vec{r'}-\vec{r})\bigg\}\phi_i(\vec{r'})\bigg]\delta \phi^*_i(\vec{r}) =f_i\delta \phi^*_i (\vec{r'})\phi_{i}(\vec{r'}) = 0.
\end{equation}
The minimization condition for the total change in $\rho$ for a change in every $\phi^*_i$ and $\phi_i$ is,
\begin{equation}\label{eq:vs6change4}
 \delta \rho_{tot}(\vec{r}) = \sum_i f_{i}\delta \phi^*_i(\vec{r}) \phi_{i}(\vec{r}) + \sum_i f_{i}\delta \phi_i(\vec{r}) \phi^*_{i}(\vec{r})= 0.
\end{equation}
So the Eq. \ref{eq:vs6} is just this same minimization condition. Since there are two parts, alternatively \cite{jou07}
\begin{equation}\label{eq:vs62}
 \sum_{i}f_{i}\psi_{i}(\vec{r})\phi^*_{i}(\vec{r}) +c.c.= 0,
\end{equation}
where,
\begin{equation}\label{eq:vs72}
\psi_{i}(\vec{r}) =  \int d^3x' \Bigg [\Bigg \{\bar{G}_{i}(\vec{r'},\vec{r})\Bigg \}\Bigg \{[v_{xci}(\vec{r'})-v_{xc}(\vec{r'})]\phi_{i}(\vec{r'})\Bigg \}\Bigg ],
\end{equation} 
Here,
\begin{equation}\label{eq:greensstar2}
\bar{G}_{i}(\vec{r'},\vec{r}) = \sum_{j\neq i} \frac{\phi_{j}^*(\vec{r'})\phi_{j}(\vec{r})}{\varepsilon_{j}-\varepsilon_{i}}.
\end{equation}
Eq. \ref{eq:vs6} or  Eq. \ref{eq:vs62} is the central OEP equation. It says that the optimal orbital independent exchange-correlation potential is the one that makes the total change in total density (where the change is measured in comparison with use of the orbital dependent exchange-correlation potential) vanish at first order \cite{k03}.
\\~\\This is solved simultaneously with,
\begin{equation}\label{eq:ks}
\hat{h}\phi_{i}(\vec{r})=\bigg[-\frac{1}{2}\nabla^2 + v_{s}(\vec{r})\bigg]\phi_{i}(\vec{r})=\varepsilon_{i}\phi_{i}(\vec{r}),
\end{equation}
with
\begin{equation}\label{eq:ks2}
v_{s}(\vec{r}) = w_{H}(\vec{r}) + v(\vec{r}) + v_{xc}(\vec{r}), 
\end{equation}
at each iteration.
%\\~\\Note that $\hat{h}_{\sigma}$ acting on the {\it ith} orbital can be called $\hat{h}_{i\sigma}$ (similarly $v_{s,\sigma} \rightarrow v_{s,i\sigma}$) since the exchange-correlation potential generally depends on the orbital being considered.
\\~\\Clearly Eq. \ref{eq:vs6} requires the orbitals to proceed and find the optimal $v_{xc}$ for a single iteration, while Eq. \ref{eq:ks} requires the potential $v_{s,}$ to find the orbitals for a single iteration.
\\~\\Thus one can with either a guess of {\it all} the orbitals or a guess of just $v_{s,o}$ to begin the iteration process.  Beginning with a guess of $v_{s,o}$ might be simplest.
\\~\\For example, one can begin with a very primary approach,
\begin{equation}\label{eq:ks3}
v_{s,o}(\vec{r}) =  v(\vec{r}),
\end{equation}
since $v(\vec{r})$ is known. 
\\~\\Or one could introduce a beginning guess for the exchange potential, using the LDA approximation,
\begin{equation}\label{eq:ks4}
v_{s,o}(\vec{r}) =  v(\vec{r}) + v_{x,o}^{LDA}(\vec{r}),
\end{equation}
since $v_{x,o}^{LDA}(\vec{r})$ is also known for given $\rho_o(\vec{r})$.
\\~\\Since one is already making a guess for $\rho$, one might as well write,
\begin{equation}\label{eq:ks5}
v_{s,o}(\vec{r}) =  w_{H,o}(\vec{r})+v(\vec{r}) + v_{x,o}^{LDA}(\vec{r}), 
\end{equation}
where one can find $w_{H,o}(\vec{r})$ from the same $\rho_o(\vec{r})$.
\\~\\From that one may obtain $\phi_{i,o}(\vec{r})$, allowing one to compute the next iteration of $w_H$ and then find $v_{xc}$. Then one can again compute $\phi_{i}(\vec{r})$ from the KS equation and continue the iteration procedure.
\\~\\The success of this technique is that for the exact exchange-only potential one finds the correct asymptotic behaviour which allows for Rydberg states \cite{engel}.
\\~\\This is a valuable achievement, since $v_{xci}$ must be known to calculate $v_{xc}$ for the KS equation. However, if one assumes $E_{xc} \rightarrow E_{x}$, then through,
\begin{equation}
 v_{xi} = \frac{\delta E_x^{HF}}{\delta \phi^*_i},
\end{equation}
which is exactly known from Eq. \ref{eq:HFEQ} for a given set of $\phi_i$, so one can find $v_{x}$ for the KS equation through Eq. \ref{eq:vs6}.
%\\~\\Consider then that $\hat{h}_{i}\phi_{i}(\vec{r})= \varepsilon_{i}\phi_{i}(\vec{r})$.
%So one might write,
%\begin{equation}
%E[\phi_{i},\phi^*_{i},\varepsilon_{i}] = \sum_{i} f_{i}\int   \phi^*_{i}(\vec{r})\varepsilon_{i}\phi_{i}(\vec{r})d^3x
%\end{equation}
%Assuming $\phi_{i}$ are orthonormal,
%\begin{equation}
%E[\phi_{i},\phi^*_{i},\varepsilon_{i}] = \sum_{i} f_{i}\varepsilon_{i} = E[\varepsilon_{i}]
%\end{equation}
%So,
%\begin{equation}
%\frac{\partial E[\varepsilon_{i}]}{\partial \varepsilon_{i} } = f_{i}
%\end{equation}
%\\~\\It follows that at any step,
%\begin{equation}
%v_{xc,\sigma}(\vec{r}) \equiv v_{s,\sigma}(\vec{r}) - v_{\sigma}(\vec{r}) - w_H(\vec{r})
%\end{equation}
\subsection{Krieger-Li-Iafrate Theory}
The Krieger-Li-Iafrate (KLI) approximation involves setting energy differences between KS eigenstates labelled by $j$ and $i$ constant,
\begin{equation}
\Delta = \varepsilon_{j}-\varepsilon_{i},
\end{equation}
so,
\begin{equation}
\bar{G}^{KLI}_{i}(\vec{r'},\vec{r}) = \sum_{j\neq i} \frac{\phi_{j}^*(\vec{r'})\phi_{j}(\vec{r})}{\Delta},
\end{equation}
or, using the completeness relation,
\begin{equation}
\delta ^{(3)}(\vec{r'}, \vec{r}) = \sum_{j} \frac{\phi_{j}^*(\vec{r'})\phi_{j}(\vec{r})}{\Delta},
\end{equation}
we have,
\begin{equation}
\bar{G}^{KLI}_{i}(\vec{r'},\vec{r}) =   \frac{\delta ^{(3)}(\vec{r'}, \vec{r})- \phi_{i}^*(\vec{r'})\phi_{i}(\vec{r})}{\Delta},
\end{equation}
or,
\begin{equation}
\bar{G}^{KLI*}_{i}(\vec{r'},\vec{r}) =  \bar{G}^{KLI}_{i}(\vec{r},\vec{r'}) =   \frac{\delta ^{(3)}(\vec{r'}, \vec{r})- \phi_{i}^*(\vec{r})\phi_{i}(\vec{r'})}{\Delta}.
\end{equation}
Putting this back in the OEP equation
\begin{equation}\label{eq:KLI1}
 \sum_{i}\int d^3x' \Bigg [\Bigg \{\frac{\delta ^{(3)}(\vec{r'}, \vec{r})- \phi_{i}^*(\vec{r})\phi_{i}(\vec{r'})}{\Delta}\phi_{i}(\vec{r})\Bigg \}\Bigg \{f_{i}[v^{KLI}_{xc}(\vec{r'})-v_{xci}(\vec{r'})]\phi^*_{i}(\vec{r'})\Bigg \} +c.c.\Bigg ] = 0,
\end{equation}
or,
\begin{align}\label{eq:KLI2}
\begin{split}
 \sum_{i}\int d^3x' \Bigg [\Bigg \{\delta ^{(3)}(\vec{r'}, \vec{r})\phi_{i}(\vec{r})\Bigg \}\Bigg \{f_{i}[v^{KLI}_{xc}(\vec{r'})-v_{xci}(\vec{r'})]\phi^*_{i}(\vec{r'})\Bigg \} +c.c.\Bigg ] 
 \\=  \sum_{i}\int d^3x' \Bigg [\Bigg \{ \phi_{i}^*(\vec{r})\phi_{i}(\vec{r'})\phi_{i}(\vec{r})\Bigg \}\Bigg \{f_{i}[v^{KLI}_{xc}(\vec{r'})-v_{xci}(\vec{r'})]\phi^*_{i}(\vec{r'})\Bigg \} +c.c.\Bigg ],
 \end{split}
\end{align}
so,
\begin{align}\label{eq:KLI3}
\begin{split}
 \sum_{i} \Bigg [\Bigg \{|\phi_{i}(\vec{r})|^2\Bigg \}\Bigg \{f_{i}[v^{KLI}_{xc}(\vec{r})-v_{xci}(\vec{r})]\Bigg \} +c.c.\Bigg ] 
 \\=  \sum_{i}|\phi_{i}(\vec{r})|^2\int d^3x' \Bigg [\Bigg \{ |\phi_{i}(\vec{r'})|^2\Bigg \}\Bigg \{f_{i}[v^{KLI}_{xc}(\vec{r'})-v_{xci}(\vec{r'})]\Bigg \} +c.c.\Bigg ],
 \end{split}
\end{align}
and,
\begin{align}\label{eq:KLI4}
\begin{split}
 v^{KLI}_{xc}(\vec{r})\sum_{i} \Bigg [\Bigg \{|\phi_{i}(\vec{r})|^2\Bigg \}\Bigg \{f_{i}\Bigg \} +c.c.\Bigg ] 
 \\= \sum_{i} \Bigg [\Bigg \{|\phi_{i}(\vec{r})|^2\Bigg \}\Bigg \{f_{i}v_{xci}(\vec{r})\Bigg \} +c.c.\Bigg ] 
 \\+  \sum_{i}|\phi_{i}(\vec{r})|^2\int d^3x' \Bigg [\Bigg \{ |\phi_{i}(\vec{r'})|^2\Bigg \}\Bigg \{f_{i}[v^{KLI}_{xc}(\vec{r'})-v_{xci}(\vec{r'})]\Bigg \} +c.c.\Bigg ],
 \end{split}
\end{align}
or,
\begin{align}\label{eq:KLI4b}
\begin{split}
 v^{KLI}_{xc}(\vec{r})\sum_{i} \Bigg \{|\phi_{i}(\vec{r})|^2\Bigg \}\Bigg \{f_{i}\Bigg \} +v^{KLI}_{xc}(\vec{r})\sum_{i} \Bigg \{|\phi_{i}(\vec{r})|^2\Bigg \}\Bigg \{f_{i}\Bigg \} 
 \\= \sum_{i} \Bigg [\Bigg \{|\phi_{i}(\vec{r})|^2\Bigg \}\Bigg \{f_{i}v_{xci}(\vec{r})\Bigg \} +c.c.\Bigg ] 
 \\+  \sum_{i}|\phi_{i}(\vec{r})|^2\int d^3x' \Bigg [\Bigg \{ |\phi_{i}(\vec{r'})|^2\Bigg \}\Bigg \{f_{i}[v^{KLI}_{xc}(\vec{r'})-v_{xci}(\vec{r'})]\Bigg \} +c.c.\Bigg ].
 \end{split}
\end{align}
This can also be written as,
\begin{align}\label{eq:KLI5}
\begin{split}
 v^{KLI}_{xc}(\vec{r})2\rho(\vec{r})
= \sum_{i} \Bigg [\Bigg \{|\phi_{i}(\vec{r})|^2\Bigg \}\Bigg \{f_{i}v_{xci}(\vec{r})\Bigg \} +c.c.\Bigg ] 
 \\+  \frac{1}{\rho(\vec{r})}\sum_{i}|\phi_{i}(\vec{r})|^2\int d^3x' \Bigg [\Bigg \{ |\phi_{i}(\vec{r'})|^2\Bigg \}\Bigg \{f_{i}[v^{KLI}_{xc}(\vec{r'})-v_{xci}(\vec{r'})]\Bigg \} +c.c.\Bigg ],
 \end{split}
\end{align}
 which can also be written as \cite{engel}
\begin{align}\label{eq:KLI6}
\begin{split}
 v^{KLI}_{xc}(\vec{r})
= \frac{1}{2\rho(\vec{r})}\sum_{i}\Bigg ( \Bigg [\Bigg \{|\phi_{i}(\vec{r})|^2\Bigg \}\Bigg \{f_{i}v_{xci}(\vec{r})\Bigg \} +c.c.\Bigg ] 
 \\+ |\phi_{i}(\vec{r})|^2\int d^3x' \Bigg [\Bigg \{ |\phi_{i}(\vec{r'})|^2\Bigg \}\Bigg \{f_{i}[v^{KLI}_{xc}(\vec{r'})-v_{xci}(\vec{r'})]\Bigg \} \Bigg ]+c.c.\Bigg).
 \end{split}
\end{align}
To calculate $v_{xc}^{KLI}$ one may start with an approximation such as the LDA approximation on the RHS. Then one can take the result of this calculation and use it on the RHS, and then continue iterating until $v_{xc}$ on the RHS and LHS after a given iteration are sufficiently close.
\bibliographystyle{apacite}
\bibliography{references1}
\end{document}